\documentclass[
amsmath,
amssymb,
aps,
pra,
reprint,
twocolumn,
superscriptaddress,
longbibliography,
showkeys,
floatfix
]{revtex4-1}

\usepackage{graphicx}
\usepackage{csquotes}
\usepackage{comment}

\usepackage{float}
\usepackage[urlcolor=blue,colorlinks=true,citecolor=blue,breaklinks]{hyperref}

\usepackage{xcolor}
\usepackage{nicefrac}
\usepackage{physics}
\usepackage{qcircuit}

\usepackage{algorithm}
\usepackage{algorithmicx}
\usepackage{algpseudocode}
\algnewcommand\algorithmicinput{\textbf{Input:}}
\algnewcommand\algorithmicoutput{\textbf{Output:}}
\algnewcommand\Input{\item[\algorithmicinput]}%
\algnewcommand\Output{\item[\algorithmicoutput]}%

\newcommand{\approptoinn}[2]{\mathrel{\vcenter{
  \offinterlineskip\halign{\hfil$##$\cr
    #1\propto\cr\noalign{\kern2pt}#1\sim\cr\noalign{\kern-2pt}}}}}

\newcommand{\appropto}{\mathpalette\approptoinn\relax}

\usepackage[english]{babel}
\makeatletter
\let\ORIbbl@fixname\bbl@fixname
\def\bbl@fixname#1{%
  \@ifundefined{languagealias@\expandafter\string#1}
    {\ORIbbl@fixname#1}
    {\edef\languagename{\@nameuse{languagealias@#1}}}%
}
\newcommand{\definelanguagealias}[2]{%
  \@namedef{languagealias@#1}{#2}%
}
\makeatother
\definelanguagealias{en}{english}

\definecolor{orange}{RGB}{255,127,0}

\newcommand{\replace}[2]{#2}

\newcommand{\defeq}{\mathrel{\mathop:}=}

\def\bra#1{\ensuremath{\mathinner{\langle{#1}|}}}
\def\ket#1{\ensuremath{\mathinner{|{#1}\rangle}}}

\def\tr{\mathrm{Tr}}

\newcommand{\eq}[1]{Eq.~\hyperref[eq:#1]{\ref*{eq:#1}}}
\renewcommand{\sec}[1]{\hyperref[sec:#1]{Section~\ref*{sec:#1}}}
\newcommand{\app}[1]{\hyperref[app:#1]{Appendix~\ref*{app:#1}}}
\newcommand{\tab}[1]{\hyperref[tab:#1]{Table~\ref*{tab:#1}}}
\newcommand{\fig}[1]{\hyperref[fig:#1]{Figure~\ref*{fig:#1}}}
\newcommand{\figa}[2]{\hyperref[fig:#1]{Figure~\ref*{fig:#1}#2}}
\newcommand{\figx}[2]{\hyperref[fig:#1]{Figure~\ref*{fig:#1}(#2)}} 
\newcommand{\thm}[1]{\hyperref[thm:#1]{Theorem~\ref*{thm:#1}}}
\newcommand{\lem}[1]{\hyperref[lem:#1]{Lemma~\ref*{lem:#1}}}
\newcommand{\cor}[1]{\hyperref[cor:#1]{Corollary~\ref*{cor:#1}}}
\newcommand{\defn}[1]{\hyperref[def:#1]{Definition~\ref*{def:#1}}}
\newcommand{\alg}[1]{\hyperref[alg:#1]{Algorithm~\ref*{alg:#1}}}

\begin{abstract}
  Contemporary quantum computers have relatively high levels of noise, making it
  difficult to use them to perform useful calculations, even with a large number
  of qubits. 
  Quantum error correction is expected to eventually enable fault-tolerant
  quantum computation at large scales, but until then it will be necessary to
  use alternative strategies to mitigate the impact of errors.
  We propose a near-term friendly strategy to mitigate errors by entangling and
  measuring \(M\) copies of a noisy state \(\rho\). 
  This enables us to estimate expectation values with respect to a state with
  dramatically reduced error, \(\rho^M/ \tr(\rho^M)\), without explicitly
  preparing it, hence the name ``virtual distillation''. 
  As \(M\) increases, this state approaches the closest pure state to \(\rho\),
  exponentially quickly. 
  We analyze the effectiveness of virtual distillation and find that it is
  governed in many regimes by the behavior of this pure state (corresponding to
  the dominant eigenvector of \(\rho\)). 
  We numerically demonstrate that virtual distillation is capable of suppressing
  errors by multiple orders of magnitude and explain how this effect is enhanced
  as the system size grows. 
  Finally, we show that this technique can improve the convergence of randomized
  quantum algorithms, even in the absence of device noise.
\end{abstract}

\begin{document}
\title{Virtual Distillation for Quantum Error Mitigation}

\author{William J.~Huggins}
\email{corresponding author: whuggins@google.com}
\affiliation{Google Quantum AI, Venice, CA 90291, United States}
\affiliation{Berkeley Quantum Information and Computation Center, Challenge Institute for Quantum Computation, and Department of Chemistry, University of California, Berkeley, CA 94720, United States}
\author{Sam McArdle}
\affiliation{Google Quantum AI, Venice, CA 90291, United States}
\affiliation{Department of Materials,
  University of Oxford, Parks Road, Oxford OX1 3PH, United Kingdom}
\author{Thomas E.~O'Brien}
\affiliation{Google Quantum AI, Venice, CA 90291, United States}
\affiliation{Instituut-Lorentz, Universiteit Leiden, 2300 RA
  Leiden, The Netherlands}
\author{Joonho Lee}
\affiliation{Department of
  Chemistry, Columbia University, New York, NY, USA}
\author{Nicholas C.~Rubin}
\affiliation{Google Quantum AI, Venice, CA 90291, United States}
\author{Sergio Boixo}
\affiliation{Google Quantum AI, Venice, CA 90291, United States}
\author{K. Birgitta Whaley}
\affiliation{Berkeley Quantum Information and Computation Center, Challenge Institute for Quantum Computation, and Department of Chemistry, University of California, Berkeley, CA 94720, United States}
\author{Ryan Babbush}
\affiliation{Google Quantum AI, Venice, CA 90291, United States}
\author{Jarrod R.~McClean}
\email{corresponding author: jmcclean@google.com}
\affiliation{Google Quantum AI, Venice, CA 90291, United States}

\date{\today}

\maketitle

\section{Introduction}

Performing meaningful calculations using near-term quantum computers is
challenging because of the relatively high error rates of these devices.
While quantum error correction promises to enable quantum computation with
arbitrarily small levels of noise, the overhead required is too large to be
currently practical~\cite{Aharonov1997-be, Fowler2012-li}. 
The most plausible paths between today's quantum computers and a fault-tolerant
device assume a modest decrease in error rates together with a large increase in
the number of qubits~\cite{Fowler2012-li, Preskill2018-po}.
We find it interesting to ask if these additional qubits can be used fruitfully
without employing the full machinery of fault-tolerance.
In this work, we explore an alternative to traditional quantum error correction
that uses multiple independently-performed copies of a computation for error
mitigation.

A variety of strategies exist to mitigate against errors on noisy
intermediate-scale quatum (NISQ) devices, i.e., to efficiently approximate the
output that would be produced in the absence of noise.
One class of approaches uses data collected at a variety of error rates to
characterize the function relating the measured value of an observable to the
error rate and extrapolate to the zero noise limit~\cite{Temme2017-hj,
  Endo2018-qs, Kandala2019-gy}. 
An alternative strategy proceeds by assuming a particular noise channel and
expressing its inverse as a quasiprobability distribution over modified copies
of the original circuit~\cite{Temme2017-hj}.
Other techniques work by comparing classically tractable simulations (tractable
because they utilize a restricted set of gates) to evaluations of the same
circuits on a noisy device~\cite{Strikis2020-ve, Czarnik2020-pb, Arute2020-ta}.
These methods aim to learn enough about the impact of the noise to predict the
noise-free expectation values for structurally similar circuits.
In Ref.~\citenum{OBrien2020-tw}, O'Brien et al.~put forward a version of quantum
phase estimation algorithm that achieves protection against errors by inverting
the state preparation procedure and verifying that the system has returned to a
reference state at the end of the computation.
Besides these methods, more specific tools have been developed for ground state
calculations~\cite{McClean2017-ta, Colless2018-af, McClean2020-pq}, for
situations when the desired state possesses certain
symmetries~\cite{Bonet-Monroig2018-od, McArdle2019-uv, Sagastizabal2019-sz,
  Huggins2019-vu, Google_AI_Quantum_and_Collaborators2020-sc}, and for treating
errors during measurement~\cite{Chen2019-sg, Maciejewski2020-ju, Bravyi2020-dn}.

Before the modern field of quantum error-correction was developed, an
alternative proposal was put forward for stabilizing quantum
computations~\cite{Berthiaume1994-ln, Barenco1997-do, Peres1999-rt}. 
The essence of this approach is to execute \(M\) redundant copies of a
computation in parallel and use a measurement to project into the symmetric
subspace between these copies. 
%
Similar measurement primitives (measurements of the swap operator and its
generalizations) have been applied to measure Renyi entanglement entropies and
other polynomial functions of the density matrix~\cite{Horodecki2002-pf,
  Ekert2002-gl, Brun2004-do, Hastings2010-tk, Islam2015-hn,
  Garcia-Escartin2013-en, Johri2017-jm, Subasi2019-dd, Banchi2016-we}.
One well-studied way to perform such a measurement is to use a Clebsch-Gordon or
Schur transform to rotate to a basis which diagonalizes the swap
operator~\cite{Bacon2006-xz}.
In Ref.~\citenum{Cotler2019-rh}, Cotler et al. 
built on these approaches to implement an idea they call ``virtual cooling.''
By performing a joint measurement on \(M\) copies of a thermal state at inverse
temperature \(\beta\) (\(\rho \propto e^{-\beta H}\)), they were able to
estimate expectation values with respect to the thermal state at inverse
temperature \(M \beta\) (\(\rho^M \propto e^{-M\beta H}\)). 
In this paper, we apply the same kind of measurement techniques to the problem
of mitigating errors in a noisy quantum computation.

Earlier work on using symmetrization to stabilize a noisy quantum computation
focused on protocols that prepared an approximately purified
state~\cite{Berthiaume1994-ln, Barenco1997-do, Peres1999-rt}.
We abandon this goal, and instead aim to reconstruct expectation values
with respect to an approximately purified state without explicitly preparing it.
We refer to this approach as virtual distillation, using the word ``virtual'' to
emphasize that we don't actually prepare a purified version of the state like a
typical distillation scheme would~\cite{Bravyi2005-vi, Knill2005-vk,
  Haah2018-tg}.
To be specific, we use collective measurements of \(M\) copies of \(\rho\)
to measure expectation values with respect to the state
\begin{equation}
  \frac{\rho^M}{\tr(\rho^M)} = \frac{\sum_{i} p_i^M \ketbra{i}}{\sum_{i} p_i^M},
  \label{eq:first_rho_M_state}
\end{equation}
where \(\rho = \sum_i p_i \ketbra{i}\) is a spectral decomposition of \(\rho\).
Under this approach, the relative weights of the non-dominant eigenvectors are
suppressed exponentially in \(M\).
This represents an improvement over approaches which demand that the
approximately purified state is prepared explicitly, which achieve a suppression
that is merely linear in \(M\) in the general case~\cite{Berthiaume1994-ln, Barenco1997-do, Peres1999-rt, Cirac1999-ir}.

Our proposed error mitigation technique offers the opportunity to make use of
additional qubits to enhance the quality of a noisy computation without the
large overhead of traditional quantum error correction. 
Furthermore, the technique is simple to use and analyze. 
If we neglect the errors that occur during measurement, it is straightforward to
obtain analytic expressions for the states whose expectation values we
effectively measure and for the variance of the resulting estimator. 
In the limit where the level of noise is small, the number of additional
measurements required by our approach goes to zero. 
Our error mitigation strategy, as we shall show, is capable of reducing the
impact of stochastic errors arising from noise on a near-term device as well as
stochastic errors inherent to randomized quantum algorithms implemented on an
error-free device.

We begin in \sec{theory} by introducing the theoretical formalism of virtual
distillation and presenting its simplest implementation.
We continue in \sec{sample_efficiency} with an analysis of the sample complexity
of the simple version of this technique along with a proof that there exist more
efficient generalizations under certain circumstances.
In \sec{analytical_performance}, we study the error mitigation performance of
virtual distillation analytically by splitting the effect of errors into two
components.
We treat the shift of the leading eigenvector of the density matrix away from
the target (error-free) state perturbatively (\sec{non_orthogonality_floor}),
and the shift of the noisy density matrix away from its dominant eigenvector
using a phenomenological model of errors (\sec{orthogonal_errors}).
Although the second effect may be exponentially suppressed by increasing the
number of states ($M$), the same is not true for the first effect, which in the
worst case limits the performance of virtual distillation to only providing a
constant-factor improvement in error rate (as a function of the underlying
physical noise rate). For purely coherent errors, this first effect is the only
consideration and virtual distillation offers no protection.
To complement this analysis, in \sec{numerical_performance} we present numerical
simulations of virtual distillation applied to various noisy quantum circuits.
We observe here that for some range of noise levels, virtual distillation
achieves a rate of error suppression exceeding the bounds suggested in
\sec{non_orthogonality_floor}.
Finally, in \sec{AlgorithmicErrors}, we consider the performance of our technique when
applied to the stochastic errors that arise during randomized algorithms for
real-time evolution.

\section{Theory}\label{sec:theory}

Virtual distillation is a protocol for using collective measurements of \(M\) copies of a state \(\rho\) to suppress incoherent errors by measuring expectation values with respect to the state \(\rho^M / \tr(\rho^M)\).
Virtual distillation approximates the error-free expectation value of \(O\) as
\begin{equation}
  \ev{O}_{\textrm{corrected}} \defeq \frac{\tr(O \rho^M)}{\tr(\rho^M)}.
  \label{eq:O_M_ev}
\end{equation}
The resulting estimator converges exponentially quickly towards the closest pure
state to \(\rho\) as \(M\) is increased.
In this section, we lay out the basic theory behind virtual distillation. 
We present the simplest implementation in \sec{measurement_by_diagonalization}
and an analysis of the measurement overhead in \sec{sample_efficiency}. 
In \alg{basic_vd} below, we present pseudocode for the implementation discussed
in more detail in \sec{measurement_by_diagonalization}.

We begin by establishing some assumptions and notation. 
Throughout this paper we deal with operations that act on multiple copies of the
same state.
We make the assumption that the noise experienced by the separate copies has the
same form and strength. 
If we relax this assumption, then we still measure an effective state that
corresponds to the product of the density matrices of the individual copies so
long as the copies are not entangled prior to virtual distillation. 
We briefly explore this more general situation in \app{different_noise_models}.

We use the letter \(N\) to indicate the number of qubits in an individual system
and the letter \(M\) to indicate the number of copies (which we sometimes refer
to as subsystems).
Superscripts with parentheses indicate an operator that acts on multiple
systems. 
For example, we shall denote the cyclic shift operator between \(M\) copies by
\(S^{(M)}\).
We use bolded superscripts without parentheses to denote which copy an operator
acts on, e.g., \(O^\textbf{1}\) indicates the operator \(O\) acting on subsystem
\(1\).
We use superscripts without a bold-faced font or parentheses to indicate
exponentiation as usual.
Subscripts are used in two different ways. 
Subscripts on an operator generally indicate which qubit within a system the
operator acts on. 
The exception is when the subscript is being used more generically as an index
in a summation, which should always be clear from the context and the presence
of the \(\sum\) symbol.

\begin{algorithm}[H]
  \caption{Virtual distillation, basic implementation (see
    \sec{measurement_by_diagonalization})}
  \label{alg:basic_vd}
  \begin{algorithmic}[0]
    \Input{ A number of measurement repetitions \(K\), \(2K\) copies of the
      \(N\) qubit state \(\rho\) (provided two at a time).}
    \Output{An error-mitigated estimate of \(\ev{Z_i}\) for each qubit in \(\rho\);
      \(\ev{Z_i}_{\textrm{corrected}} \approx \frac{\tr(Z_i
        \rho^2)}{\tr(\rho^2)}\).}
    \\
    \State Set \(E_i = 0\) for each qubit \(i \in 1..N\).
    \State Set \(D = 0\).
    \For{\(k \in 1..K\) }
    \State Perform any SWAP operations necessary to make it possible to couple
      each qubit in the first copy of \(\rho\) with the corresponding qubit in
      the second copy.
    \State Apply the two-qubit gate \(B^{(2)}_i\) (defined below in \eq{beamsplitter_gate_matrix} of
      \sec{measurement_by_diagonalization}) between each qubit \(i\) in the first copy and the corresponding
      qubit in the second copy.
    \State Measure both states in the computational basis. 
    \State Let \(z_i^\textbf{1}\) and \(z_i^\textbf{2}\) denote the measurement
outcomes for the \(i\)th qubits in the first and second copies of \(\rho\)
respectively.
\For{\(i \in 1..N\)}
\State \(E_i \mathrel{+\!\!=} \frac{1}{2^N}\big( z_i^\textbf{1} + z_i^\textbf{2}
\big)\prod_{j \neq i} 1 + z_j^\textbf{1} - z_j^\textbf{2} +
z_j^\textbf{1}z_j^\textbf{2}\)
\EndFor
\State \(D \mathrel{+\!\!=} \frac{1}{2^N}
\prod_{j=1}^{N} 1 + z_j^\textbf{1} - z_j^\textbf{2} +
z_j^\textbf{1}z_j^\textbf{2}\)
\EndFor
\State \Return \(\big\{ \ev{Z_i}_{\textrm{corrected}} \defeq \frac{E_i}{D} \big\} \)
\end{algorithmic}

\end{algorithm}

\begin{figure}[t]
  \includegraphics[width=.48\textwidth]{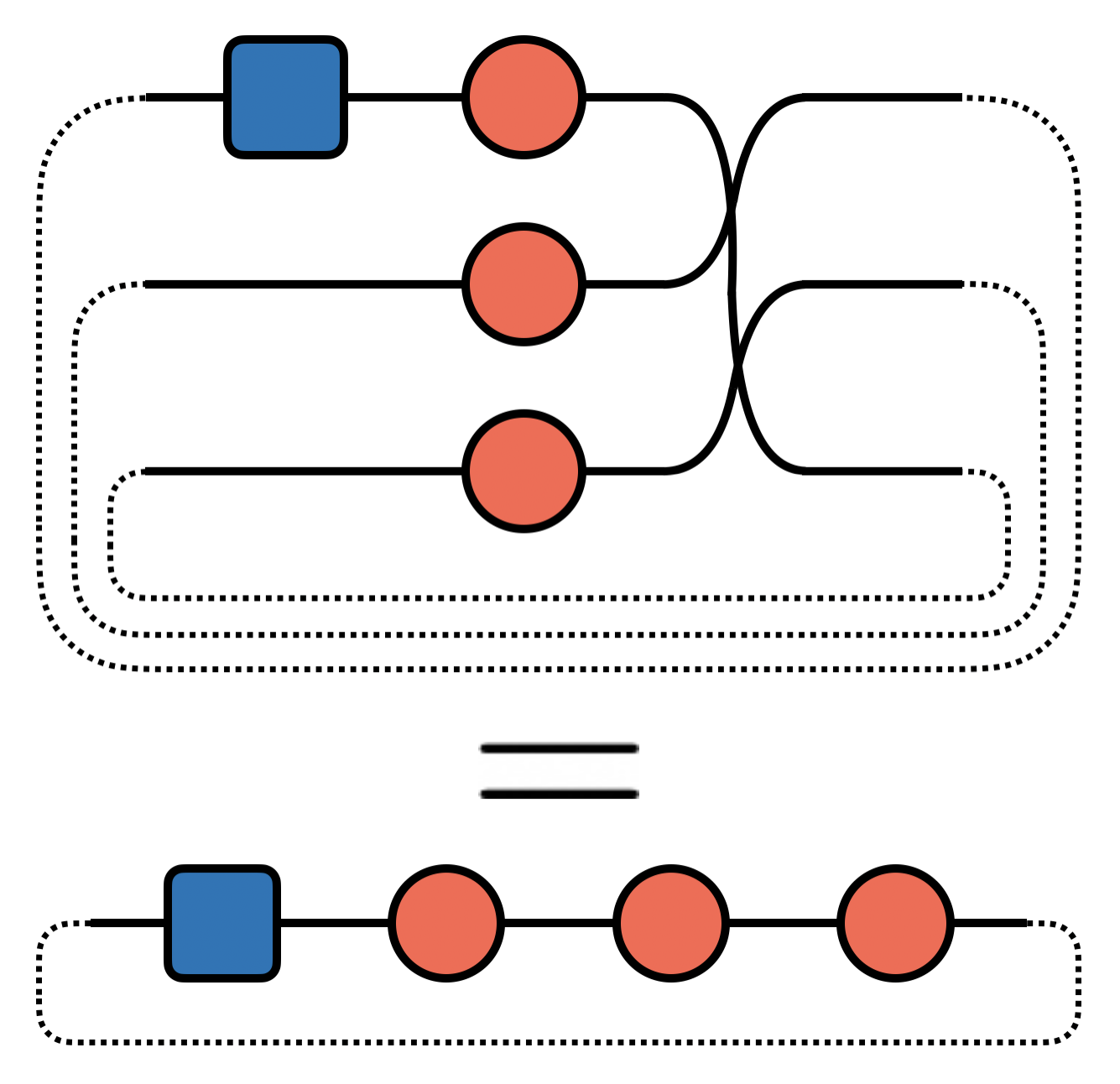}
  \caption{A diagrammatic representation of \eq{fundamental_trace_equality} with
    \(M=3\) and \(i=1\) using tensor network notation~\cite{Biamonte2017-zx,
      Bridgeman2017-kg, Orus2014-xx}. 
    The blue square represents the operator \(O^{\textbf{1}}\), each red circle
    represent a copy of the state \(\rho\), and the connections between the
    shapes indicate indices which are summed over. 
    The cyclic shift operator \(S^{(3)}\) is naturally represented as a product
    of two swap operators, which are themselves indicated by the crossed wires. 
    Note that the top diagram actually corresponds to the expression
    \(\tr(O^{\textbf{1}} \rho^{\otimes 3} S^{(3)})\); we commuted \(\rho^{\otimes
      3}\) with \(S^{(3)}\) before producing the figure. 
    Rearranging the wires to yield the bottom diagram is equivalent to the
    simplification of the summation in \eq{cyclic_shift_proof}.}
  \label{fig:cyclic_shift_trace_diagram}
\end{figure}

In order to evaluate the numerator and denominator of \eq{first_rho_M_state}, we can make
use of the following equality~\cite{Ekert2002-gl, Brun2004-do, Cotler2019-rh},
\begin{equation}
  \tr(O \rho^M) = \tr(O^\textbf{i} S^{(M)} \rho^{\otimes M}).
  \label{eq:fundamental_trace_equality}
\end{equation}
Here, \(O^\textbf{i}\) indicates the observable \(O\) acting on (an arbitrary)
subsystem \(i\) and \(S^{(M)}\) indicates the cyclic shift operator on \(M\)
systems, i.e.,
\begin{align}
  & O^\textbf{i} \defeq \mathbb{I} \otimes \mathbb{I} \cdots O \cdots \mathbb{I}, \nonumber \\
  & S^{(M)}\ket{\psi_1} \otimes \ket{\psi_2} \cdots \ket{\psi_M} \defeq
    \ket{\psi_2}\otimes \ket{\psi_3} \cdots \ket{\psi_{1}}.
\end{align}
This identity can be proven by expanding the right-hand side, carefully keeping
track of the indices. 
Without loss of generality we choose \(i = 1\), yielding
\begin{gather}
  \tr(O^\textbf{1} S^{(M)} \rho^{\otimes M}) = \nonumber \\
  \sum_{i_1, i_2, ...i_M, j_1, j_2, \cdots j_M, k} O_{k, j_1} \delta_{j_2, i_1}
  \cdots \delta_{j_1, i_M}
  \rho_{i_1, k} \cdots \rho_{i_M, j_M} = \nonumber \\
  \label{eq:cyclic_shift_proof}
  \sum_{i_1, i_2, ...i_M, k}
  \rho_{i_1, k}  O_{k, i_M} \rho_{i_{M}, i_{M-1}} \cdots \rho_{i_2, i_1} = \\
  \tr(O \rho^M). 
  \nonumber
\end{gather}
In \fig{cyclic_shift_trace_diagram} we present a diagrammatic representation of
\eq{fundamental_trace_equality} for the case where \(M=3\) (note that we have
commuted \(\rho^{\otimes 3}\) with \(S^{(3)}\) in the diagram).

\begin{figure}[t]
  \includegraphics[width=.48\textwidth]{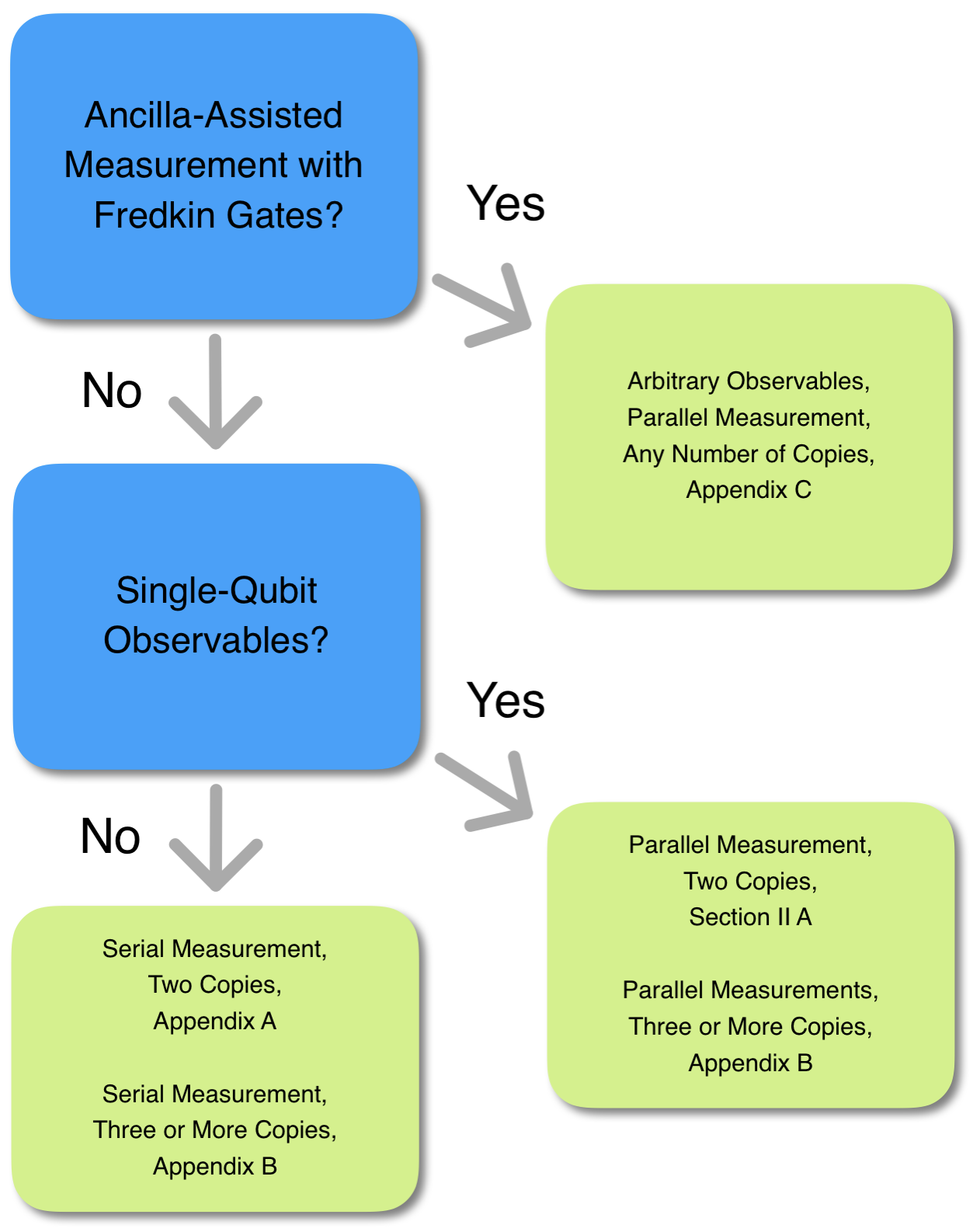}
  \caption{A flowchart that describes the choices involved in selecting between
    the different variants of virtual distillation presented in this work. 
    Blue boxes denote questions for the experimentalist to answer about the
    available quantum resources and problem to be studied, green boxes link to
    the relevant sections in the text and briefly summarize the main features of
    each variant. 
    The flowchart provides direction to the most flexible variant given the
    answers provided in the blue boxes but the actual experimental performance
    will depend on many factors.}
  \label{fig:algorithm_flowchart}
\end{figure}

The quantities in the numerator and denominator of \eq{O_M_ev} can be evaluated
in a number of different ways. 
For simplicity, we focus our presentation one such approach
\sec{measurement_by_diagonalization}.
In that section, we roughly follow the work of Ref.~\citenum{Cotler2019-rh},
except that we use the language of qubits rather than bosonic systems.
We discuss a variety of alternative protocols in \app{diagonalize_multi_qubit},
\app{diagonalize_three_or_more}, and \app{ancilla_assisted_measurement}. 
\fig{algorithm_flowchart} summarizes the differences between these variants.
The practical utility of these techniques as error-mitigation tools will be
partly determined by the number of samples necessary to evaluate the corrected
expectation values to within some target precision \(\epsilon\).
We address this issue in \sec{sample_efficiency} and also show that their exists
generalizations of our approach that can further reduce the number of circuit
repetitions for a desired precision.

\subsection{Measurement by Diagonalization}
\label{sec:measurement_by_diagonalization}

\begin{figure}[t]
  \centering
  \includegraphics[width=.48\textwidth]{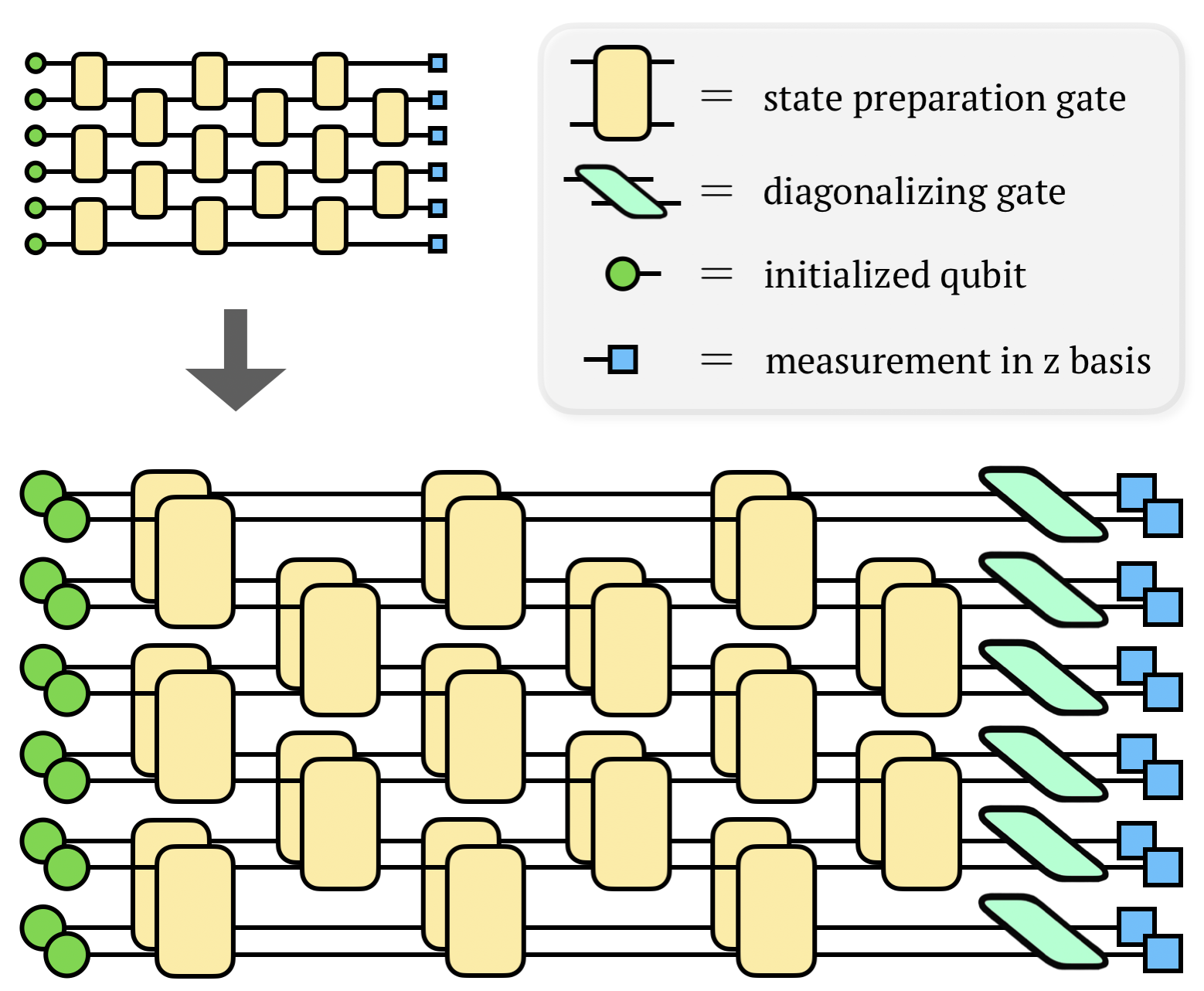}
  \caption{A circuit diagram of our approach applied to a six-qubit circuit.
    We use twice the number of qubits to independently perform two copies of the
    original circuit.
    We then apply a single layer of the two-qubit gates specified in
    \eq{beamsplitter_gate_matrix} before measuring each qubit in the
    computational basis. 
    This allows us to estimate the error-mitigated expectation values for all
    single-site \(Z\) operators.}
  \label{fig:circuit_diagram_transformation}
\end{figure}

In this section, we present a straightforward strategy applicable when the
operator \(O\) is the Pauli \(Z\) operator acting on a single qubit and \(M=2\).
Other single-qubit observables can be accessed by applying the appropriate
single-qubit rotations before the virtual distillation procedure.
This realization of our error mitigation technique requires only a single
additional layer of two-qubit gates followed by measurement in the computational
basis. 
We present a schematic of this approach in \fig{circuit_diagram_transformation}.

Rather than using the relation in \eq{fundamental_trace_equality} directly, we
instead define a symmetrized version of our observable,
\begin{equation}
  O^{(M)} = \frac{1}{M} \sum_{i=1}^M O^\textbf{i}.
  \label{eq:symmetrized_O}
\end{equation}
For the specific case we consider here, that means we take
\begin{equation}
  Z^{(2)}_k = \frac{1}{2}(Z^\textbf{1}_k + Z^\textbf{2}_k).
\end{equation}
It is straightforward to use \eq{fundamental_trace_equality} to show that
\begin{equation}
  \frac{\tr(O \rho^M)}{\tr(\rho^M)} = \frac{\tr(O^{(M)} S^{(M)} \rho^{\otimes M})}{\tr(S^{(M)} \rho^{\otimes M})}.
  \label{eq:symmetrized_O_M_ev}
\end{equation}
Using the symmetrized observable is advantageous because
\begin{equation}
  [O^{(M)}, S^{(M)}] = 0,
\end{equation}
or, in our case, \([Z^{(2)}_k, S^{(2)}] = 0\).

Both \(S^{(2)}\) and \(Z^{(2)}_k\) factorize into tensor products of operators
that act separately on each pair of qubits, where the \(i\)th pair consists of
the \(i\)th qubit from each system.
Therefore, we may simultaneously diagonalize \(S^{(2)}\) and \(Z^{(2)}_k
S^{(2)}\) using an operator that factorizes with the same structure. 
We denote the two-qubit unitary that performs this diagonalization on the
\(i\)th pair \(B^{(2)}_i\).
We give a matrix representation for this gate below, noting that
there is some freedom in the choice of phases for the matrix elements,
\begin{equation}
  B^{(2)}_i \defeq
  \begin{bmatrix}
    1 & 0 & 0 & 0 \\
    0 & \frac{\sqrt{2}}{2} & -\frac{\sqrt{2}}{2} & 0 \\
    0 & \frac{\sqrt{2}}{2} &  \frac{\sqrt{2}}{2} & 0 \\
    0 & 0 & 0 & 1
  \end{bmatrix}.
  \label{eq:beamsplitter_gate_matrix}
\end{equation}
We then define
\begin{equation}
  B^{(2)} \defeq \bigotimes_{i=1}^M B^{(2)}_i.
\end{equation}

As desired, this unitary diagonalized the individual factors that make up the
observables,
\begin{align}
  B^{(2)} S^{(2)}_i B^{(2) \dagger} \rightarrow& \frac{1}{2}(1 + Z_i^\textbf{1} - Z_i^\textbf{2} + Z_i^\textbf{1}Z_i^\textbf{2}), \\
  B^{(2)} Z^{(2)}_k S^{(2)}_k B^{(2) \dagger} \rightarrow& \frac{1}{2}(Z_k^\textbf{1} + Z_k^\textbf{2}).
\end{align}
This diagonalization is particularly easy to implement when each qubit from the
first copy of \(\rho\) is adjacent to the corresponding qubit from the second
copy. 
The procedure for measuring the observables required to estimate the numerator
and denominator of \eq{symmetrized_O_M_ev} then reduces to applying a single
layer of \(N\) two-qubit gates in parallel and measuring in the computational
basis. 
In fact, because \(B^{(2)}\) diagonalizes \(Z^{(2)}_k S^{(2)}\) for all \(N\)
values of \(k\), we naturally collect the data required to estimate the
error-mitigated expectation values for all \(N\) of the operators \(Z_k\)
simultaneously. By applying the appropriate single-qubit rotations before performing virtual distillation, we could instead access an arbitrary single-qubit observable on each qubit.
We capture this process diagrammatically in
\fig{circuit_diagram_transformation}.

In order to develop some intuition, it is helpful to express
\(\rho^{\otimes 2}\) using a spectral decomposition of \(\rho\) and consider two
separate components of the resulting sum,
\begin{align}
  \rho^{\otimes 2} &= \sum_{ij} p_i p_j \ketbra{i}\otimes\ketbra{j} \\
                   &=
                     \sum_{i} p_i^2 \ketbra{i} \otimes \ketbra{i} + \sum_{i \neq j} p_i p_j \ketbra{i} \otimes \ketbra{j}.
                     \nonumber
\end{align}
The calculation of measurement probabilities and expectation values is a linear
operation on the density matrix; we can therefore consider these two components
separately.
The component of the state with \(i=j\) is in the \(+1\) eigenspace of
\(S^{(2)}\) and leads to measurements of \(S^{(2)}\) which yield the \(+1\)
eigenvalue with probability \(p = \sum_{i} p_i^2 = \tr(\rho^2)\). 
In the case where \(i \neq j\), \(\ketbra{i}\otimes\ketbra{j}\) is an even
superposition of symmetric and anti-symmetric states,
\begin{equation}
  \ket{i}\ket{j} = 
  \frac{1}{2}
  \big(\ket{i}\ket{j} + \ket{j}\ket{i}\big) +
  \frac{1}{2}
  \big(\ket{i}\ket{j} - \ket{j}\ket{i}\big).
\end{equation}
For this component of the state, measurements of \(S^{(2)}\) yield \(+1\) and
\(-1\) with equal probability and \(\ev{S^{(2)}} = 0\). 
Combining these two cases, we have the expected equality, \(\tr(S^{(2)}
\rho^{\otimes 2}) = \tr(\rho^2)\). 
Measurements of \(S^{(2)}O^{(2)}\) follow a similar pattern.

We find it interesting to contrast this behavior with the stabilizer theory of
quantum error correction. 
In the stabilizer formalism, errors are detected by projecting through measurement into the \(-1\)
eigenspace of one or more symmetries. 
In our approach, we instead rely on errors being equally supported on the
eigenspaces of the symmetry we measure.

\subsection{Sample Efficiency}
\label{sec:sample_efficiency}

The number of circuit repetitions required to determine the error-mitigated
expectation values within a precision \(\epsilon\) depends on the variance of our
estimator.
In this section, we present expressions for this variance.
We focus on the \(M=2\) case and the methods discussed in
\sec{measurement_by_diagonalization}.
The calculations are also applicable to the variant protocol we present in
\app{ancilla_assisted_measurement}.
We also show how their exists an extension to our protocol that makes more
efficient use of multiple copies when the noise level is sufficiently high.

We'd like to determine the variance of our estimator for the error-mitigated expectation value
\begin{equation}
  \label{eq:error_mitigated_ev_1}
  \ev{O}_\textrm{corrected} = \frac{\tr(O^{(2)} S^{(2)} \rho^{\otimes 2})}{\tr(S^{(2)} \rho^{\otimes 2})}.
\end{equation}
We leave the derivation to \app{estimator_variance} and
simply give an (approximate) expression for the variance,
\begin{align}
  & \textrm{Var}(\ev{O}_\textrm{corrected}) \approx \nonumber \\
  &\frac{1}{R}\Big(\frac{1}{\tr(\rho^2)^2}\big(\frac{1}{2}\tr(\rho O^2) + \frac{1}{2}\tr(\rho O)^2 - \tr(\rho^2 O)^2\big)
    \nonumber \\
  \label{eq:rho_2_estimator_variance_intro}
  &- 2 \frac{\tr(\rho^2 O)}{\tr(\rho^2)^3}\big(\tr(\rho O) - \tr(\rho^2 O)\tr(\rho^2)\big)
  \\
  &+ \frac{\tr(\rho^2 O)^2}{\tr(\rho^2)^4}\big(1 - \tr(\rho^2)^2\big)\Big),
    \nonumber
\end{align}
where \(R\) refers to the number of measurement repetitions. 
It's useful to consider what happens in the limit where \(\rho\) is a pure
state. 
In that case, the second and third lines are zero and the variance reduces to
\begin{equation}
  \textrm{Var}(\ev{O}_\textrm{corrected}) = \frac{1}{2R}\big(\tr(\rho O^2) - \tr(\rho O)^2 \big),
\end{equation}
exactly what one would expect when averaging \(2R\) independent measurements of
\(O\). 
As the purity of \(\rho\) decreases, the variance, and the number of circuit
repetitions, increases.

The rest of this section focuses on laying the groundwork to improve the sample
efficiency of these techniques. 
This is an important goal because the number of samples required can grow large
given sufficiently noisy circuits. 
At high enough error rates, we are highly likely to find ourselves in a
situation where
\begin{equation}
  \tr(\rho^3) \ll \tr(\rho^2) \ll 1.
\end{equation}
We now make the assumption that the level of error mitigation offered by
measuring \(\rho^M\) is sufficient but we have \(2K \gg M\) copies of \(\rho\)
available. 
For simplicity, we focus on the case where \(M=2\) and \(O\) is a Pauli operator
acting on one or more qubits.
We present a generalization of our approach involving a collective measurement
of all \(2K\) copies of \(\rho\) that performs better than a naive
parallelization.

The naive approach we hope to beat consists of taking \(K\) pairs and running the protocol described above in parallel, averaging
the results. 
For simplicity, we focus on the variance of our
estimator for the quantity that appears in the numerator of
\eq{symmetrized_O_M_ev} rather than the ratio itself.
In \app{estimator_variance} we show that the variance of our estimator for
\(S^{(2)}O^{(2)}\) is \(\frac{1}{2}\tr(\rho O^2) + \frac{1}{2}\tr(\rho O)^2 -
\tr(\rho^2 O) ^2 \). 
Therefore, the variance obtained when using \(2K\) copies in parallel is exactly
\begin{equation}
  \textrm{Var}(\ev{S^{(2)}O^{(2)}}) = \frac{1}{2K} \big(\tr(\rho O^2) + \tr(\rho O)^2 - 2 \tr(\rho^2 O)^2\big).
\end{equation}
We prove below that it is possible in some situations to obtain a more
sample-efficient estimator for the corrected expectation value by performing a
joint measurement on all \(2K\) copies. 
We do so by providing an operator \(\tilde{O}\) with the desired
expectation value and calculating its variance.

First, we define the operator
\begin{equation}
  \tilde{O} = \frac{1}{\binom{2K}{2}}\sum_{i=1}^{2K} \sum_{j > i}\frac{1}{2}( O^\textbf{i} + O^\textbf{j})S^{{(i, j)}},
\end{equation}
where we use \(S^{(i, j)}\) to denote the swap operator specifically between subsystems \(i\)
and \(j\).
It is simple to show that
\begin{equation}
  \tr(\tilde{O} \rho^{\otimes 2K}) = \tr(O \rho^2).
\end{equation}
We compute the variance of \(\tilde{O}\) with respect to the state
\(\rho^{\otimes 2K}\) in \app{collective_measurement_variance}, finding that
\begin{equation}
  \label{eq:variance_inequality_win}
  \textrm{Var}(\ev{\tilde{O}}) \leq \frac{1 + 7 (K-1) \tr(\rho^3)}{K(2K-1)}.
\end{equation}

When \(\tr(\rho^3)\) is small, the second term in \eq{variance_inequality_win} is suppressed and there is a
regime where the variance of this operator shrinks quadratically with \(K\).
The naive approach, where we perform \(K\) independent
calculations on separate pairs results in an estimator whose variance is
suppressed only linearly in \(K\).
We do not suggest a particular strategy, let alone one that is NISQ-friendly,
for implementing the measurement of \(\tilde{O}\).
We hope that future work can address this issue.
Furthermore, while we have established that generalizations of the simplest
virtual distillation procedure can outperform a naive parallel strategy, we have
not established a comprehensive theory on the limitations of virtual
distillation.
It would be useful to quantify the minimum number of samples required to resolve
\(Tr(O\rho^M)\) given access to a large number of copies of \(\rho\) under various
assumptions about the spectrum of the density matrix.

\section{Performance Under Different Noise Models}
\label{sec:analytical_performance}
In the numerical studies, we will present evidence that the performance of virtual
distillation can be essentially predicted by the combination of two
contributions. Here we find it instructive to consider them separately using
simple analytical models. 
To understand the potential benefit 
of our approach using the minimal setup, we consider the
fidelity of
\begin{align}
  \rho_\textrm{corrected} \defeq \frac{\rho^2}{\tr( \rho^2 )}
\end{align}
with the ideal state generated by noiseless evolution (neglecting error
introduced by the measurement procedure). 
We first consider the performance under noise that maps the ideal state to
states orthogonal to it, leaving the dominant eigenvector of the density matrix
as the ideal state.
We then turn towards the effect of errors that lead to states non-orthogonal to
the ideal state, causing a drift in the dominant eigenvector of the density
matrix.
The essential behavior of virtual distillation is to remove errors of the first
kind rapidly, while converging to a floor determined by the drift in the
dominant eigenvector that enables a large constant factor improvement over the
erred state.

\subsection{Orthogonal Errors}
\label{sec:orthogonal_errors}
We first consider idealized errors that leave the dominant eigenvector as the
ideal state.
We consider a phenomenological error model motivated by the assumption that we
can think of errors as discrete events that occur locally in space and time with
some probability. 
For simplicity, we model every gate as a stochastic quantum map where with
probability $p$ an error occurs, and we assume that every new error sends the
quantum evolution to a new orthogonal state. 
The resulting density matrix for a circuit with $G$ gates is
\begin{align}
\rho &= (1-p)^G\rho_0 + (1-p)^{G-1}p \sum_{j=1}^G \rho_{j_1} \nonumber \\
&+ (1-p)^{G-2}p^2\sum_{j_1 \ne j_2} \rho_{j_1,j_2}\nonumber \\
&+ (1-p)^{G-3}p^3\sum_{j_1 \ne j_2 \ne j_3} \rho_{j_1,j_2,j_3} + \ldots
\end{align}
The operator for $\rho^2$ is similar with all the coefficients squared, as all
the states are assumed to be orthogonal. 
Therefore,
\begin{align}
    \tr(\rho^2) = ((1-p)^2 + p^2)^G\;.
\end{align}
The fidelity with the ideal state $\rho_0$ is 
\begin{align}
    \frac{\tr( \rho_0 \rho^2 )}{\tr( \rho^2 )} &= \frac{(1-p)^{2 G}}{((1-p)^2 + p^2)^G} \\
    &\simeq 1 - G p^2 + O( G p^3 )\;.
\end{align}
Therefore we expect a quadratic suppression of errors in the most favorable case. 

The result is similar in the case of $M$ copies:
\begin{align}
    \frac{\tr( \rho_0 \rho^M )}{\tr( \rho^M )} &= \frac{(1-p)^{M G}}{((1-p)^M + p^M)^G} \\
&\simeq 1 - G p^M + O( Gp^{M+1} )\;.
\end{align}

The other factor affecting the performance of virtual distillation besides the fidelity is the sample complexity. We analyze the general case in more detail in \app{estimator_variance}, but it is instructive to briefly consider the performance under this simplified model of errors.
For simplicity, we assume that we aim to measure the error-mitigated expectation value of a Pauli operator \(O\) at the \(M = 2\) level using \(R\) independent experiments to estimate the numerator and denominator of \eq{error_mitigated_ev_1} (for a total of \(2R\) experiments). Then the variance of our estimators for the numerator and denominator are upper bounded by \(1\), and we have
\begin{equation}
    \textrm{Var}(\ev{O}_\textrm{corrected}) \lessapprox \frac{1}{R} \Big( \frac{1}{\tr(\rho^2)^2} + \frac{\tr(O \rho^2)^2}{\tr(\rho^2)^4}\Big).
\end{equation}
Because of our assumption that \(O\) is a Pauli operator, and therefore \(||O|| = 1\), we have \(\tr(O\rho^2) \leq \tr(\rho^2)\), implying \(\tr(O\rho^2)^2 \leq \tr(\rho^2)^2\).
Therefore, 
\begin{equation}
    \textrm{Var}(\ev{O}_\textrm{corrected}) \lessapprox  \frac{2}{R\tr(\rho^2)^2} = \frac{2}{R((1-p)^2 + p^2)^{2G}}.
\end{equation}
When \(p\) is small, we can neglect the \(p^2\) term in the denominator. Therefore,
taking
\begin{equation}
    R \appropto (1-p)^{-4G}
\end{equation}
is sufficient to estimate \(\ev{O_\textrm{corrected}}\) to within a fixed additive error.

\subsection{Non-Orthogonal Error Floor}
\label{sec:non_orthogonality_floor}
The analysis of the previous section made the simplifying assumption that the
dominant eigenvector of the density matrix, \(\rho_0 = \ketbra{0}\), corresponds
exactly to the ideal state generated by noiseless evolution. 
In practice, errors will lead to population in states that may not be orthogonal
to the target state, leading to a drift in the dominant eigenvector of the
density matrix. 
We will see in our numerical studies that this drift limits the maximum
potential upside of virtual distillation.
In this second, we develop an understanding of this drift by using perturbation
theory to consider the first-order change in the dominant eigenvector of the
density matrix.

Let us consider a state $\rho$ in the middle of a noisy preparation circuit,
allowing for $\rho$ to already be somewhat distorted by noise.
Writing $\rho$ in its eigenbasis, we have
\begin{equation}
  \rho = \sum_i \lambda_i \ketbra{i},
\end{equation}
where we order the eigenvalues in descending order.
Note that we use the symbol \(\lambda_i\) for the \(i\)th eigenvector of the density matrix rather than \(p_i\) throughout this section, reserving the symbol \(p\) for the coefficients associated with a Kraus operator decomposition of our noise channel.
We wish to consider the impact of a subsequent noise channel defined in terms of
a set of Kraus operators,
\begin{equation}
  \rho \rightarrow p_0 \rho + \sum_{j \neq 0} p_j K_j \rho K_j^\dagger.
\end{equation}
Note that we have demanded a representation of the channel where \(K_0\) is the
identity matrix in order to simplify our analysis.
Now let \(\Delta V\) denote the change in the density matrix induced by this channel
(\(\rho \rightarrow \rho + \Delta V\)),
\begin{equation}
  \Delta V \defeq (p_0 - 1) \rho + \sum_{j \neq 0} p_j K_j \rho K_j^\dagger,
\end{equation}
where we define the scale \(\Delta\) by taking \(||V||\) to be \(O(1)\). 

Now we make the assumption that we are in the low-error regime. Specifically, we assume that \(\lambda_0 \gg \lambda_1\) and that \(\Delta \ll |\lambda_0 - \lambda_1|\). Under this assumption, we satisfy the necessary conditions for applying matrix perturbation theory to the dominant eigenvector~\cite{Kato2013-tz}. We can therefore proceed by expressing the dominant eigenvector of \(\rho + \Delta V\) as a convergent power series in \(\Delta\).
This yields
\begin{equation}
      \ket{0} = \ket{0^{(0)}} + \Delta \ket{0^{(1)}} +\Delta^2 \ket{0^{(2)}} + O(\Delta^3), \\
\end{equation}
where \(\ket{0}\) denotes the dominant eigenvector of \(\rho + \Delta V\), \(\ket{0^{(0)}}\) denotes the dominant eigenvector of the unperturbed \(\rho\), and \(\ket{0^{(i)}}\) denotes the correction at \(i\)th order.
Likewise, we can also express the eigenvalue corresponding to the dominant eigenvector as a power series in \(\Delta\),
\begin{equation}
    \lambda_0 = \lambda_0^{(0)} + \Delta \lambda_0^{(1)} +  \Delta^2 \lambda_0^{(2)} +  O(\Delta ^3).
\end{equation}

We can then proceed in the usual way, expanding the eigenvalue equation,
\begin{equation}
    (\rho + \Delta V) \ket{0} = \lambda_0 \ket{0},
\end{equation}
and equating terms order by order. This leads to a familiar expression for the first order correction to the dominant eigenvector in terms of the zeroth order eigenvalues and eigenvectors,
\begin{equation}
  \ket{0^{(1)}} = \sum_{i \neq 0} \frac{\bra{i^{(0)}} V \ket{0^{(0)}}}{\lambda^{(0)}_0 - \lambda_i^{(0)}} \ket{i^{(0)}}.
\end{equation}
At this point, it's useful to carefully consider the normalization of \(\ket{0}\). Let \(\ket{D}\) denote the normalized form of \(\ket{0}\),
\begin{align}
    \ket{D} \defeq& \frac{\ket{0^{(0)}} + \Delta \ket{0^{(1)}} + \Delta^2 \ket{0^{(2)}} + O(\Delta^3) }{\sqrt{1 + \Delta^2 \braket{0^{(1)}} + O(\Delta^3)}} \nonumber \\
    =& \ket{0^{(0)}} + \Delta \ket{0^{(1)}} \nonumber \\ &+ \Delta^2 \ket{0^{(2)}} - \frac{\Delta^2}{2} \braket{0^{(1)}} \ket{0^{(0)}} + O(\Delta^3),
\end{align}
where we have made use of the fact that the first and second order corrections are both orthogonal to the unperturbed eigenvector.

We can now
compute the trace distance between \(\ket{D}\) and the dominant eigenvector of the unperturbed state,
\begin{widetext}
\begin{align}
  T(&\ket{D}, \ket{0^{(0)}}) \nonumber \\ &= \frac{1}{2} \tr\Big(\sqrt{(\Delta \ketbra{0^{(0)}}{0^{(1)}} + \Delta\ketbra{0^{(1)}}{0^{(0)}} + \Delta^2 \ketbra{0^{(0)}}{0^{(2)}} + \Delta^2\ketbra{0^{(2)}}{0^{(0)}} - \Delta^2 \braket{0^{(1)}} \ketbra{0^{(0)}} + O(\Delta^3))^2} \Big)\nonumber\\
                      &= \frac{1}{2} \tr\Big(\sqrt{\Delta^2 \braket{0^{(1)}}\ketbra{0^{(0)}} + \Delta^2\ketbra{0^{(1)}} + O(\Delta^3)}\Big)\nonumber\\
                      &= \frac{1}{2} \tr\Big(\sqrt{\Delta^2\braket{0^{(1)}}\ketbra{0^{(0)}} + \Delta^2\braket{0^{(1)}} \frac{\ketbra{0^{(1)}}}{\braket{0^{(1)}}} + O(\Delta^3) }\Big) \nonumber\\
                      &= \Delta \sqrt{\frac{\braket{0^{(1)}}}{2}} + O(\Delta^2).
\end{align}
\end{widetext}
Now let us expand \(\braket{0^{(1)}}\) in terms of the Kraus operators of our noise
model. 
\begin{align}
  \braket{0^{(1)}} &= \sum_{i\neq 0} \frac{1}{(\lambda_0 - \lambda_i)^2}\bra{0}V^\dagger\ketbra{i}V\ket{0} \nonumber \\
             &= \sum_{i \neq 0}\frac{1}{(\lambda_0 - \lambda_i)^2} \nonumber \\
             &\;\;\;\;\;\;\;
               \Big|\bra{i}
               \Big((p_0 - 1)\mathbb{I} + \sum_{j \neq 0}p_j K_j \ketbra{0} K_j^\dagger \Big)
               \ket{0} \Big|^2
               \nonumber \\
             &= \sum_{i \neq 0}\frac{1}{(\lambda_0 - \lambda_i)^2}
               \Big|\sum_{j \neq 0} p_j \bra{i}
               K_j \ketbra{0} K_j^\dagger 
               \ket{0}\Big|^2,
\end{align}
where we omit the \((0)\) superscripts of the eigenvalues and eigenvectors on the right-hand side for readability.

We can see that, in the general case, we expect a non-zero contribution to the
trace distance at first order in \(\Delta\). 
Because \(\rho^2 / \tr(\rho^2) \approx \ketbra{D}\) in the low-noise regime,
this will effectively set a floor for how well our method can correct errors. 
Therefore, without further constraints on the state, the noise model, or the
observables being measured, our method will not achieve a quadratic suppression
in errors in the low noise limit but rather a constant factor improvement whose
magnitude depends on the typical size of a quantity we denote by the symbol \(\gamma\), 
\begin{equation}
  \gamma \defeq \Big|\sum_{j \neq 0} p_j \bra{i}
  K_j \ketbra{0} K_j^\dagger 
  \ket{0}\Big|.
  \label{eq:thing_that_must_be_small}
\end{equation}

Interestingly, when we examine the data from our numerical simulations, we do
obtain an improvement consistent with a quadratic suppression of errors at
intermediate error rates.
Additionally, \(\gamma\) has no lower bound;
it can in some cases be zero, in which case we expect to recover the quadratic
suppression of error predicted from \sec{orthogonal_errors}.
As the trace distance is an upper bound for the error in any observable,
particular observables of particular states may recover this performance even
when \(\gamma \neq 0\).

In order to shed some light on the error floor set by the drift in the dominant eigenvector of the density matrix, it can be helpful to ask when we might expect \(\gamma\) to be near zero.
It is clear that this quantity must be zero if \replace{
\(K_j\ket{0} \propto \ket{0}\) or \(\ev{K_j^\dagger}{0} = 0\) for all \(j\).}{one of two conditions hold:}
\begin{align}
    K_j\ket{0} &\propto \ket{0}\label{eq:quad_condition_1}\\
    \ev{K_j^\dagger}{0} &= 0\label{eq:quad_condition_2}.
\end{align}
One way that this can occur is if the state and the circuit have a natural set of symmetries. The first condition holds if the error is drawn from such a symmetry group, while the second is satisfied if it violates it strictly.
For an example of the second case, consider a bit-flip or amplitude-damping error channel acting on a state
with a definite number of excitations. 
There are other situations where the second equality is approximately satisfied.  For example, in circuits
exhibiting the limits of quantum chaos, apart from a small light cone at the end of the circuit, any local errors lead to a state nearly indistinguishable from a Haar random state. Therefore, the matrix elements in \replace{the second condition}{Eq.~\ref{eq:quad_condition_2}} are exponentially small in the number of qubits.
This sensitivity to local perturbations in random circuits is used in the cross-entropy benchmarking technique~\cite{Arute2019-jy}, and explains the improved behavior of our technique in numerical tests on random circuits.

\section{Numerical Experiments}
\label{sec:numerical_performance}

In this section, we present numerical simulations of virtual distillation
applied to three model systems. 
We first consider two classes of random circuits, chosen because they are simple limits where the behaviour of virtual distillation is easy to analyze.
We then turn towards the application of virtual distillation to the simulation of the
dynamics of a one dimensional spin chain following a quantum quench. 
This example allows us to study the behaviour of virtual distillation in the context of quantum simulation, an application which is a promising candidate for the eventual demonstration of practical quantum advantage in the NISQ era. 
We choose to focus on time evolution rather than the ground state problem mainly because ground states have additional structure which enables specialized error mitigation techniques and we are interested in how virtual distillation behaves in the absence of this structure.

We find it illuminating to characterize the effectiveness of our approach as a
function of the expected number of errors in a particular circuit.
This tends to allow more universal prediction of performance when trading
between error rate per gate and number of gates.
We consider a noise model that focuses on stochastic errors in two-qubit gates. 
Specifically, after each two-qubit gate, we apply a single-qubit depolarizing
channel to both qubits acted on by the gate.
The expected number of errors (\(E\)) can be expressed simply as a function of
the number of two-qubit gates in the circuit (\(G\)) and the single-qubit
depolarizing probability (\(p\), defined in the usual way in
\eq{single_qubit_depolarizing}),
\begin{equation}
  \label{eq:expected_num_errors_depolarizing}
    E = 2pG.
\end{equation}
To quantify the error, we focus mainly on the trace distance between the ideal
state that would be obtained with noise-free evolution and the effective state
accessed by virtual distillation
The trace distance leads to a natural bound in the error for the expectation
value of an arbitrary observable,
\begin{equation}
  \label{eq:trace_distance_bound}
  |\tr(\rho O) - \ev{O}{\psi_{\mathrm{ideal}}}| \leq 2 || O || T(\rho, \ket{\psi_{\mathrm{ideal}}}),
\end{equation}
where \(O\) is an observable with operator norm \(||O||\), and \(T(-, -)\) denotes
the trace distance.

\subsection{Scrambling Circuits}
\label{sec:scrambling_circuits}

Both classes of random circuits that we simulate are related to the scrambling
circuits used to demonstrate beyond classical computation in
Ref.~\citenum{Arute2019-jy}.
The first class is essentially a one-dimensional version of the circuit family
considered in that work.
The second class of circuits is exactly the same as the first class, except that
we remove the two-qubit gates.
We provide some additional details in \app{scrambling_circuit_details}.
For these non-entangling random circuits, we still perform the noisy simulations
of these circuits by applying single-qubit depolarizing channels in the same
locations where the two-qubit gates would have been. 

Because the behaviour of the non-entangling random circuits is particularly
simple to understand, we consider this class of circuits first. 
In the absence of entangling gates, we can commute the applications of the
single-qubit depolarizing channel to the end of the circuit. 
We can then combine them together into a single application per qubit with a
larger effective error rate. 
We carry this procedure out analytically in \app{scrambling_circuit_details},
showing that the dominant eigenvector of the density matrix corresponds exactly
the ideal state.
This leads us to expect behaviour similar to that of the phenomenological noise
model we considered in \sec{orthogonal_errors}.

\begin{figure}[t]
  \centering
  \includegraphics[width=.48\textwidth]{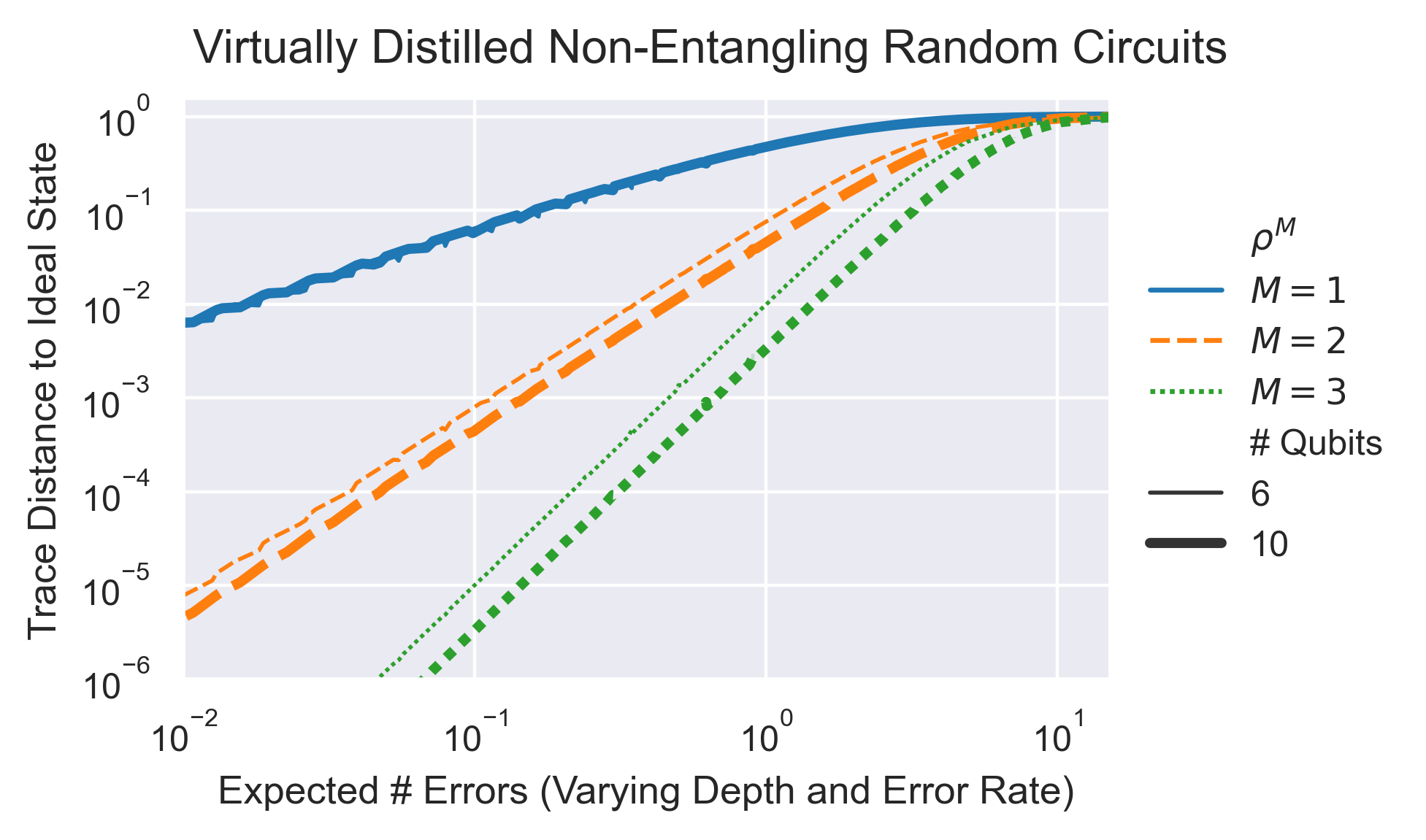}
  \caption{The error in the unmitigated noisy states (\(M=1\)) and the states
    accessed by virtual distillation (\(M=2,3\)) for a variety of
    non-entangling random circuits at two different system sizes (differentiated
    by thickness of markers). 
    We plot the error, quantified by the trace distance to the state obtained
    from noiseless evolution, as a function of the expected number of
    single-qubit depolarizing errors, resulting from varying both the error rate
    and number of gates.  Unlike other cases, for these non-entangling circuits, 
    the eigenvalue floor vanishes and we see exponential suppression in the
    number of copies.
    \label{fig:tensor_product_scrambler_scaling}}
\end{figure}

In \fig{tensor_product_scrambler_scaling}, we plot the trace distances between
the ideal states generated by noiseless evolution and the states obtained by
noisy evolution of these non-entangling random circuits (blue curve). 
We consider a variety of different circuit depths and error rates for both
six-qubit systems (thin curves) and ten qubit systems (thick curves). 
For each of these simulations, we also calculate the trace distance between the
ideal state and the states we are effectively accessing by using virtual
distillation with \(M=2\) (orange dotted curve, \(\rho^2 / \tr(\rho^2)\)) or
\(M=3\) (green dotted curve, \(\rho^3 / \tr(\rho^3)\)) copies.
For each particular setting of circuit depth and error rate, we consider a
single randomly chosen member from the ensemble of non-entangling scrambling
circuits described above. 

We see that the data from this variety of simulations collapses together when we
plot the error (in terms of trace distance) as a function of the expected number
of gate errors.
When the expected number of errors is not too large, the curves for \(M=1\),
\(M=2\), and \(M=3\) are nearly linear with slopes \(1\), \(2\), and \(3\)
respectively. 
Although the noise model in this case does not exactly match the
phenomenological model of \sec{orthogonal_errors}, the results are broadly
consistent.
For these non-entangling random circuits, we observe a level of error
suppression that is exponential in \(M\).

\begin{figure}[t]
  \centering
  \includegraphics[width=.48\textwidth]{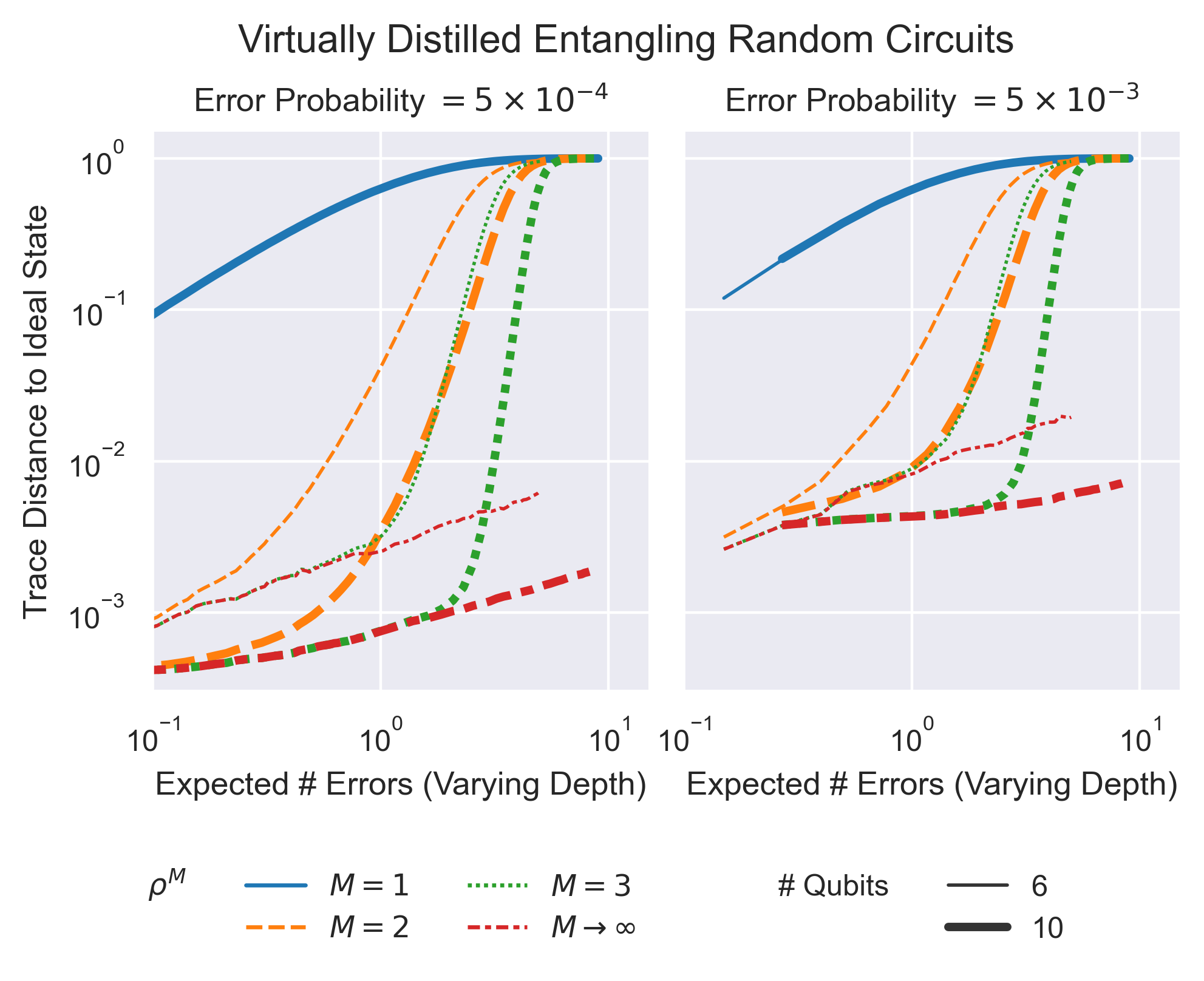}
  \caption{ The error in the unmitigated noisy states (\(M=1\)), the states
    accessed by virtual distillation (\(M=2,3\)), and the dominant eigenvector
    of the density matrices (\(M \rightarrow \infty\)) for a variety of entangling
    random circuits.
    We plot the trace distance to the state obtained from noiseless evolution as
    a function of the expected number of single-qubit depolarizing errors for
    two different system sizes (represented by thickness of marker). 
    Here we vary the expected number of errors by varying the number of gates,
    fixing the single-qubit depolarizing probabilities to \(5 \times 10^{-4}\)
    (left panel) or \(5 \times 10^{-3}\) (right panel).
    We see that the dominant eigenvector determines the noise floor beyond which
    we cannot improve, independent of the number of copies, and that this floor
    drops as the size of the system increases.
  } \label{fig:1d_scrambler_depth_scaling}
\end{figure}

\begin{figure}[t]
  \centering
  \includegraphics[width=.48\textwidth]{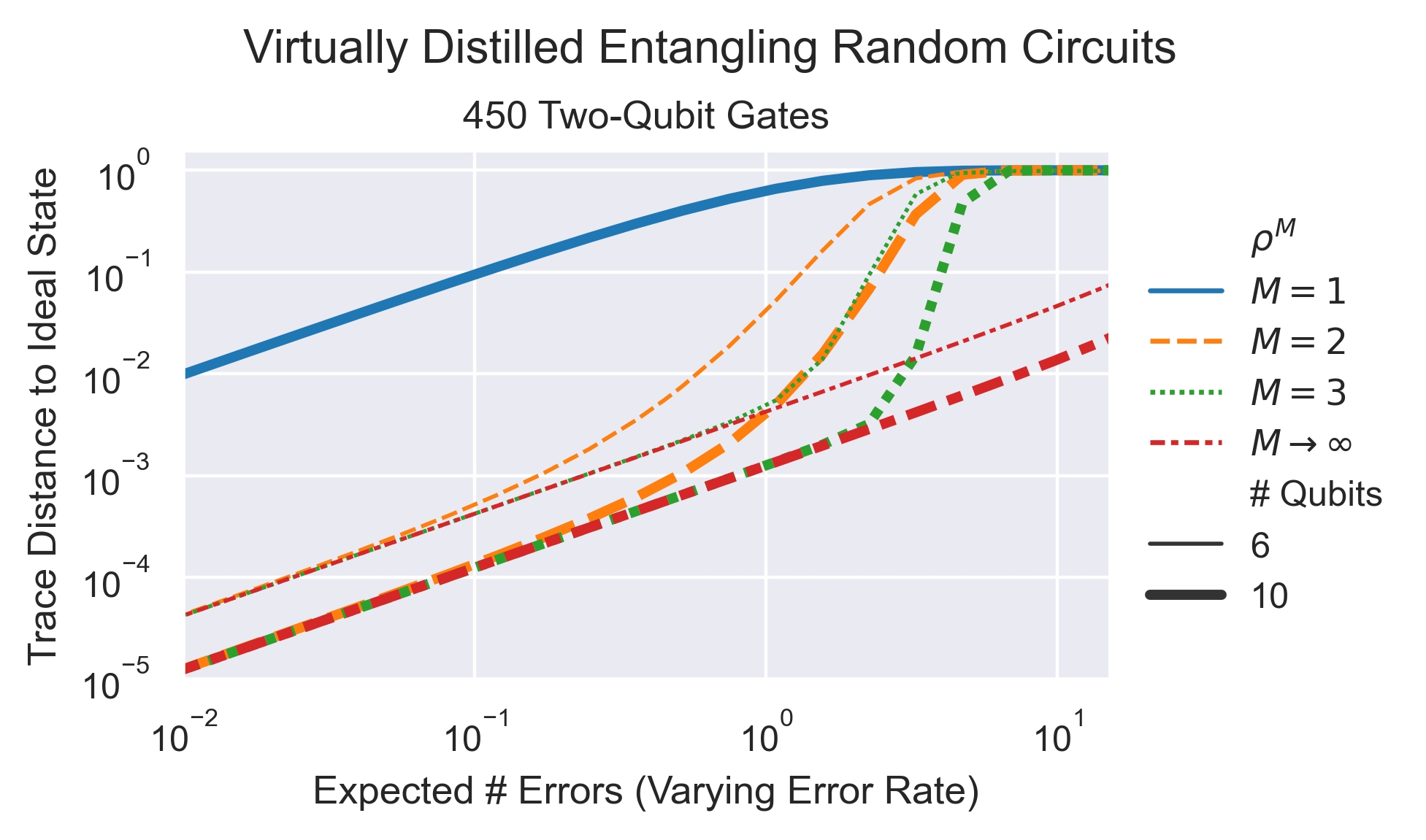}
  \caption{ The error in the unmitigated noisy states (\(M=1\)), the states
    accessed by virtual distillation (\(M=2,3\)), and the dominant eigenvector
    of the density matrices (\(M \rightarrow \infty\)) for a variety of
    entangling random circuits.
    We plot the trace distance to the state obtained from noiseless evolution,
    as a function of the expected number of single-qubit depolarizing errors,
    for 6 and 10 qubit systems (demarcated by the thickness of the symbols).
    We vary the expected number of errors by varying the error rate per-gate,
    fixing the number of two-qubit gates to be \(450\). 
    It's clear that there is a maximum number of expected errors for which the
    technique is effective, and below a certain error rate, the achievable
    improvement is fixed by the drift in the dominant eigenvector ($M\rightarrow
    \infty)$. 
  }
    \label{fig:1d_scrambler_error_rate_scaling}

\end{figure}

In \fig{1d_scrambler_depth_scaling} and \fig{1d_scrambler_error_rate_scaling},
we present plots that explore the behaviour of entangling random circuits on a
one-dimensional line of qubits. 
When we considered the non-entangling random circuits, we found that error
(quantified by the trace distance to the ideal state) depended mostly on the
system size and the expected number of gate errors.
This was true regardless of whether or the expected number of errors was varied
by changing the circuit depth or by changing the error rate per-gate. 
Here we observe slightly different behaviour between these two cases, and
therefore consider them separately.
These two figures also differ from \fig{tensor_product_scrambler_scaling} in
that they include a red dashed curve corresponding to trace distance between the
dominant eigenvector of the density matrix (\(\lim_{M\rightarrow \infty} \rho^M
/ \tr(\rho^M)\)) and the ideal state, a quantity which is non-zero for the
richer family of circuits we now consider.

\fig{1d_scrambler_depth_scaling} presents two plots that show the effects of
varying the circuit depth at two different fixed error rates.
We see that the error in the dominant eigenvector effectively sets a floor for
the minimal error achievable by virtual distillation for any value of \(M\). 
This floor grows slowly with increasing circuit depth.
Furthermore, both the absolute magnitude and the rate of growth appear to be
suppressed with system size. 
In \sec{non_orthogonality_floor} we showed that the leading order contributions
to the trace distance between the dominant eigenvector and the ideal state can
be understood in terms of the matrix elements of the Kraus operators (see
\eq{thing_that_must_be_small}). 
As the circuit depth of the random circuit increases, we expect a 1D random
circuit to approach a Haar random circuit at a depth proportional to the number
of qubits $N$. 
Once this approximation is sufficient, all but a small fraction of errors in the
lightcone of the observable at the end of the circuit will lead to matrix
elements that contribute to the drift in the dominant eigenvector that are
exponentially small in the number of qubits.  
This observation may explain the scaling we see in
\fig{1d_scrambler_depth_scaling}. 

In \fig{1d_scrambler_error_rate_scaling} we plot the error in terms of trace
distance as we vary the expected number of gate errors by varying the per-gate
error rate for a fixed circuit.
At low error rates, we see that the errors in the dominant eigenvectors (red
dashed curves) scale linearly with the error rate but are orders of magnitude
smaller than the errors in the unmitigated state.
This matches the behaviour we would expect from the analysis of
\sec{non_orthogonality_floor}. 
As in \fig{1d_scrambler_depth_scaling}, the error in the dominant eigenvector
sets a floor for the performance of our method at finite \(M\) and that this
floor is suppressed as the system size increases. 

\subsection{Heisenberg Quench}
\label{sec:heisenberg_evolution}

The properties of random circuits can be somewhat unique in their ability to
scramble errors.  It is thus important to consider how the approach
works for other circuits of interest, like the quantum simulation of physical systems.
In this section, we explore the performance of our approach applied to the
simulation of time evolution following a quantum quench in a spin model.
We initialize the system in an antiferromagnetic state, e.g., \(\ket{0101}\), and
simulate the time evolution under the Hamiltonian
\begin{align}
  H =& \sum_{i=1}^{N-1} \Big( J_x X_iX_{i+1} + J_y Y_iY_{i+1} + J_z Z_iZ_{i+1}\Big) \\ &+ \sum_{i=1}^{N} h X_i.
                                                                                                                    \nonumber
\end{align}
Here we have chosen the parameters, \(J_x = J_y = 1.0\), \(J_z = 1.5\), \(h =
1.0\), in order to match a previously studied family of non-integrable
models~\cite{Dmitriev2004-de}, although we take open boundary conditions rather
than periodic ones.
We approximate the time evolution under this Hamiltonian by Trotterization with
a timestep of \(\Delta t = 0.2\). 
Specifically, we use alternating layers of single-qubit gates, two-qubit gates
between odd-even pairs of qubits, and two-qubit gates between even-odd pairs of
qubits. 
As above, we simulate the resulting circuits with single-qubit depolarizing
noise applied after every two-qubit gate.

\begin{figure}[t]
  \centering
  \includegraphics[width=.48\textwidth]{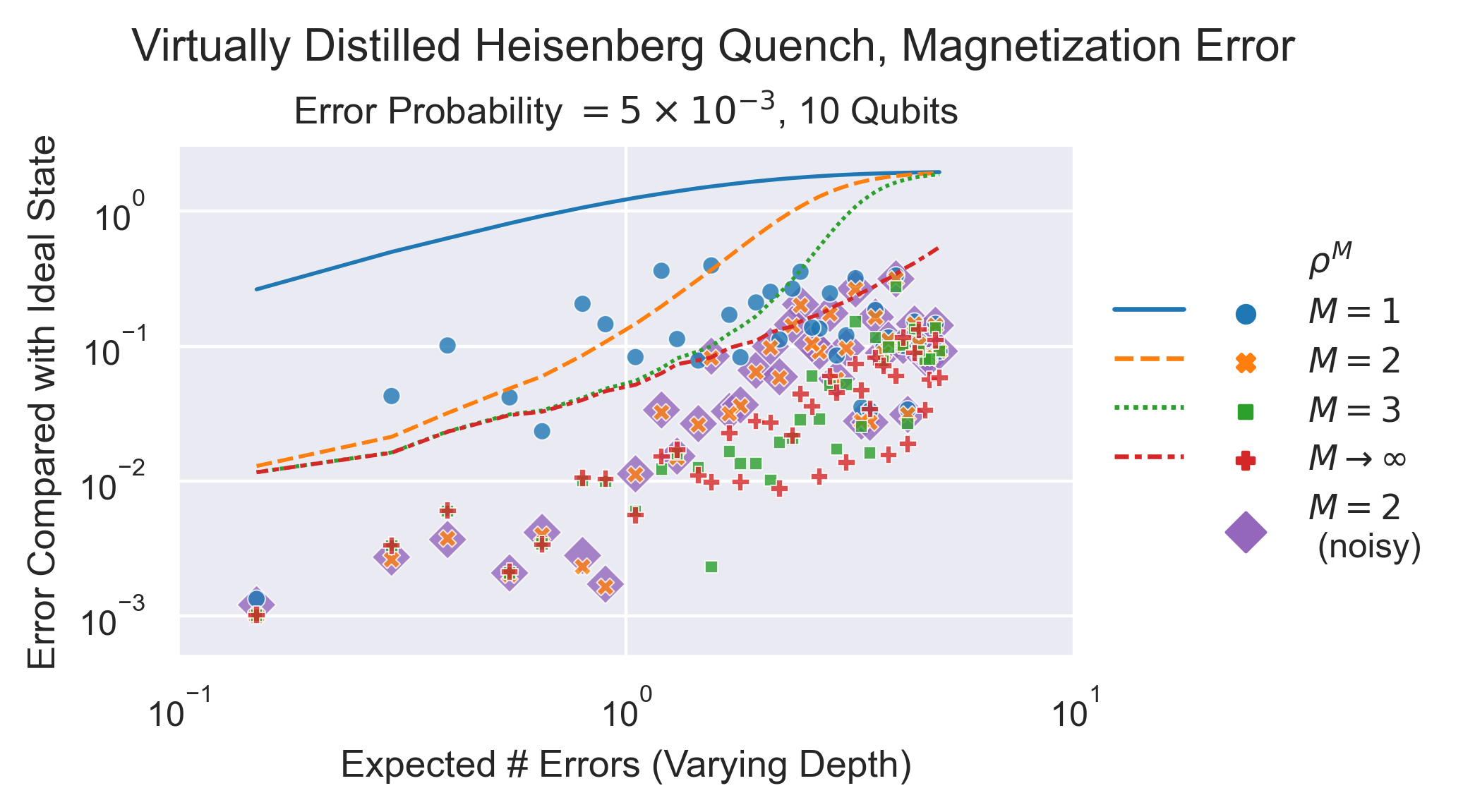}
  \caption{The average error in the single-site magnetization in the unmitigated
    noisy states (\(M=1\)), the states accessed by our error mitigation
    technique (\(M=2,3\)), and the dominant eigenvectors of the density matrices
    (\(M \rightarrow \infty\)) for states generated by the Trotterized time
    evolution of a Heisenberg model.
    We plot the actual average errors we calculate using the blue dots
    (\(M=1\)), orange crosses (\(M=2\)), green squares (\(M=3\)), red diamonds
    (\(M \rightarrow \infty\)), and large purple diamonds (\(M=2\), noisy
    distillation) alongside bounds determined by the trace distance to the ideal
    state using \eq{trace_distance_bound} (various curves).
    We plot these quantities as a function of the expected number of
    single-qubit depolarizing errors, which we vary by varying the number of
    gates, fixing the single-qubit depolarizing probability to \(5 \times
    10^{-3}\). 
    We see that for this specific observable, the trace distance bounds are
    pessimistic by roughly an order of magnitude, though generally respect the
    behavior of the eigenvector floor. 
    Furthermore, we notice an almost perfect coincidence between the orange
    crosses and purple diamonds, indicating that performing the virtual
    distillation circuits of \sec{measurement_by_diagonalization} with noise
    has a negligible effect on the corrected expectation value. 
    \label{fig:heisenberg_depth_error_scatter_scaling}}
\end{figure}

In \fig{heisenberg_depth_error_scatter_scaling}, we plot the bounds on the error
of an arbitrary observable (normalized so that \(||O|| = 1\)) derived from the
trace distance to the noiseless state. We vary the expected number of errors by varying the circuit depth of a six qubit system with a fixed single-qubit depolarizing probability of \(5 \times 10^{-3}\). Alongside these bounds (plotted using solid and dashed curves) we also plot the
actual average error in the single-site magnetization (averaged over the \(6\)
sites) at various points throughout the circuit. 
For the two-copy (\(M=2\)) version of our proposal, we plot the error calculated
directly from the state \(\rho^2 / \tr(\rho^2)\) using yellow crosses and
the error we would obtain by applying the destructive measurement described in
\sec{measurement_by_diagonalization} using large purple diamonds. 
For this second calculation, we simulate the application of the six two-qubit
gates (\eq{beamsplitter_gate_matrix}) required to diagonalize the
observables using the same noise model as the rest of the circuit. 
From the nearly perfect overlap of the yellow crosses with the purple diamonds,
we can see that circuit noise during the diagonalization step has barely any
effect on the reconstructed expectation values. 
It is also apparent that, although the average error in the magnetization does
not saturate the bounds implied by the trace distance, our approach suppresses
the errors in the actual expectation values to a similar degree that it
suppresses the trace distance.

\begin{figure}[t]
  \centering
  \includegraphics[width=.48\textwidth]{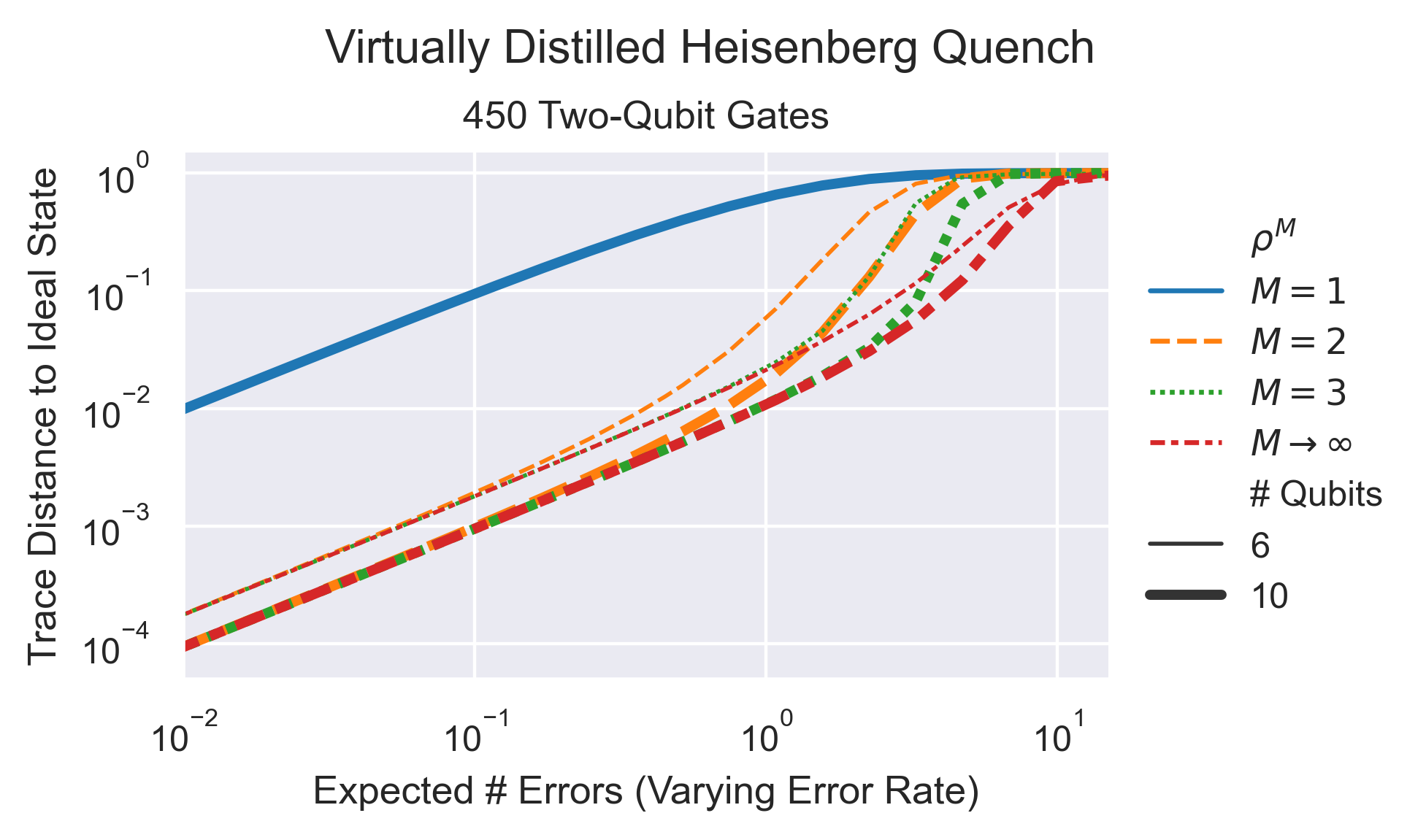}
  \caption{The error in the unmitigated noisy states (\(M=1\)), the states
    accessed by virtual distillation (\(M=2,3\)), and the dominant eigenvectors
    of the density matrices (\(M \rightarrow \infty\)), for states generated by
    the Trotterized time evolution of a Heisenberg model.
    We plot the trace distance to the state obtained from noiseless evolution as
    a function of the expected number of single-qubit depolarizing errors.
    The expected number of errors is varied by changing the error rate per-gate,
    fixing the number of two-qubit gates to be \(450\).
    We show this data for 6 and 10 qubit systems (differentiated by the size of
    the markers). 
    As we increase the system size from 6 qubits to 10, we observe that the
    error (quantified by the trace distance to the ideal state) decreases for
    the error-mitigated states (\(M>1\)).}
  \label{fig:heisenberg_error_rate_scaling}
\end{figure}

\fig{heisenberg_error_rate_scaling} offers a different look at the same system.
As with \fig{1d_scrambler_error_rate_scaling}, in this figure we fix the number
of two-qubit gates to be \(450\) and we vary the expected number of gate errors
by sweeping over a range of per-gate error rates. 
Here we clearly see the impact of the floor set by the drift in the dominant
eigenvector (red dotted curve).
At low gate error rates, the error (in terms of trace distance to the ideal
state) for the virtually distilled states (\(M>1\)) is suppressed by a constant
factor relative to the error in the unmitigated state (\(M=1\)).
The constant factor improvement is substantial and appears to increase with
system size. 
Both the size of the improvement and its sensitivity to the system size are
smaller than we observed for the one-dimensional scrambling circuits of
\fig{1d_scrambler_error_rate_scaling}.

\begin{figure}[t]
  \centering
  \includegraphics[width=.48\textwidth]{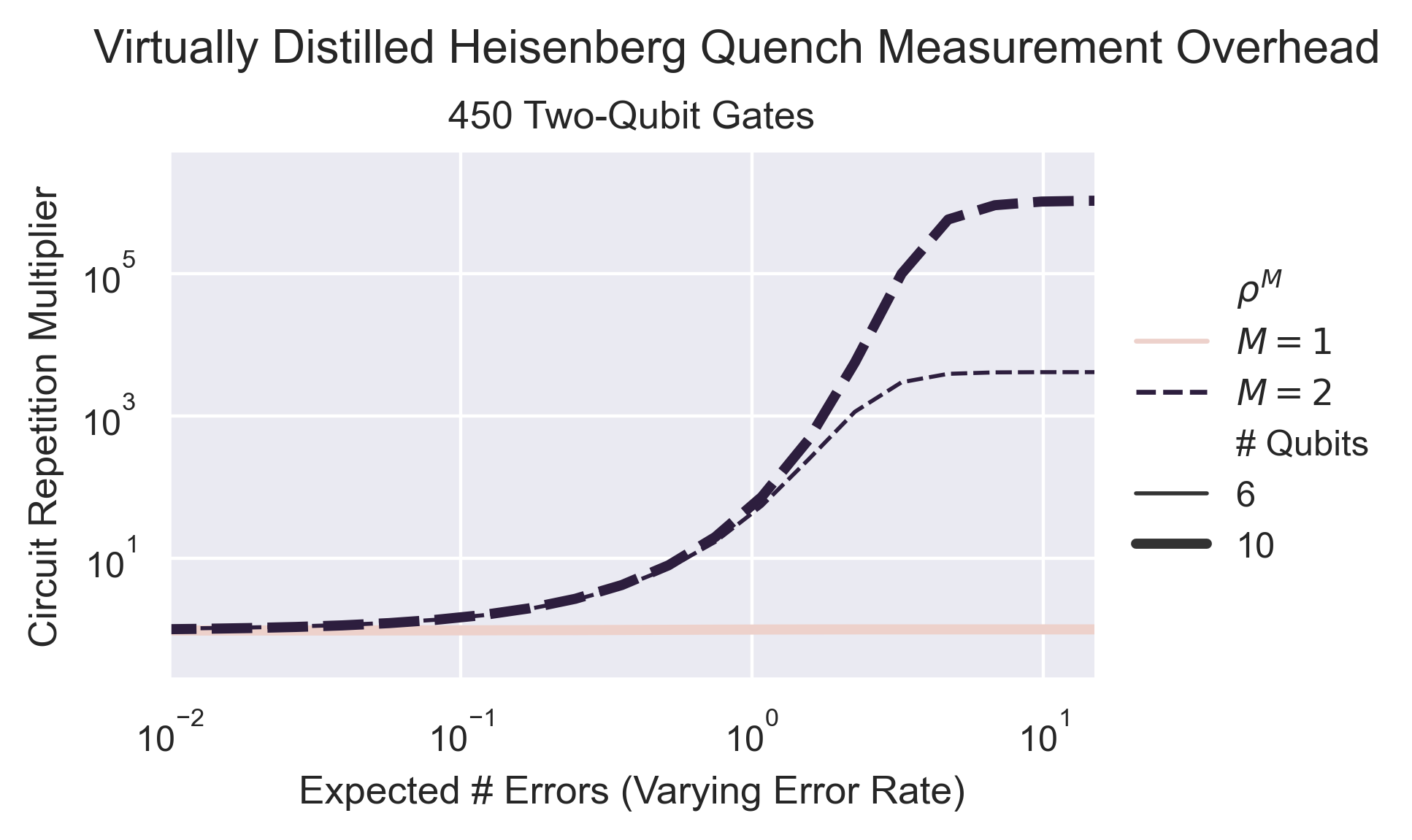}
  \caption{The average overhead in the number of measurement repetitions
    required to measure the single-site magnetization (of a Heisenberg model) with virtual distillation (\(M=2\)). 
    The overhead is calculated by taking the ratio of the time required using
    our technique to achieve a fixed target precision and the time required when
    measuring the same quantity to the same precision with respect to the
    unmitigated noisy state (\(M=1\)).     We vary the expected number of errors 
    by varying the error rate per-gate, fixing the number of two-qubit gates to be \(450\), and 
    plot the overhead as a function of the expected number of single-qubit
    depolarizing errors.  We show data for 6 and 10 qubit systems, denoting the larger system using larger markers.
    As the error rate grows beyond $O(1)$ errors, the overhead increases
    dramatically, with the larger system seeing the greatest inflation.
  }
 \label{fig:heisenberg_measurement_overhead}
\end{figure}

In \fig{heisenberg_measurement_overhead}, we consider the cost of performing the
two-copy (\(M=2\)) version of virtual distillation for the same systems
considered in \fig{heisenberg_error_rate_scaling}. 
We do this using the expression for the variance presented in
\eq{rho_2_estimator_variance_intro} and derived in \app{estimator_variance}. 
Using this expression, we calculate the average variance of our error-mitigated
estimators for the single-site magnetization, \(\{Z_i\}\).
We consider the ratio of this average variance (for the error-mitigated
expectation values) with the average variance of the same measurements without
error mitigation.
Because the number of measurements required for some fixed precision scales
linearly with the variance, this ratio is also the ratio between the number of
measurements required to use virtual distillation and the number of measurements
required to measure the unmitigated expectation values (assuming the same target
precision). 
This quantity therefore encapsulates the overhead incurred by our scheme.

When the expected number of errors is small, we see that virtual distillation
barely increases the number of measurements required. 
It is only as the number of errors grows larger than one that the measurement
cost rises dramatically.
We note that \eq{rho_2_estimator_variance_intro} implicitly assumes that we
perform a number of measurements \(R\) such that
\begin{equation}
  R \gg \frac{1}{\tr(\rho^2)^2}.
\end{equation}
This assumption will break down when the target precision is low and the
expected number of errors in \fig{heisenberg_measurement_overhead} is large, but the
qualitative conclusion remains the same.

\section{Mitigating Algorithmic Errors}
\label{sec:AlgorithmicErrors}
To date, most error mitigation methods have focused on the reduction of errors
caused by imperfections in a device implementation, such as decoherence or
control errors. 
Here explore the idea that some of these techniques can be applied to
algorithmic errors incurred during otherwise noise-free implementations of
randomized algorithms. 
Previous works have used extrapolation~\cite{Endo2019-rr} or randomized symmetry
application~\cite{Tran2020-ts} to mitigate coherent errors in
evolution; we extend this concept to incoherent errors.

\replace{In particular, r}{R}ecent developments in Hamiltonian simulation have
led to the development of randomized evolution methods such as
qDRIFT~\cite{Campbell2019-id}, randomized Trotter~\cite{Childs2019-dv}, and
combinations thereof~\cite{Ouyang2020-pp}, which have benefits in some
situations over their deterministic counterparts. 
As these methods are randomized, they output mixed states rather than pure
states, even in the absence of noise. 
Moreover, they depend on an approximation parameter with a natural limit in
which they converge to the pure state generated by exact evolution.
In this section, we show numerically that virtual distillation applied to qDRIFT
can suppress this deviation from the exact evolution.
For the particular model system we consider, we find that virtual distillation
can reduce the coherent space-time volume required to reach a particular
accuracy threshold by a factor of $8$ or more compared with the standard qDRIFT.

\subsection{qDRIFT}\label{sec:BackgroundqDRIFT}
We briefly introduce some background on the qDRIFT method.
qDRIFT simulates time evolution under a Hamiltonian $H$, by constructing product
formulae using a randomized selection rule. 
Terms are chosen from $H$ at random, with a selection probability proportional
to their interaction strength in the Hamiltonian. 
One then evolves the system forwards in time under this Hamiltonian term, for a
fixed timestep and repeats this process a number of times, generating a product formula that provides an approximation to the time evolution operator. 
When averaged over the classical randomness (in the choice of interaction
terms), qDRIFT generates a quantum channel that closely approximates the exact
evolution more closely than the individual product formulae.
Importantly, unlike most deterministic Trotter methods, the scaling of this
approach does not depend explicitly on the number of terms in the Hamiltonian,
but rather than $1$-norm of the coefficients.

More precisely, we consider a Hamiltonian that we may decompose as $H=\sum_i h_i
H_i$, where all $h_i$ are made real and positive by absorbing signs into $H_i$,
and the spectral norm of $H_i$ is bounded by $1$. 
Defining $\lambda=\sum_i h_i$, the diamond norm distance between the qDRIFT
channel and the true time evolution is bounded by
\begin{align}
  \epsilon = \frac{2 \lambda^2 t^2}{\eta},
\end{align}
where $\eta$ is the number of qDRIFT selection steps performed to generate each
instance of the qDRIFT channel, and hence controls the amount of coherent
evolution required. 
As $\eta$ increases, the resulting quantum channel converges to the unitary
corresponding to the exact evolution.
It will be our aim to understand how our virtual distillation technique can
reduce the coherent space-time volume required, by reducing this factor $\eta$
required to achieve the same error in practice.

\begin{figure}[t!]
  \centering \includegraphics[width=.48\textwidth]{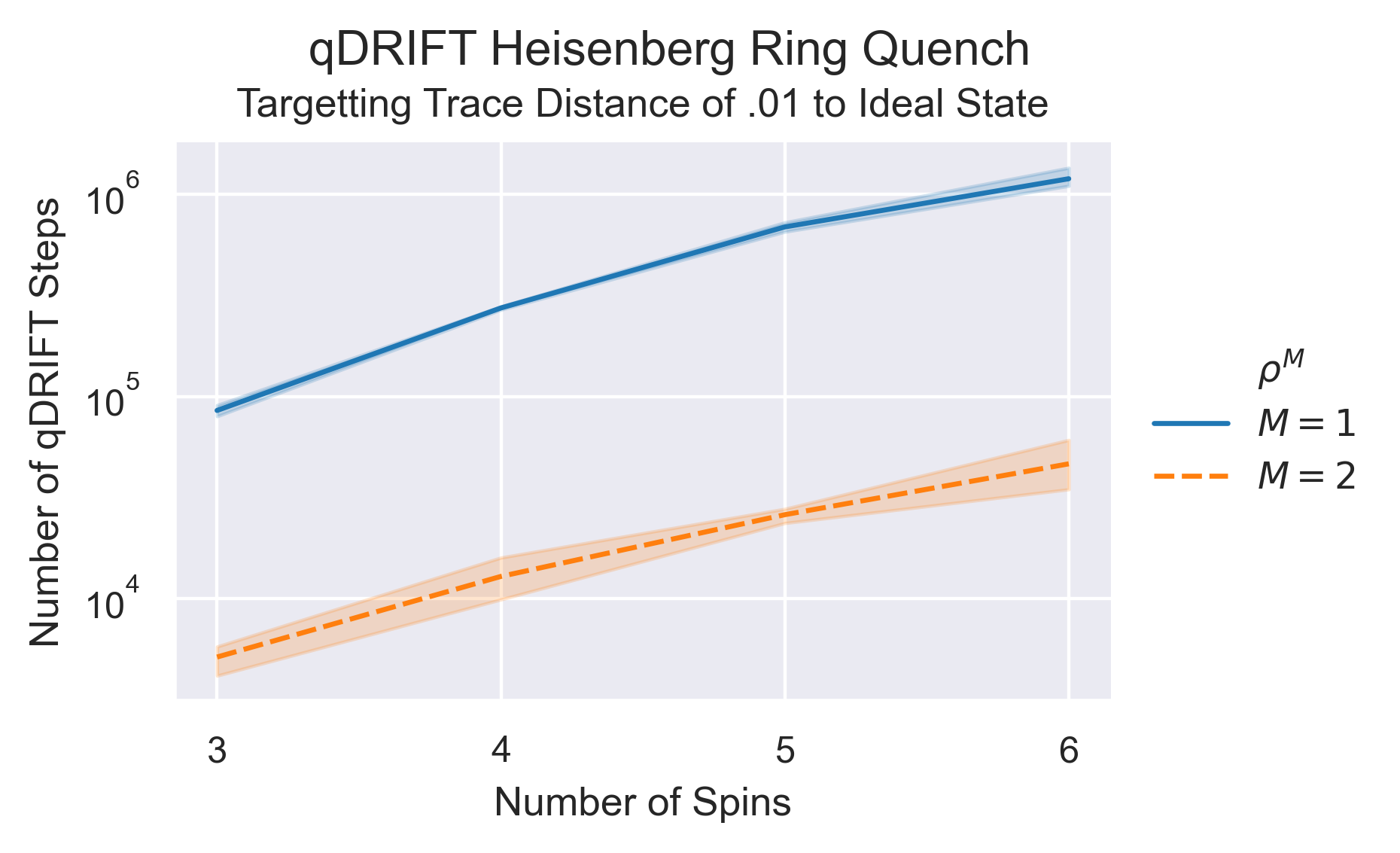}
  \caption{qDRIFT coherent cost reduction through virtual distillation in the Heisenberg model.  Here we show the number of coherent qDRIFT steps required to reach a target trace distance, with and without virtual distillation using 2 copies.  We see that there is a consistent reduction of at least 16x in the number of required steps. When accounting for the overhead of using two copies, this amounts to a 8x reduction in the coherent space-time volume used to reach the same error rate.
    \label{fig:qdrift_ring_scaling}}
\end{figure}

\subsection{Virtual distillation applied to qDRIFT}\label{sec:qDRIFT}
Here we study the application of virtual distillation to qDRIFT numerically. 
Specifically, we investigate how virtual distillation can impact the number of
coherent steps, $\eta$, required to reach a target accuracy. 
For this, we choose a Heisenberg Hamiltonian with up to 6 qubits per copy. 
The Hamiltonian is given by
\begin{align}
  H = \sum_{i=1}^{N} X_i X_{i+1} + Y_i Y_{i+1} + Z_i Z_{i+1} + h_i Z_i,
\end{align}
where $h_i \in \{-h, h\}$ are randomly chosen $Z$ magnetic field strengths, and
periodic boundary conditions are applied such that site $N+1$ is site $1$. 
For our studies here, we choose a time evolution length of $t=N$ and let $h=1$. 
We numerically investigate the number of coherent qDRIFT steps required to
achieve a trace distance of $0.01$ to the ideal state for evolutions under such
a Heisenberg model. 
The results of this analysis are shown in Fig.~\ref{fig:qdrift_ring_scaling}. 
We see for this system that virtual distillation consistently reduces the
required number of coherent steps to achieve the desired trace distance by a
factor of more than $16$x. 
If we account for the space overhead of using two copies, this still amounts to
a space-time advantage of $8$x. 
These results suggest that the use of error mitigation techniques may be further
developed to yield practical algorithmic improvements for real systems,
especially in the NISQ regime.

\section{Conclusion}

In this work, we showed how techniques for using multiple copies of a state to
access polynomials of the density matrix can be used to mitigate incoherent
errors.
We studied the effectiveness of this approach for two analytically tractable
noise models and characterized its limit in terms of the dominant eigenvector of
the noisy density matrix. 
We numerically demonstrated reductions in the error (quantified by the trace
distance to the noise-free state) of up to three orders of magnitude for a
collection of small model systems. 
Furthermore, we showed that this error suppression is enhanced as the system
size or the speed of information scrambling grows.
We also considered the application of our error mitigation approach to the
incoherent algorithmic error that arises when approximating time-evolution using
the qDRIFT algorithm, finding a substantial constant factor improvement.

Our proposed strategy for error mitigation, which we refer to as virtual
distillation, is simple to use and analyze. 
It provides a natural way to take advantage of the surplus of qubits that we
expect to have available as the NISQ era continues to suppress the effects of
incoherent errors.
We expect that our technique will prove complementary to other error mitigation
and calibration techniques, especially those capable of addressing coherent
errors.
The effective state accessed by virtual distillation approaches the dominant
eigenvector of the density matrix exponentially quickly as the number of copies,
\(M\), increases.
Therefore, the utility of our technique depends mainly on the error in the
dominant eigenvector and the number of samples required.
In particular, we have devoted a signficant amount of attention to the question
of sample complexity. 
This is due to the fact that proposed NISQ applications, especially in quantum
simulation, already face a daunting cost in this regard~\cite{Huggins2019-vu}.
Our analytical work and numerical simulations indicate that our strategy is most
effective and affordable when the number of errors expected in the circuit is
\(O(1)\).
The reach of our approach will therefore naturally grow throughout the NISQ era
as hardware platforms continue to improve. 

There are several other important considerations relevant to the performance of
this technique in the NISQ era besides the drift in the dominant eigenvector and
the sample complexity.
First of all, virtual distillation requires collective measurements that couple
the qubits of one copy with the corresponding qubits of additional copies.
For hardware platforms based on a 2D grid of qubits, these measurements are
easiest to perform when the connectivity of the original circuit is linear and
would require a substantial number of additional gates in the general case.
Fortunately, a linearly connected array of qubits is known to be sufficient to
achieve the optimal gate complexity in some cases, including certain approaches
for the simulation of quantum chemistry~\cite{Kivlichan2018-yz, OGorman2019-jj}. 
A second important consideration is that we're often interested in measuring
observables with support on more than one qubit. 
We discuss some options for performing such measurements in
\app{diagonalize_multi_qubit}, \app{diagonalize_three_or_more}, and
\app{ancilla_assisted_measurement}, but these options come with their own
overheads in gate complexity or the number of measurements repetitions.
Thirdly, and perhaps most importantly, the assumption that we're able to measure
expectation values with respect to \(\rho^M / \tr(\rho^M)\) is violated when the
individual copies are not in the same state or when errors occur during
execution of the measurements. We discuss some aspects of the breakdown of this
assumption in \app{different_noise_models}. In general, we do not expect virtual
distillation to be able to correct errors during the measurement process but we
note that we saw a reasonable level of robustness to such errors in our
numerical simulations (see \fig{heisenberg_depth_error_scatter_scaling}).

Even as we begin to leave the NISQ era behind and approach devices that start to incorporate quantum error correction,
in the early days the desire to use as many logical qubits as possible means we
may perform some computations that still have an appreciable number of logical
errors. 
Given that our technique can provide a substantial improvement in error at
negligible overhead compared to traditional quantum error correction techniques,
there may be some advantageous interplay between the two, where small distance
codes are used in conjunction with this technique before more qubits are
available. 
We explore this opportunity in more detail in Appendix ~\ref{app:surface_code}.

Our technique builds upon a long tradition of work that uses the symmetric group
for stabilizing quantum computations and mitigating errors. 
We believe that this research direction continues to hold promise, and we
identify a few directions that we find particularly intriguing. 
Virtual distillation is based on a simple collective measurement of \(M\) copies
of \(\rho\). 
In \sec{sample_efficiency}, we show that there exists more sophisticated
collective measurements whose sample complexity improves quadratically upon our
approach in some regimes. 
It would be interesting to investigate this further, both from a fundamental
perspective, and with an eye towards practical implementation. 
Besides this potential improvement, another clear question arises from our work. 
The drift in the dominant eigenvector of the density matrix is a coherent error.
As with the coherent errors that occur directly in the execution of circuits on
a NISQ device, we are hopeful that variational optimization or other,
complementary, error mitigation techniques may prove useful in addressing this
additional source of coherent error.
Studying this in the context of the real noise experienced in hardware will be
especially important and illuminating.

In the process of preparing this work, a related paper,
Ref.~\citenum{Koczor2020-ub} appeared in the literature. 
Our work highlights different conclusions than theirs in error scaling due to
our explicit consideration of errors that induce a drift in the dominant
eigenvector of the density matrix.


\subsection*{Acknowledgements}

The authors are extremely grateful to Dave Bacon for key discussions around the
symmetric subspace and diagrammatic derivations, Gian-Luca Anselmetti for suggestions around performing virtual distillation with two distinct states, and Nathan Wiebe for helpful
advice at various stages. WJH and KBW acknowledge support from the NSF QLCI program through grant number OMA-2016245.


\begin{thebibliography}{62}%
\makeatletter
\providecommand \@ifxundefined [1]{%
 \@ifx{#1\undefined}
}%
\providecommand \@ifnum [1]{%
 \ifnum #1\expandafter \@firstoftwo
 \else \expandafter \@secondoftwo
 \fi
}%
\providecommand \@ifx [1]{%
 \ifx #1\expandafter \@firstoftwo
 \else \expandafter \@secondoftwo
 \fi
}%
\providecommand \natexlab [1]{#1}%
\providecommand \enquote  [1]{``#1''}%
\providecommand \bibnamefont  [1]{#1}%
\providecommand \bibfnamefont [1]{#1}%
\providecommand \citenamefont [1]{#1}%
\providecommand \href@noop [0]{\@secondoftwo}%
\providecommand \href [0]{\begingroup \@sanitize@url \@href}%
\providecommand \@href[1]{\@@startlink{#1}\@@href}%
\providecommand \@@href[1]{\endgroup#1\@@endlink}%
\providecommand \@sanitize@url [0]{\catcode `\\12\catcode `\$12\catcode
  `\&12\catcode `\#12\catcode `\^12\catcode `\_12\catcode `\%12\relax}%
\providecommand \@@startlink[1]{}%
\providecommand \@@endlink[0]{}%
\providecommand \url  [0]{\begingroup\@sanitize@url \@url }%
\providecommand \@url [1]{\endgroup\@href {#1}{\urlprefix }}%
\providecommand \urlprefix  [0]{URL }%
\providecommand \Eprint [0]{\href }%
\providecommand \doibase [0]{http://dx.doi.org/}%
\providecommand \selectlanguage [0]{\@gobble}%
\providecommand \bibinfo  [0]{\@secondoftwo}%
\providecommand \bibfield  [0]{\@secondoftwo}%
\providecommand \translation [1]{[#1]}%
\providecommand \BibitemOpen [0]{}%
\providecommand \bibitemStop [0]{}%
\providecommand \bibitemNoStop [0]{.\EOS\space}%
\providecommand \EOS [0]{\spacefactor3000\relax}%
\providecommand \BibitemShut  [1]{\csname bibitem#1\endcsname}%
\let\auto@bib@innerbib\@empty
\bibitem [{\citenamefont {Aharonov}\ and\ \citenamefont
  {Ben-Or}(1997)}]{Aharonov1997-be}%
  \BibitemOpen
  \bibfield  {author} {\bibinfo {author} {\bibfnamefont {D}~\bibnamefont
  {Aharonov}}\ and\ \bibinfo {author} {\bibfnamefont {M}~\bibnamefont
  {Ben-Or}},\ }\bibfield  {title} {\enquote {\bibinfo {title} {Fault-tolerant
  quantum computation with constant error},}\ }in\ \href {\doibase
  10.1145/258533.258579} {\emph {\bibinfo {booktitle} {Proceedings of the
  twenty-ninth annual {ACM} symposium on Theory of computing}}},\ \bibinfo
  {series and number} {STOC '97}\ (\bibinfo  {publisher} {Association for
  Computing Machinery},\ \bibinfo {address} {New York, NY, USA},\ \bibinfo
  {year} {1997})\ pp.\ \bibinfo {pages} {176--188}\BibitemShut {NoStop}%
\bibitem [{\citenamefont {Fowler}\ \emph {et~al.}(2012)\citenamefont {Fowler},
  \citenamefont {Mariantoni}, \citenamefont {Martinis},\ and\ \citenamefont
  {Cleland}}]{Fowler2012-li}%
  \BibitemOpen
  \bibfield  {author} {\bibinfo {author} {\bibfnamefont {Austin~G}\
  \bibnamefont {Fowler}}, \bibinfo {author} {\bibfnamefont {Matteo}\
  \bibnamefont {Mariantoni}}, \bibinfo {author} {\bibfnamefont {John~M}\
  \bibnamefont {Martinis}}, \ and\ \bibinfo {author} {\bibfnamefont {Andrew~N}\
  \bibnamefont {Cleland}},\ }\bibfield  {title} {\enquote {\bibinfo {title}
  {Surface codes: Towards practical large-scale quantum computation},}\ }\href
  {\doibase 10.1103/PhysRevA.86.032324} {\bibfield  {journal} {\bibinfo
  {journal} {Phys. Rev. A}\ }\textbf {\bibinfo {volume} {86}},\ \bibinfo
  {pages} {032324} (\bibinfo {year} {2012})}\BibitemShut {NoStop}%
\bibitem [{\citenamefont {Preskill}(2018)}]{Preskill2018-po}%
  \BibitemOpen
  \bibfield  {author} {\bibinfo {author} {\bibfnamefont {John}\ \bibnamefont
  {Preskill}},\ }\bibfield  {title} {\enquote {\bibinfo {title} {Quantum
  computing in the {NISQ} era and beyond},}\ }\href
  {https://quantum-journal.org/papers/q-2018-08-06-79/} {\bibfield  {journal}
  {\bibinfo  {journal} {Quantum}\ }\textbf {\bibinfo {volume} {2}},\ \bibinfo
  {pages} {79} (\bibinfo {year} {2018})}\BibitemShut {NoStop}%
\bibitem [{\citenamefont {Temme}\ \emph {et~al.}(2017)\citenamefont {Temme},
  \citenamefont {Bravyi},\ and\ \citenamefont {Gambetta}}]{Temme2017-hj}%
  \BibitemOpen
  \bibfield  {author} {\bibinfo {author} {\bibfnamefont {Kristan}\ \bibnamefont
  {Temme}}, \bibinfo {author} {\bibfnamefont {Sergey}\ \bibnamefont {Bravyi}},
  \ and\ \bibinfo {author} {\bibfnamefont {Jay~M}\ \bibnamefont {Gambetta}},\
  }\bibfield  {title} {{\selectlanguage {en}\enquote {\bibinfo {title} {Error
  mitigation for {Short-Depth} quantum circuits},}\ }}\href {\doibase
  10.1103/PhysRevLett.119.180509} {\bibfield  {journal} {\bibinfo  {journal}
  {Phys. Rev. Lett.}\ }\textbf {\bibinfo {volume} {119}},\ \bibinfo {pages}
  {180509} (\bibinfo {year} {2017})}\BibitemShut {NoStop}%
\bibitem [{\citenamefont {Endo}\ \emph {et~al.}(2018)\citenamefont {Endo},
  \citenamefont {Benjamin},\ and\ \citenamefont {Li}}]{Endo2018-qs}%
  \BibitemOpen
  \bibfield  {author} {\bibinfo {author} {\bibfnamefont {Suguru}\ \bibnamefont
  {Endo}}, \bibinfo {author} {\bibfnamefont {Simon~C}\ \bibnamefont
  {Benjamin}}, \ and\ \bibinfo {author} {\bibfnamefont {Ying}\ \bibnamefont
  {Li}},\ }\bibfield  {title} {\enquote {\bibinfo {title} {Practical quantum
  error mitigation for {Near-Future} applications},}\ }\href {\doibase
  10.1103/PhysRevX.8.031027} {\bibfield  {journal} {\bibinfo  {journal} {Phys.
  Rev. X}\ }\textbf {\bibinfo {volume} {8}},\ \bibinfo {pages} {031027}
  (\bibinfo {year} {2018})}\BibitemShut {NoStop}%
\bibitem [{\citenamefont {Kandala}\ \emph {et~al.}(2019)\citenamefont
  {Kandala}, \citenamefont {Temme}, \citenamefont {C{\'o}rcoles}, \citenamefont
  {Mezzacapo}, \citenamefont {Chow},\ and\ \citenamefont
  {Gambetta}}]{Kandala2019-gy}%
  \BibitemOpen
  \bibfield  {author} {\bibinfo {author} {\bibfnamefont {Abhinav}\ \bibnamefont
  {Kandala}}, \bibinfo {author} {\bibfnamefont {Kristan}\ \bibnamefont
  {Temme}}, \bibinfo {author} {\bibfnamefont {Antonio~D}\ \bibnamefont
  {C{\'o}rcoles}}, \bibinfo {author} {\bibfnamefont {Antonio}\ \bibnamefont
  {Mezzacapo}}, \bibinfo {author} {\bibfnamefont {Jerry~M}\ \bibnamefont
  {Chow}}, \ and\ \bibinfo {author} {\bibfnamefont {Jay~M}\ \bibnamefont
  {Gambetta}},\ }\bibfield  {title} {{\selectlanguage {en}\enquote {\bibinfo
  {title} {Error mitigation extends the computational reach of a noisy quantum
  processor},}\ }}\href {\doibase 10.1038/s41586-019-1040-7} {\bibfield
  {journal} {\bibinfo  {journal} {Nature}\ }\textbf {\bibinfo {volume} {567}},\
  \bibinfo {pages} {491--495} (\bibinfo {year} {2019})}\BibitemShut {NoStop}%
\bibitem [{\citenamefont {Strikis}\ \emph {et~al.}(2020)\citenamefont
  {Strikis}, \citenamefont {Qin}, \citenamefont {Chen}, \citenamefont
  {Benjamin},\ and\ \citenamefont {Li}}]{Strikis2020-ve}%
  \BibitemOpen
  \bibfield  {author} {\bibinfo {author} {\bibfnamefont {Armands}\ \bibnamefont
  {Strikis}}, \bibinfo {author} {\bibfnamefont {Dayue}\ \bibnamefont {Qin}},
  \bibinfo {author} {\bibfnamefont {Yanzhu}\ \bibnamefont {Chen}}, \bibinfo
  {author} {\bibfnamefont {Simon~C}\ \bibnamefont {Benjamin}}, \ and\ \bibinfo
  {author} {\bibfnamefont {Ying}\ \bibnamefont {Li}},\ }\bibfield  {title}
  {\enquote {\bibinfo {title} {Learning-based quantum error mitigation},}\
  }\href {http://arxiv.org/abs/2005.07601} {\bibfield  {journal} {\bibinfo
  {journal} {arXiv:2005.07601}\ } (\bibinfo {year} {2020})}\BibitemShut
  {NoStop}%
\bibitem [{\citenamefont {Czarnik}\ \emph {et~al.}(2020)\citenamefont
  {Czarnik}, \citenamefont {Arrasmith}, \citenamefont {Coles},\ and\
  \citenamefont {Cincio}}]{Czarnik2020-pb}%
  \BibitemOpen
  \bibfield  {author} {\bibinfo {author} {\bibfnamefont {Piotr}\ \bibnamefont
  {Czarnik}}, \bibinfo {author} {\bibfnamefont {Andrew}\ \bibnamefont
  {Arrasmith}}, \bibinfo {author} {\bibfnamefont {Patrick~J}\ \bibnamefont
  {Coles}}, \ and\ \bibinfo {author} {\bibfnamefont {Lukasz}\ \bibnamefont
  {Cincio}},\ }\bibfield  {title} {\enquote {\bibinfo {title} {Error mitigation
  with clifford quantum-circuit data},}\ }\href
  {http://arxiv.org/abs/2005.10189} {\bibfield  {journal} {\bibinfo  {journal}
  {arXiv:2005.10189}\ } (\bibinfo {year} {2020})}\BibitemShut {NoStop}%
\bibitem [{\citenamefont {Arute}\ \emph {et~al.}(2020)\citenamefont {Arute},
  \citenamefont {Arya}, \citenamefont {Babbush}, \citenamefont {Bacon},
  \citenamefont {Bardin}, \citenamefont {Barends}, \citenamefont {Bengtsson},
  \citenamefont {Boixo}, \citenamefont {Broughton}, \citenamefont {Buckley},
  \citenamefont {Buell}, \citenamefont {Burkett}, \citenamefont {Bushnell},
  \citenamefont {Chen}, \citenamefont {Chen}, \citenamefont {Chen},
  \citenamefont {Chiaro}, \citenamefont {Collins}, \citenamefont {Cotton},
  \citenamefont {Courtney}, \citenamefont {Demura}, \citenamefont {Derk},
  \citenamefont {Dunsworth}, \citenamefont {Eppens}, \citenamefont {Eckl},
  \citenamefont {Erickson}, \citenamefont {Farhi}, \citenamefont {Fowler},
  \citenamefont {Foxen}, \citenamefont {Gidney}, \citenamefont {Giustina},
  \citenamefont {Graff}, \citenamefont {Gross}, \citenamefont {Habegger},
  \citenamefont {Harrigan}, \citenamefont {Ho}, \citenamefont {Hong},
  \citenamefont {Huang}, \citenamefont {Huggins}, \citenamefont {Ioffe},
  \citenamefont {Isakov}, \citenamefont {Jeffrey}, \citenamefont {Jiang},
  \citenamefont {Jones}, \citenamefont {Kafri}, \citenamefont {Kechedzhi},
  \citenamefont {Kelly}, \citenamefont {Kim}, \citenamefont {Klimov},
  \citenamefont {Korotkov}, \citenamefont {Kostritsa}, \citenamefont
  {Landhuis}, \citenamefont {Laptev}, \citenamefont {Lindmark}, \citenamefont
  {Lucero}, \citenamefont {Marthaler}, \citenamefont {Martin}, \citenamefont
  {Martinis}, \citenamefont {Marusczyk}, \citenamefont {McArdle}, \citenamefont
  {McClean}, \citenamefont {McCourt}, \citenamefont {McEwen}, \citenamefont
  {Megrant}, \citenamefont {Mejuto-Zaera}, \citenamefont {Mi}, \citenamefont
  {Mohseni}, \citenamefont {Mruczkiewicz}, \citenamefont {Mutus}, \citenamefont
  {Naaman}, \citenamefont {Neeley}, \citenamefont {Neill}, \citenamefont
  {Neven}, \citenamefont {Newman}, \citenamefont {Niu}, \citenamefont
  {O'Brien}, \citenamefont {Ostby}, \citenamefont {Pat{\'o}}, \citenamefont
  {Petukhov}, \citenamefont {Putterman}, \citenamefont {Quintana},
  \citenamefont {Reiner}, \citenamefont {Roushan}, \citenamefont {Rubin},
  \citenamefont {Sank}, \citenamefont {Satzinger}, \citenamefont {Smelyanskiy},
  \citenamefont {Strain}, \citenamefont {Sung}, \citenamefont {Schmitteckert},
  \citenamefont {Szalay}, \citenamefont {Tubman}, \citenamefont {Vainsencher},
  \citenamefont {White}, \citenamefont {Vogt}, \citenamefont {Jamie~Yao},
  \citenamefont {Yeh}, \citenamefont {Zalcman},\ and\ \citenamefont
  {Zanker}}]{Arute2020-ta}%
  \BibitemOpen
  \bibfield  {author} {\bibinfo {author} {\bibfnamefont {Frank}\ \bibnamefont
  {Arute}}, \bibinfo {author} {\bibfnamefont {Kunal}\ \bibnamefont {Arya}},
  \bibinfo {author} {\bibfnamefont {Ryan}\ \bibnamefont {Babbush}}, \bibinfo
  {author} {\bibfnamefont {Dave}\ \bibnamefont {Bacon}}, \bibinfo {author}
  {\bibfnamefont {Joseph~C}\ \bibnamefont {Bardin}}, \bibinfo {author}
  {\bibfnamefont {Rami}\ \bibnamefont {Barends}}, \bibinfo {author}
  {\bibfnamefont {Andreas}\ \bibnamefont {Bengtsson}}, \bibinfo {author}
  {\bibfnamefont {Sergio}\ \bibnamefont {Boixo}}, \bibinfo {author}
  {\bibfnamefont {Michael}\ \bibnamefont {Broughton}}, \bibinfo {author}
  {\bibfnamefont {Bob~B}\ \bibnamefont {Buckley}}, \bibinfo {author}
  {\bibfnamefont {David~A}\ \bibnamefont {Buell}}, \bibinfo {author}
  {\bibfnamefont {Brian}\ \bibnamefont {Burkett}}, \bibinfo {author}
  {\bibfnamefont {Nicholas}\ \bibnamefont {Bushnell}}, \bibinfo {author}
  {\bibfnamefont {Yu}~\bibnamefont {Chen}}, \bibinfo {author} {\bibfnamefont
  {Zijun}\ \bibnamefont {Chen}}, \bibinfo {author} {\bibfnamefont {Yu-An}\
  \bibnamefont {Chen}}, \bibinfo {author} {\bibfnamefont {Ben}\ \bibnamefont
  {Chiaro}}, \bibinfo {author} {\bibfnamefont {Roberto}\ \bibnamefont
  {Collins}}, \bibinfo {author} {\bibfnamefont {Stephen~J}\ \bibnamefont
  {Cotton}}, \bibinfo {author} {\bibfnamefont {William}\ \bibnamefont
  {Courtney}}, \bibinfo {author} {\bibfnamefont {Sean}\ \bibnamefont {Demura}},
  \bibinfo {author} {\bibfnamefont {Alan}\ \bibnamefont {Derk}}, \bibinfo
  {author} {\bibfnamefont {Andrew}\ \bibnamefont {Dunsworth}}, \bibinfo
  {author} {\bibfnamefont {Daniel}\ \bibnamefont {Eppens}}, \bibinfo {author}
  {\bibfnamefont {Thomas}\ \bibnamefont {Eckl}}, \bibinfo {author}
  {\bibfnamefont {Catherine}\ \bibnamefont {Erickson}}, \bibinfo {author}
  {\bibfnamefont {Edward}\ \bibnamefont {Farhi}}, \bibinfo {author}
  {\bibfnamefont {Austin}\ \bibnamefont {Fowler}}, \bibinfo {author}
  {\bibfnamefont {Brooks}\ \bibnamefont {Foxen}}, \bibinfo {author}
  {\bibfnamefont {Craig}\ \bibnamefont {Gidney}}, \bibinfo {author}
  {\bibfnamefont {Marissa}\ \bibnamefont {Giustina}}, \bibinfo {author}
  {\bibfnamefont {Rob}\ \bibnamefont {Graff}}, \bibinfo {author} {\bibfnamefont
  {Jonathan~A}\ \bibnamefont {Gross}}, \bibinfo {author} {\bibfnamefont
  {Steve}\ \bibnamefont {Habegger}}, \bibinfo {author} {\bibfnamefont
  {Matthew~P}\ \bibnamefont {Harrigan}}, \bibinfo {author} {\bibfnamefont
  {Alan}\ \bibnamefont {Ho}}, \bibinfo {author} {\bibfnamefont {Sabrina}\
  \bibnamefont {Hong}}, \bibinfo {author} {\bibfnamefont {Trent}\ \bibnamefont
  {Huang}}, \bibinfo {author} {\bibfnamefont {William}\ \bibnamefont
  {Huggins}}, \bibinfo {author} {\bibfnamefont {Lev~B}\ \bibnamefont {Ioffe}},
  \bibinfo {author} {\bibfnamefont {Sergei~V}\ \bibnamefont {Isakov}}, \bibinfo
  {author} {\bibfnamefont {Evan}\ \bibnamefont {Jeffrey}}, \bibinfo {author}
  {\bibfnamefont {Zhang}\ \bibnamefont {Jiang}}, \bibinfo {author}
  {\bibfnamefont {Cody}\ \bibnamefont {Jones}}, \bibinfo {author}
  {\bibfnamefont {Dvir}\ \bibnamefont {Kafri}}, \bibinfo {author}
  {\bibfnamefont {Kostyantyn}\ \bibnamefont {Kechedzhi}}, \bibinfo {author}
  {\bibfnamefont {Julian}\ \bibnamefont {Kelly}}, \bibinfo {author}
  {\bibfnamefont {Seon}\ \bibnamefont {Kim}}, \bibinfo {author} {\bibfnamefont
  {Paul~V}\ \bibnamefont {Klimov}}, \bibinfo {author} {\bibfnamefont
  {Alexander~N}\ \bibnamefont {Korotkov}}, \bibinfo {author} {\bibfnamefont
  {Fedor}\ \bibnamefont {Kostritsa}}, \bibinfo {author} {\bibfnamefont {David}\
  \bibnamefont {Landhuis}}, \bibinfo {author} {\bibfnamefont {Pavel}\
  \bibnamefont {Laptev}}, \bibinfo {author} {\bibfnamefont {Mike}\ \bibnamefont
  {Lindmark}}, \bibinfo {author} {\bibfnamefont {Erik}\ \bibnamefont {Lucero}},
  \bibinfo {author} {\bibfnamefont {Michael}\ \bibnamefont {Marthaler}},
  \bibinfo {author} {\bibfnamefont {Orion}\ \bibnamefont {Martin}}, \bibinfo
  {author} {\bibfnamefont {John~M}\ \bibnamefont {Martinis}}, \bibinfo {author}
  {\bibfnamefont {Anika}\ \bibnamefont {Marusczyk}}, \bibinfo {author}
  {\bibfnamefont {Sam}\ \bibnamefont {McArdle}}, \bibinfo {author}
  {\bibfnamefont {Jarrod~R}\ \bibnamefont {McClean}}, \bibinfo {author}
  {\bibfnamefont {Trevor}\ \bibnamefont {McCourt}}, \bibinfo {author}
  {\bibfnamefont {Matt}\ \bibnamefont {McEwen}}, \bibinfo {author}
  {\bibfnamefont {Anthony}\ \bibnamefont {Megrant}}, \bibinfo {author}
  {\bibfnamefont {Carlos}\ \bibnamefont {Mejuto-Zaera}}, \bibinfo {author}
  {\bibfnamefont {Xiao}\ \bibnamefont {Mi}}, \bibinfo {author} {\bibfnamefont
  {Masoud}\ \bibnamefont {Mohseni}}, \bibinfo {author} {\bibfnamefont
  {Wojciech}\ \bibnamefont {Mruczkiewicz}}, \bibinfo {author} {\bibfnamefont
  {Josh}\ \bibnamefont {Mutus}}, \bibinfo {author} {\bibfnamefont {Ofer}\
  \bibnamefont {Naaman}}, \bibinfo {author} {\bibfnamefont {Matthew}\
  \bibnamefont {Neeley}}, \bibinfo {author} {\bibfnamefont {Charles}\
  \bibnamefont {Neill}}, \bibinfo {author} {\bibfnamefont {Hartmut}\
  \bibnamefont {Neven}}, \bibinfo {author} {\bibfnamefont {Michael}\
  \bibnamefont {Newman}}, \bibinfo {author} {\bibfnamefont {Murphy~Yuezhen}\
  \bibnamefont {Niu}}, \bibinfo {author} {\bibfnamefont {Thomas~E}\
  \bibnamefont {O'Brien}}, \bibinfo {author} {\bibfnamefont {Eric}\
  \bibnamefont {Ostby}}, \bibinfo {author} {\bibfnamefont {B{\'a}lint}\
  \bibnamefont {Pat{\'o}}}, \bibinfo {author} {\bibfnamefont {Andre}\
  \bibnamefont {Petukhov}}, \bibinfo {author} {\bibfnamefont {Harald}\
  \bibnamefont {Putterman}}, \bibinfo {author} {\bibfnamefont {Chris}\
  \bibnamefont {Quintana}}, \bibinfo {author} {\bibfnamefont {Jan-Michael}\
  \bibnamefont {Reiner}}, \bibinfo {author} {\bibfnamefont {Pedram}\
  \bibnamefont {Roushan}}, \bibinfo {author} {\bibfnamefont {Nicholas~C}\
  \bibnamefont {Rubin}}, \bibinfo {author} {\bibfnamefont {Daniel}\
  \bibnamefont {Sank}}, \bibinfo {author} {\bibfnamefont {Kevin~J}\
  \bibnamefont {Satzinger}}, \bibinfo {author} {\bibfnamefont {Vadim}\
  \bibnamefont {Smelyanskiy}}, \bibinfo {author} {\bibfnamefont {Doug}\
  \bibnamefont {Strain}}, \bibinfo {author} {\bibfnamefont {Kevin~J}\
  \bibnamefont {Sung}}, \bibinfo {author} {\bibfnamefont {Peter}\ \bibnamefont
  {Schmitteckert}}, \bibinfo {author} {\bibfnamefont {Marco}\ \bibnamefont
  {Szalay}}, \bibinfo {author} {\bibfnamefont {Norm~M}\ \bibnamefont {Tubman}},
  \bibinfo {author} {\bibfnamefont {Amit}\ \bibnamefont {Vainsencher}},
  \bibinfo {author} {\bibfnamefont {Theodore}\ \bibnamefont {White}}, \bibinfo
  {author} {\bibfnamefont {Nicolas}\ \bibnamefont {Vogt}}, \bibinfo {author}
  {\bibfnamefont {Z}~\bibnamefont {Jamie~Yao}}, \bibinfo {author}
  {\bibfnamefont {Ping}\ \bibnamefont {Yeh}}, \bibinfo {author} {\bibfnamefont
  {Adam}\ \bibnamefont {Zalcman}}, \ and\ \bibinfo {author} {\bibfnamefont
  {Sebastian}\ \bibnamefont {Zanker}},\ }\bibfield  {title} {\enquote {\bibinfo
  {title} {Observation of separated dynamics of charge and spin in the
  {Fermi-Hubbard} model},}\ }\href {http://arxiv.org/abs/2010.07965} {\bibfield
   {journal} {\bibinfo  {journal} {arXiv:2010.07965}\ } (\bibinfo {year}
  {2020})}\BibitemShut {NoStop}%
\bibitem [{\citenamefont {O'Brien}\ \emph {et~al.}(2020)\citenamefont
  {O'Brien}, \citenamefont {Polla}, \citenamefont {Rubin}, \citenamefont
  {Huggins}, \citenamefont {McArdle}, \citenamefont {Boixo}, \citenamefont
  {McClean},\ and\ \citenamefont {Babbush}}]{OBrien2020-tw}%
  \BibitemOpen
  \bibfield  {author} {\bibinfo {author} {\bibfnamefont {Thomas~E}\
  \bibnamefont {O'Brien}}, \bibinfo {author} {\bibfnamefont {Stefano}\
  \bibnamefont {Polla}}, \bibinfo {author} {\bibfnamefont {Nicholas~C}\
  \bibnamefont {Rubin}}, \bibinfo {author} {\bibfnamefont {William~J}\
  \bibnamefont {Huggins}}, \bibinfo {author} {\bibfnamefont {Sam}\ \bibnamefont
  {McArdle}}, \bibinfo {author} {\bibfnamefont {Sergio}\ \bibnamefont {Boixo}},
  \bibinfo {author} {\bibfnamefont {Jarrod~R}\ \bibnamefont {McClean}}, \ and\
  \bibinfo {author} {\bibfnamefont {Ryan}\ \bibnamefont {Babbush}},\ }\bibfield
   {title} {\enquote {\bibinfo {title} {Error mitigation via verified phase
  estimation},}\ }\href {http://arxiv.org/abs/2010.02538} {\bibfield  {journal}
  {\bibinfo  {journal} {arXiv:2010.02538}\ } (\bibinfo {year}
  {2020})}\BibitemShut {NoStop}%
\bibitem [{\citenamefont {McClean}\ \emph {et~al.}(2017)\citenamefont
  {McClean}, \citenamefont {Kimchi-Schwartz}, \citenamefont {Carter},\ and\
  \citenamefont {de~Jong}}]{McClean2017-ta}%
  \BibitemOpen
  \bibfield  {author} {\bibinfo {author} {\bibfnamefont {Jarrod~R}\
  \bibnamefont {McClean}}, \bibinfo {author} {\bibfnamefont {Mollie~E}\
  \bibnamefont {Kimchi-Schwartz}}, \bibinfo {author} {\bibfnamefont {Jonathan}\
  \bibnamefont {Carter}}, \ and\ \bibinfo {author} {\bibfnamefont {Wibe~A}\
  \bibnamefont {de~Jong}},\ }\bibfield  {title} {\enquote {\bibinfo {title}
  {Hybrid quantum-classical hierarchy for mitigation of decoherence and
  determination of excited states},}\ }\href {\doibase
  10.1103/PhysRevA.95.042308} {\bibfield  {journal} {\bibinfo  {journal} {Phys.
  Rev. A}\ }\textbf {\bibinfo {volume} {95}},\ \bibinfo {pages} {042308}
  (\bibinfo {year} {2017})}\BibitemShut {NoStop}%
\bibitem [{\citenamefont {Colless}\ \emph {et~al.}(2018)\citenamefont
  {Colless}, \citenamefont {Ramasesh}, \citenamefont {Dahlen}, \citenamefont
  {Blok}, \citenamefont {Kimchi-Schwartz}, \citenamefont {McClean},
  \citenamefont {Carter}, \citenamefont {de~Jong},\ and\ \citenamefont
  {Siddiqi}}]{Colless2018-af}%
  \BibitemOpen
  \bibfield  {author} {\bibinfo {author} {\bibfnamefont {J~I}\ \bibnamefont
  {Colless}}, \bibinfo {author} {\bibfnamefont {V~V}\ \bibnamefont {Ramasesh}},
  \bibinfo {author} {\bibfnamefont {D}~\bibnamefont {Dahlen}}, \bibinfo
  {author} {\bibfnamefont {M~S}\ \bibnamefont {Blok}}, \bibinfo {author}
  {\bibfnamefont {M~E}\ \bibnamefont {Kimchi-Schwartz}}, \bibinfo {author}
  {\bibfnamefont {J~R}\ \bibnamefont {McClean}}, \bibinfo {author}
  {\bibfnamefont {J}~\bibnamefont {Carter}}, \bibinfo {author} {\bibfnamefont
  {W~A}\ \bibnamefont {de~Jong}}, \ and\ \bibinfo {author} {\bibfnamefont
  {I}~\bibnamefont {Siddiqi}},\ }\bibfield  {title} {\enquote {\bibinfo {title}
  {Computation of molecular spectra on a quantum processor with an
  {Error-Resilient} algorithm},}\ }\href {\doibase 10.1103/PhysRevX.8.011021}
  {\bibfield  {journal} {\bibinfo  {journal} {Phys. Rev. X}\ }\textbf {\bibinfo
  {volume} {8}},\ \bibinfo {pages} {011021} (\bibinfo {year}
  {2018})}\BibitemShut {NoStop}%
\bibitem [{\citenamefont {McClean}\ \emph {et~al.}(2020)\citenamefont
  {McClean}, \citenamefont {Jiang}, \citenamefont {Rubin}, \citenamefont
  {Babbush},\ and\ \citenamefont {Neven}}]{McClean2020-pq}%
  \BibitemOpen
  \bibfield  {author} {\bibinfo {author} {\bibfnamefont {Jarrod~R}\
  \bibnamefont {McClean}}, \bibinfo {author} {\bibfnamefont {Zhang}\
  \bibnamefont {Jiang}}, \bibinfo {author} {\bibfnamefont {Nicholas~C}\
  \bibnamefont {Rubin}}, \bibinfo {author} {\bibfnamefont {Ryan}\ \bibnamefont
  {Babbush}}, \ and\ \bibinfo {author} {\bibfnamefont {Hartmut}\ \bibnamefont
  {Neven}},\ }\bibfield  {title} {{\selectlanguage {en}\enquote {\bibinfo
  {title} {Decoding quantum errors with subspace expansions},}\ }}\href
  {\doibase 10.1038/s41467-020-14341-w} {\bibfield  {journal} {\bibinfo
  {journal} {Nat. Commun.}\ }\textbf {\bibinfo {volume} {11}},\ \bibinfo
  {pages} {636} (\bibinfo {year} {2020})}\BibitemShut {NoStop}%
\bibitem [{\citenamefont {Bonet-Monroig}\ \emph {et~al.}(2018)\citenamefont
  {Bonet-Monroig}, \citenamefont {Sagastizabal}, \citenamefont {Singh},\ and\
  \citenamefont {O'Brien}}]{Bonet-Monroig2018-od}%
  \BibitemOpen
  \bibfield  {author} {\bibinfo {author} {\bibfnamefont {X}~\bibnamefont
  {Bonet-Monroig}}, \bibinfo {author} {\bibfnamefont {R}~\bibnamefont
  {Sagastizabal}}, \bibinfo {author} {\bibfnamefont {M}~\bibnamefont {Singh}},
  \ and\ \bibinfo {author} {\bibfnamefont {T~E}\ \bibnamefont {O'Brien}},\
  }\bibfield  {title} {\enquote {\bibinfo {title} {Low-cost error mitigation by
  symmetry verification},}\ }\href {\doibase 10.1103/PhysRevA.98.062339}
  {\bibfield  {journal} {\bibinfo  {journal} {Phys. Rev. A}\ }\textbf {\bibinfo
  {volume} {98}},\ \bibinfo {pages} {062339} (\bibinfo {year}
  {2018})}\BibitemShut {NoStop}%
\bibitem [{\citenamefont {McArdle}\ \emph {et~al.}(2019)\citenamefont
  {McArdle}, \citenamefont {Yuan},\ and\ \citenamefont
  {Benjamin}}]{McArdle2019-uv}%
  \BibitemOpen
  \bibfield  {author} {\bibinfo {author} {\bibfnamefont {Sam}\ \bibnamefont
  {McArdle}}, \bibinfo {author} {\bibfnamefont {Xiao}\ \bibnamefont {Yuan}}, \
  and\ \bibinfo {author} {\bibfnamefont {Simon}\ \bibnamefont {Benjamin}},\
  }\bibfield  {title} {{\selectlanguage {en}\enquote {\bibinfo {title}
  {{Error-Mitigated} digital quantum simulation},}\ }}\href {\doibase
  10.1103/PhysRevLett.122.180501} {\bibfield  {journal} {\bibinfo  {journal}
  {Phys. Rev. Lett.}\ }\textbf {\bibinfo {volume} {122}},\ \bibinfo {pages}
  {180501} (\bibinfo {year} {2019})}\BibitemShut {NoStop}%
\bibitem [{\citenamefont {Sagastizabal}\ \emph {et~al.}(2019)\citenamefont
  {Sagastizabal}, \citenamefont {Bonet-Monroig}, \citenamefont {Singh},
  \citenamefont {Rol}, \citenamefont {Bultink}, \citenamefont {Fu},
  \citenamefont {Price}, \citenamefont {Ostroukh}, \citenamefont
  {Muthusubramanian}, \citenamefont {Bruno}, \citenamefont {Beekman},
  \citenamefont {Haider}, \citenamefont {O'Brien},\ and\ \citenamefont
  {DiCarlo}}]{Sagastizabal2019-sz}%
  \BibitemOpen
  \bibfield  {author} {\bibinfo {author} {\bibfnamefont {R}~\bibnamefont
  {Sagastizabal}}, \bibinfo {author} {\bibfnamefont {X}~\bibnamefont
  {Bonet-Monroig}}, \bibinfo {author} {\bibfnamefont {M}~\bibnamefont {Singh}},
  \bibinfo {author} {\bibfnamefont {M~A}\ \bibnamefont {Rol}}, \bibinfo
  {author} {\bibfnamefont {C~C}\ \bibnamefont {Bultink}}, \bibinfo {author}
  {\bibfnamefont {X}~\bibnamefont {Fu}}, \bibinfo {author} {\bibfnamefont
  {C~H}\ \bibnamefont {Price}}, \bibinfo {author} {\bibfnamefont {V~P}\
  \bibnamefont {Ostroukh}}, \bibinfo {author} {\bibfnamefont {N}~\bibnamefont
  {Muthusubramanian}}, \bibinfo {author} {\bibfnamefont {A}~\bibnamefont
  {Bruno}}, \bibinfo {author} {\bibfnamefont {M}~\bibnamefont {Beekman}},
  \bibinfo {author} {\bibfnamefont {N}~\bibnamefont {Haider}}, \bibinfo
  {author} {\bibfnamefont {T~E}\ \bibnamefont {O'Brien}}, \ and\ \bibinfo
  {author} {\bibfnamefont {L}~\bibnamefont {DiCarlo}},\ }\bibfield  {title}
  {\enquote {\bibinfo {title} {Experimental error mitigation via symmetry
  verification in a variational quantum eigensolver},}\ }\href {\doibase
  10.1103/PhysRevA.100.010302} {\bibfield  {journal} {\bibinfo  {journal}
  {Phys. Rev. A}\ }\textbf {\bibinfo {volume} {100}},\ \bibinfo {pages}
  {010302} (\bibinfo {year} {2019})}\BibitemShut {NoStop}%
\bibitem [{\citenamefont {Huggins}\ \emph {et~al.}(2019)\citenamefont
  {Huggins}, \citenamefont {McClean}, \citenamefont {Rubin}, \citenamefont
  {Jiang}, \citenamefont {Wiebe}, \citenamefont {Birgitta~Whaley},\ and\
  \citenamefont {Babbush}}]{Huggins2019-vu}%
  \BibitemOpen
  \bibfield  {author} {\bibinfo {author} {\bibfnamefont {William~J}\
  \bibnamefont {Huggins}}, \bibinfo {author} {\bibfnamefont {Jarrod}\
  \bibnamefont {McClean}}, \bibinfo {author} {\bibfnamefont {Nicholas}\
  \bibnamefont {Rubin}}, \bibinfo {author} {\bibfnamefont {Zhang}\ \bibnamefont
  {Jiang}}, \bibinfo {author} {\bibfnamefont {Nathan}\ \bibnamefont {Wiebe}},
  \bibinfo {author} {\bibfnamefont {K}~\bibnamefont {Birgitta~Whaley}}, \ and\
  \bibinfo {author} {\bibfnamefont {Ryan}\ \bibnamefont {Babbush}},\ }\bibfield
   {title} {\enquote {\bibinfo {title} {Efficient and noise resilient
  measurements for quantum chemistry on {Near-Term} quantum computers},}\
  }\href {http://arxiv.org/abs/1907.13117} {\bibfield  {journal} {\bibinfo
  {journal} {arXiv:1907.13117}\ } (\bibinfo {year} {2019})}\BibitemShut
  {NoStop}%
\bibitem [{\citenamefont {{Google AI Quantum and
  Collaborators}}(2020)}]{Google_AI_Quantum_and_Collaborators2020-sc}%
  \BibitemOpen
  \bibfield  {author} {\bibinfo {author} {\bibnamefont {{Google AI Quantum and
  Collaborators}}},\ }\bibfield  {title} {{\selectlanguage {en}\enquote
  {\bibinfo {title} {{Hartree-Fock} on a superconducting qubit quantum
  computer},}\ }}\href {\doibase 10.1126/science.abb9811} {\bibfield  {journal}
  {\bibinfo  {journal} {Science}\ }\textbf {\bibinfo {volume} {369}},\ \bibinfo
  {pages} {1084--1089} (\bibinfo {year} {2020})}\BibitemShut {NoStop}%
\bibitem [{\citenamefont {Chen}\ \emph {et~al.}(2019)\citenamefont {Chen},
  \citenamefont {Farahzad}, \citenamefont {Yoo},\ and\ \citenamefont
  {Wei}}]{Chen2019-sg}%
  \BibitemOpen
  \bibfield  {author} {\bibinfo {author} {\bibfnamefont {Yanzhu}\ \bibnamefont
  {Chen}}, \bibinfo {author} {\bibfnamefont {Maziar}\ \bibnamefont {Farahzad}},
  \bibinfo {author} {\bibfnamefont {Shinjae}\ \bibnamefont {Yoo}}, \ and\
  \bibinfo {author} {\bibfnamefont {Tzu-Chieh}\ \bibnamefont {Wei}},\
  }\bibfield  {title} {\enquote {\bibinfo {title} {Detector tomography on {IBM}
  quantum computers and mitigation of an imperfect measurement},}\ }\href
  {\doibase 10.1103/PhysRevA.100.052315} {\bibfield  {journal} {\bibinfo
  {journal} {Phys. Rev. A}\ }\textbf {\bibinfo {volume} {100}},\ \bibinfo
  {pages} {052315} (\bibinfo {year} {2019})}\BibitemShut {NoStop}%
\bibitem [{\citenamefont {Maciejewski}\ \emph {et~al.}(2020)\citenamefont
  {Maciejewski}, \citenamefont {Zimbor{\'a}s},\ and\ \citenamefont
  {Oszmaniec}}]{Maciejewski2020-ju}%
  \BibitemOpen
  \bibfield  {author} {\bibinfo {author} {\bibfnamefont {Filip~B}\ \bibnamefont
  {Maciejewski}}, \bibinfo {author} {\bibfnamefont {Zolt{\'a}n}\ \bibnamefont
  {Zimbor{\'a}s}}, \ and\ \bibinfo {author} {\bibfnamefont {Micha{\l}}\
  \bibnamefont {Oszmaniec}},\ }\bibfield  {title} {\enquote {\bibinfo {title}
  {Mitigation of readout noise in near-term quantum devices by classical
  post-processing based on detector tomography},}\ }\href {\doibase
  10.22331/q-2020-04-24-257} {\bibfield  {journal} {\bibinfo  {journal}
  {Quantum}\ }\textbf {\bibinfo {volume} {4}},\ \bibinfo {pages} {257}
  (\bibinfo {year} {2020})}\BibitemShut {NoStop}%
\bibitem [{\citenamefont {Bravyi}\ \emph {et~al.}(2020)\citenamefont {Bravyi},
  \citenamefont {Sheldon}, \citenamefont {Kandala}, \citenamefont {Mckay},\
  and\ \citenamefont {Gambetta}}]{Bravyi2020-dn}%
  \BibitemOpen
  \bibfield  {author} {\bibinfo {author} {\bibfnamefont {Sergey}\ \bibnamefont
  {Bravyi}}, \bibinfo {author} {\bibfnamefont {Sarah}\ \bibnamefont {Sheldon}},
  \bibinfo {author} {\bibfnamefont {Abhinav}\ \bibnamefont {Kandala}}, \bibinfo
  {author} {\bibfnamefont {David~C}\ \bibnamefont {Mckay}}, \ and\ \bibinfo
  {author} {\bibfnamefont {Jay~M}\ \bibnamefont {Gambetta}},\ }\bibfield
  {title} {\enquote {\bibinfo {title} {Mitigating measurement errors in
  multi-qubit experiments},}\ }\href {http://arxiv.org/abs/2006.14044} {\
  (\bibinfo {year} {2020})},\ \Eprint {http://arxiv.org/abs/2006.14044}
  {arXiv:2006.14044 [quant-ph]} \BibitemShut {NoStop}%
\bibitem [{\citenamefont {Berthiaume}\ \emph {et~al.}(1994)\citenamefont
  {Berthiaume}, \citenamefont {Deutsch},\ and\ \citenamefont
  {Jozsa}}]{Berthiaume1994-ln}%
  \BibitemOpen
  \bibfield  {author} {\bibinfo {author} {\bibfnamefont {A}~\bibnamefont
  {Berthiaume}}, \bibinfo {author} {\bibfnamefont {D}~\bibnamefont {Deutsch}},
  \ and\ \bibinfo {author} {\bibfnamefont {R}~\bibnamefont {Jozsa}},\
  }\bibfield  {title} {\enquote {\bibinfo {title} {The stabilisation of quantum
  computations},}\ }in\ \href {\doibase 10.1109/PHYCMP.1994.363698} {\emph
  {\bibinfo {booktitle} {Proceedings Workshop on Physics and Computation.
  {PhysComp} '94}}}\ (\bibinfo  {publisher} {ieeexplore.ieee.org},\ \bibinfo
  {year} {1994})\ pp.\ \bibinfo {pages} {60--62}\BibitemShut {NoStop}%
\bibitem [{\citenamefont {Barenco}\ \emph {et~al.}(1997)\citenamefont
  {Barenco}, \citenamefont {Berthiaume}, \citenamefont {Deutsch}, \citenamefont
  {Ekert}, \citenamefont {Jozsa},\ and\ \citenamefont
  {Macchiavello}}]{Barenco1997-do}%
  \BibitemOpen
  \bibfield  {author} {\bibinfo {author} {\bibfnamefont {Adriano}\ \bibnamefont
  {Barenco}}, \bibinfo {author} {\bibfnamefont {Andr{\'e}}\ \bibnamefont
  {Berthiaume}}, \bibinfo {author} {\bibfnamefont {David}\ \bibnamefont
  {Deutsch}}, \bibinfo {author} {\bibfnamefont {Artur}\ \bibnamefont {Ekert}},
  \bibinfo {author} {\bibfnamefont {Richard}\ \bibnamefont {Jozsa}}, \ and\
  \bibinfo {author} {\bibfnamefont {Chiara}\ \bibnamefont {Macchiavello}},\
  }\bibfield  {title} {\enquote {\bibinfo {title} {Stabilization of quantum
  computations by symmetrization},}\ }\href {\doibase
  10.1137/S0097539796302452} {\bibfield  {journal} {\bibinfo  {journal} {SIAM
  J. Comput.}\ }\textbf {\bibinfo {volume} {26}},\ \bibinfo {pages}
  {1541--1557} (\bibinfo {year} {1997})}\BibitemShut {NoStop}%
\bibitem [{\citenamefont {Peres}(1999)}]{Peres1999-rt}%
  \BibitemOpen
  \bibfield  {author} {\bibinfo {author} {\bibfnamefont {Asher}\ \bibnamefont
  {Peres}},\ }\bibfield  {title} {\enquote {\bibinfo {title} {Error
  symmetrization in quantum computers},}\ }\href {\doibase
  10.1023/A:1026648717079} {\bibfield  {journal} {\bibinfo  {journal} {Int. J.
  Theor. Phys.}\ }\textbf {\bibinfo {volume} {38}},\ \bibinfo {pages}
  {799--805} (\bibinfo {year} {1999})}\BibitemShut {NoStop}%
\bibitem [{\citenamefont {Horodecki}\ and\ \citenamefont
  {Ekert}(2002)}]{Horodecki2002-pf}%
  \BibitemOpen
  \bibfield  {author} {\bibinfo {author} {\bibfnamefont {Pawe{\l}}\
  \bibnamefont {Horodecki}}\ and\ \bibinfo {author} {\bibfnamefont {Artur}\
  \bibnamefont {Ekert}},\ }\bibfield  {title} {{\selectlanguage {en}\enquote
  {\bibinfo {title} {Method for direct detection of quantum entanglement},}\
  }}\href {\doibase 10.1103/PhysRevLett.89.127902} {\bibfield  {journal}
  {\bibinfo  {journal} {Phys. Rev. Lett.}\ }\textbf {\bibinfo {volume} {89}},\
  \bibinfo {pages} {127902} (\bibinfo {year} {2002})}\BibitemShut {NoStop}%
\bibitem [{\citenamefont {Ekert}\ \emph {et~al.}(2002)\citenamefont {Ekert},
  \citenamefont {Alves}, \citenamefont {Oi}, \citenamefont {Horodecki},
  \citenamefont {Horodecki},\ and\ \citenamefont {Kwek}}]{Ekert2002-gl}%
  \BibitemOpen
  \bibfield  {author} {\bibinfo {author} {\bibfnamefont {Artur~K}\ \bibnamefont
  {Ekert}}, \bibinfo {author} {\bibfnamefont {Carolina~Moura}\ \bibnamefont
  {Alves}}, \bibinfo {author} {\bibfnamefont {Daniel K~L}\ \bibnamefont {Oi}},
  \bibinfo {author} {\bibfnamefont {Micha{\l}}\ \bibnamefont {Horodecki}},
  \bibinfo {author} {\bibfnamefont {Pawe{\l}}\ \bibnamefont {Horodecki}}, \
  and\ \bibinfo {author} {\bibfnamefont {L~C}\ \bibnamefont {Kwek}},\
  }\bibfield  {title} {{\selectlanguage {en}\enquote {\bibinfo {title} {Direct
  estimations of linear and nonlinear functionals of a quantum state},}\
  }}\href {\doibase 10.1103/PhysRevLett.88.217901} {\bibfield  {journal}
  {\bibinfo  {journal} {Phys. Rev. Lett.}\ }\textbf {\bibinfo {volume} {88}},\
  \bibinfo {pages} {217901} (\bibinfo {year} {2002})}\BibitemShut {NoStop}%
\bibitem [{\citenamefont {Brun}(2004)}]{Brun2004-do}%
  \BibitemOpen
  \bibfield  {author} {\bibinfo {author} {\bibfnamefont {Todd~A}\ \bibnamefont
  {Brun}},\ }\bibfield  {title} {\enquote {\bibinfo {title} {Measuring
  polynomial functions of states},}\ }\href
  {http://arxiv.org/abs/quant-ph/0401067} {\bibfield  {journal} {\bibinfo
  {journal} {arXiv:quant-ph/0401067}\ } (\bibinfo {year} {2004})}\BibitemShut
  {NoStop}%
\bibitem [{\citenamefont {Hastings}\ \emph {et~al.}(2010)\citenamefont
  {Hastings}, \citenamefont {Gonz{\'a}lez}, \citenamefont {Kallin},\ and\
  \citenamefont {Melko}}]{Hastings2010-tk}%
  \BibitemOpen
  \bibfield  {author} {\bibinfo {author} {\bibfnamefont {Matthew~B}\
  \bibnamefont {Hastings}}, \bibinfo {author} {\bibfnamefont {Iv{\'a}n}\
  \bibnamefont {Gonz{\'a}lez}}, \bibinfo {author} {\bibfnamefont {Ann~B}\
  \bibnamefont {Kallin}}, \ and\ \bibinfo {author} {\bibfnamefont {Roger~G}\
  \bibnamefont {Melko}},\ }\bibfield  {title} {{\selectlanguage {en}\enquote
  {\bibinfo {title} {Measuring renyi entanglement entropy in quantum monte
  carlo simulations},}\ }}\href {\doibase 10.1103/PhysRevLett.104.157201}
  {\bibfield  {journal} {\bibinfo  {journal} {Phys. Rev. Lett.}\ }\textbf
  {\bibinfo {volume} {104}},\ \bibinfo {pages} {157201} (\bibinfo {year}
  {2010})}\BibitemShut {NoStop}%
\bibitem [{\citenamefont {Islam}\ \emph {et~al.}(2015)\citenamefont {Islam},
  \citenamefont {Ma}, \citenamefont {Preiss}, \citenamefont {Tai},
  \citenamefont {Lukin}, \citenamefont {Rispoli},\ and\ \citenamefont
  {Greiner}}]{Islam2015-hn}%
  \BibitemOpen
  \bibfield  {author} {\bibinfo {author} {\bibfnamefont {Rajibul}\ \bibnamefont
  {Islam}}, \bibinfo {author} {\bibfnamefont {Ruichao}\ \bibnamefont {Ma}},
  \bibinfo {author} {\bibfnamefont {Philipp~M}\ \bibnamefont {Preiss}},
  \bibinfo {author} {\bibfnamefont {M~Eric}\ \bibnamefont {Tai}}, \bibinfo
  {author} {\bibfnamefont {Alexander}\ \bibnamefont {Lukin}}, \bibinfo {author}
  {\bibfnamefont {Matthew}\ \bibnamefont {Rispoli}}, \ and\ \bibinfo {author}
  {\bibfnamefont {Markus}\ \bibnamefont {Greiner}},\ }\bibfield  {title}
  {{\selectlanguage {en}\enquote {\bibinfo {title} {Measuring entanglement
  entropy in a quantum many-body system},}\ }}\href {\doibase
  10.1038/nature15750} {\bibfield  {journal} {\bibinfo  {journal} {Nature}\
  }\textbf {\bibinfo {volume} {528}},\ \bibinfo {pages} {77--83} (\bibinfo
  {year} {2015})}\BibitemShut {NoStop}%
\bibitem [{\citenamefont {Garcia-Escartin}\ and\ \citenamefont
  {Chamorro-Posada}(2013)}]{Garcia-Escartin2013-en}%
  \BibitemOpen
  \bibfield  {author} {\bibinfo {author} {\bibfnamefont {Juan~Carlos}\
  \bibnamefont {Garcia-Escartin}}\ and\ \bibinfo {author} {\bibfnamefont
  {Pedro}\ \bibnamefont {Chamorro-Posada}},\ }\bibfield  {title} {\enquote
  {\bibinfo {title} {{SWAP} test and {Hong-Ou-Mandel} effect are equivalent},}\
  }\href {\doibase 10.1103/PhysRevA.87.052330} {\bibfield  {journal} {\bibinfo
  {journal} {Phys. Rev. A}\ }\textbf {\bibinfo {volume} {87}},\ \bibinfo
  {pages} {052330} (\bibinfo {year} {2013})}\BibitemShut {NoStop}%
\bibitem [{\citenamefont {Johri}\ \emph {et~al.}(2017)\citenamefont {Johri},
  \citenamefont {Steiger},\ and\ \citenamefont {Troyer}}]{Johri2017-jm}%
  \BibitemOpen
  \bibfield  {author} {\bibinfo {author} {\bibfnamefont {Sonika}\ \bibnamefont
  {Johri}}, \bibinfo {author} {\bibfnamefont {Damian~S}\ \bibnamefont
  {Steiger}}, \ and\ \bibinfo {author} {\bibfnamefont {Matthias}\ \bibnamefont
  {Troyer}},\ }\bibfield  {title} {\enquote {\bibinfo {title} {Entanglement
  spectroscopy on a quantum computer},}\ }\href {\doibase
  10.1103/PhysRevB.96.195136} {\bibfield  {journal} {\bibinfo  {journal} {Phys.
  Rev. B Condens. Matter}\ }\textbf {\bibinfo {volume} {96}},\ \bibinfo {pages}
  {195136} (\bibinfo {year} {2017})}\BibitemShut {NoStop}%
\bibitem [{\citenamefont {Suba{\c s}{\i}}\ \emph {et~al.}(2019)\citenamefont
  {Suba{\c s}{\i}}, \citenamefont {Cincio},\ and\ \citenamefont
  {Coles}}]{Subasi2019-dd}%
  \BibitemOpen
  \bibfield  {author} {\bibinfo {author} {\bibfnamefont {Yi{\u g}it}\
  \bibnamefont {Suba{\c s}{\i}}}, \bibinfo {author} {\bibfnamefont {Lukasz}\
  \bibnamefont {Cincio}}, \ and\ \bibinfo {author} {\bibfnamefont {Patrick~J}\
  \bibnamefont {Coles}},\ }\bibfield  {title} {{\selectlanguage {en}\enquote
  {\bibinfo {title} {Entanglement spectroscopy with a depth-two quantum
  circuit},}\ }}\href {\doibase 10.1088/1751-8121/aaf54d} {\bibfield  {journal}
  {\bibinfo  {journal} {J. Phys. A: Math. Theor.}\ }\textbf {\bibinfo {volume}
  {52}},\ \bibinfo {pages} {044001} (\bibinfo {year} {2019})}\BibitemShut
  {NoStop}%
\bibitem [{\citenamefont {Banchi}\ \emph {et~al.}(2016)\citenamefont {Banchi},
  \citenamefont {Bayat},\ and\ \citenamefont {Bose}}]{Banchi2016-we}%
  \BibitemOpen
  \bibfield  {author} {\bibinfo {author} {\bibfnamefont {Leonardo}\
  \bibnamefont {Banchi}}, \bibinfo {author} {\bibfnamefont {Abolfazl}\
  \bibnamefont {Bayat}}, \ and\ \bibinfo {author} {\bibfnamefont {Sougato}\
  \bibnamefont {Bose}},\ }\bibfield  {title} {\enquote {\bibinfo {title}
  {Entanglement entropy scaling in solid-state spin arrays via capacitance
  measurements},}\ }\href {\doibase 10.1103/PhysRevB.94.241117} {\bibfield
  {journal} {\bibinfo  {journal} {Phys. Rev. B Condens. Matter}\ }\textbf
  {\bibinfo {volume} {94}},\ \bibinfo {pages} {241117} (\bibinfo {year}
  {2016})}\BibitemShut {NoStop}%
\bibitem [{\citenamefont {Bacon}\ \emph {et~al.}(2006)\citenamefont {Bacon},
  \citenamefont {Chuang},\ and\ \citenamefont {Harrow}}]{Bacon2006-xz}%
  \BibitemOpen
  \bibfield  {author} {\bibinfo {author} {\bibfnamefont {Dave}\ \bibnamefont
  {Bacon}}, \bibinfo {author} {\bibfnamefont {Isaac~L}\ \bibnamefont {Chuang}},
  \ and\ \bibinfo {author} {\bibfnamefont {Aram~W}\ \bibnamefont {Harrow}},\
  }\bibfield  {title} {{\selectlanguage {en}\enquote {\bibinfo {title}
  {Efficient quantum circuits for schur and {Clebsch-Gordan} transforms},}\
  }}\href {\doibase 10.1103/PhysRevLett.97.170502} {\bibfield  {journal}
  {\bibinfo  {journal} {Phys. Rev. Lett.}\ }\textbf {\bibinfo {volume} {97}},\
  \bibinfo {pages} {170502} (\bibinfo {year} {2006})}\BibitemShut {NoStop}%
\bibitem [{\citenamefont {Cotler}\ \emph {et~al.}(2019)\citenamefont {Cotler},
  \citenamefont {Choi}, \citenamefont {Lukin}, \citenamefont {Gharibyan},
  \citenamefont {Grover}, \citenamefont {Tai}, \citenamefont {Rispoli},
  \citenamefont {Schittko}, \citenamefont {Preiss}, \citenamefont {Kaufman},
  \citenamefont {Greiner}, \citenamefont {Pichler},\ and\ \citenamefont
  {Hayden}}]{Cotler2019-rh}%
  \BibitemOpen
  \bibfield  {author} {\bibinfo {author} {\bibfnamefont {Jordan}\ \bibnamefont
  {Cotler}}, \bibinfo {author} {\bibfnamefont {Soonwon}\ \bibnamefont {Choi}},
  \bibinfo {author} {\bibfnamefont {Alexander}\ \bibnamefont {Lukin}}, \bibinfo
  {author} {\bibfnamefont {Hrant}\ \bibnamefont {Gharibyan}}, \bibinfo {author}
  {\bibfnamefont {Tarun}\ \bibnamefont {Grover}}, \bibinfo {author}
  {\bibfnamefont {M~Eric}\ \bibnamefont {Tai}}, \bibinfo {author}
  {\bibfnamefont {Matthew}\ \bibnamefont {Rispoli}}, \bibinfo {author}
  {\bibfnamefont {Robert}\ \bibnamefont {Schittko}}, \bibinfo {author}
  {\bibfnamefont {Philipp~M}\ \bibnamefont {Preiss}}, \bibinfo {author}
  {\bibfnamefont {Adam~M}\ \bibnamefont {Kaufman}}, \bibinfo {author}
  {\bibfnamefont {Markus}\ \bibnamefont {Greiner}}, \bibinfo {author}
  {\bibfnamefont {Hannes}\ \bibnamefont {Pichler}}, \ and\ \bibinfo {author}
  {\bibfnamefont {Patrick}\ \bibnamefont {Hayden}},\ }\bibfield  {title}
  {\enquote {\bibinfo {title} {Quantum virtual cooling},}\ }\href {\doibase
  10.1103/PhysRevX.9.031013} {\bibfield  {journal} {\bibinfo  {journal} {Phys.
  Rev. X}\ }\textbf {\bibinfo {volume} {9}},\ \bibinfo {pages} {031013}
  (\bibinfo {year} {2019})}\BibitemShut {NoStop}%
\bibitem [{\citenamefont {Bravyi}\ and\ \citenamefont
  {Kitaev}(2005)}]{Bravyi2005-vi}%
  \BibitemOpen
  \bibfield  {author} {\bibinfo {author} {\bibfnamefont {Sergey}\ \bibnamefont
  {Bravyi}}\ and\ \bibinfo {author} {\bibfnamefont {Alexei}\ \bibnamefont
  {Kitaev}},\ }\bibfield  {title} {\enquote {\bibinfo {title} {Universal
  quantum computation with ideal clifford gates and noisy ancillas},}\ }\href
  {\doibase 10.1103/PhysRevA.71.022316} {\bibfield  {journal} {\bibinfo
  {journal} {Phys. Rev. A}\ }\textbf {\bibinfo {volume} {71}},\ \bibinfo
  {pages} {022316} (\bibinfo {year} {2005})}\BibitemShut {NoStop}%
\bibitem [{\citenamefont {Knill}(2005)}]{Knill2005-vk}%
  \BibitemOpen
  \bibfield  {author} {\bibinfo {author} {\bibfnamefont {E}~\bibnamefont
  {Knill}},\ }\bibfield  {title} {{\selectlanguage {en}\enquote {\bibinfo
  {title} {Quantum computing with realistically noisy devices},}\ }}\href
  {\doibase 10.1038/nature03350} {\bibfield  {journal} {\bibinfo  {journal}
  {Nature}\ }\textbf {\bibinfo {volume} {434}},\ \bibinfo {pages} {39--44}
  (\bibinfo {year} {2005})}\BibitemShut {NoStop}%
\bibitem [{\citenamefont {Haah}\ and\ \citenamefont
  {Hastings}(2018)}]{Haah2018-tg}%
  \BibitemOpen
  \bibfield  {author} {\bibinfo {author} {\bibfnamefont {Jeongwan}\
  \bibnamefont {Haah}}\ and\ \bibinfo {author} {\bibfnamefont {Matthew~B}\
  \bibnamefont {Hastings}},\ }\bibfield  {title} {\enquote {\bibinfo {title}
  {Codes and protocols for distilling {T} , controlled- {S} , and toffoli
  gates},}\ }\href {\doibase 10.22331/q-2018-06-07-71} {\bibfield  {journal}
  {\bibinfo  {journal} {Quantum}\ }\textbf {\bibinfo {volume} {2}},\ \bibinfo
  {pages} {71} (\bibinfo {year} {2018})}\BibitemShut {NoStop}%
\bibitem [{\citenamefont {Cirac}\ \emph {et~al.}(1999)\citenamefont {Cirac},
  \citenamefont {Ekert},\ and\ \citenamefont {Macchiavello}}]{Cirac1999-ir}%
  \BibitemOpen
  \bibfield  {author} {\bibinfo {author} {\bibfnamefont {J~I}\ \bibnamefont
  {Cirac}}, \bibinfo {author} {\bibfnamefont {A~K}\ \bibnamefont {Ekert}}, \
  and\ \bibinfo {author} {\bibfnamefont {C}~\bibnamefont {Macchiavello}},\
  }\bibfield  {title} {\enquote {\bibinfo {title} {Optimal purification of
  single qubits},}\ }\href {\doibase 10.1103/PhysRevLett.82.4344} {\bibfield
  {journal} {\bibinfo  {journal} {Phys. Rev. Lett.}\ }\textbf {\bibinfo
  {volume} {82}},\ \bibinfo {pages} {4344--4347} (\bibinfo {year}
  {1999})}\BibitemShut {NoStop}%
\bibitem [{\citenamefont {Biamonte}\ and\ \citenamefont
  {Bergholm}(2017)}]{Biamonte2017-zx}%
  \BibitemOpen
  \bibfield  {author} {\bibinfo {author} {\bibfnamefont {Jacob}\ \bibnamefont
  {Biamonte}}\ and\ \bibinfo {author} {\bibfnamefont {Ville}\ \bibnamefont
  {Bergholm}},\ }\bibfield  {title} {\enquote {\bibinfo {title} {Tensor
  networks in a nutshell},}\ }\href {http://arxiv.org/abs/1708.00006}
  {\bibfield  {journal} {\bibinfo  {journal} {arXiv:1708.00006}\ } (\bibinfo
  {year} {2017})}\BibitemShut {NoStop}%
\bibitem [{\citenamefont {Bridgeman}\ and\ \citenamefont
  {Chubb}(2017)}]{Bridgeman2017-kg}%
  \BibitemOpen
  \bibfield  {author} {\bibinfo {author} {\bibfnamefont {Jacob~C}\ \bibnamefont
  {Bridgeman}}\ and\ \bibinfo {author} {\bibfnamefont {Christopher~T}\
  \bibnamefont {Chubb}},\ }\bibfield  {title} {{\selectlanguage {en}\enquote
  {\bibinfo {title} {Hand-waving and interpretive dance: an introductory course
  on tensor networks},}\ }}\href {\doibase 10.1088/1751-8121/aa6dc3} {\bibfield
   {journal} {\bibinfo  {journal} {J. Phys. A: Math. Theor.}\ }\textbf
  {\bibinfo {volume} {50}},\ \bibinfo {pages} {223001} (\bibinfo {year}
  {2017})}\BibitemShut {NoStop}%
\bibitem [{\citenamefont {Orus}(2014)}]{Orus2014-xx}%
  \BibitemOpen
  \bibfield  {author} {\bibinfo {author} {\bibfnamefont {Roman}\ \bibnamefont
  {Orus}},\ }\bibfield  {title} {\enquote {\bibinfo {title} {A practical
  introduction to tensor networks: Matrix product states and projected
  entangled pair states},}\ }\href@noop {} {\bibfield  {journal} {\bibinfo
  {journal} {Ann. Phys.}\ }\textbf {\bibinfo {volume} {349}},\ \bibinfo {pages}
  {117--158} (\bibinfo {year} {2014})}\BibitemShut {NoStop}%
\bibitem [{\citenamefont {Kato}(2013)}]{Kato2013-tz}%
  \BibitemOpen
  \bibfield  {author} {\bibinfo {author} {\bibfnamefont {Tosio}\ \bibnamefont
  {Kato}},\ }\href
  {https://play.google.com/store/books/details?id=k-7nCAAAQBAJ}
  {{\selectlanguage {en}\emph {\bibinfo {title} {Perturbation theory for linear
  operators}}}}\ (\bibinfo  {publisher} {Springer Science \& Business Media},\
  \bibinfo {year} {2013})\BibitemShut {NoStop}%
\bibitem [{\citenamefont {Arute}\ \emph {et~al.}(2019)\citenamefont {Arute},
  \citenamefont {Arya}, \citenamefont {Babbush}, \citenamefont {Bacon},
  \citenamefont {Bardin}, \citenamefont {Barends}, \citenamefont {Biswas},
  \citenamefont {Boixo}, \citenamefont {Brandao}, \citenamefont {Buell},
  \citenamefont {Burkett}, \citenamefont {Chen}, \citenamefont {Chen},
  \citenamefont {Chiaro}, \citenamefont {Collins}, \citenamefont {Courtney},
  \citenamefont {Dunsworth}, \citenamefont {Farhi}, \citenamefont {Foxen},
  \citenamefont {Fowler}, \citenamefont {Gidney}, \citenamefont {Giustina},
  \citenamefont {Graff}, \citenamefont {Guerin}, \citenamefont {Habegger},
  \citenamefont {Harrigan}, \citenamefont {Hartmann}, \citenamefont {Ho},
  \citenamefont {Hoffmann}, \citenamefont {Huang}, \citenamefont {Humble},
  \citenamefont {Isakov}, \citenamefont {Jeffrey}, \citenamefont {Jiang},
  \citenamefont {Kafri}, \citenamefont {Kechedzhi}, \citenamefont {Kelly},
  \citenamefont {Klimov}, \citenamefont {Knysh}, \citenamefont {Korotkov},
  \citenamefont {Kostritsa}, \citenamefont {Landhuis}, \citenamefont
  {Lindmark}, \citenamefont {Lucero}, \citenamefont {Lyakh}, \citenamefont
  {Mandr{\`a}}, \citenamefont {McClean}, \citenamefont {McEwen}, \citenamefont
  {Megrant}, \citenamefont {Mi}, \citenamefont {Michielsen}, \citenamefont
  {Mohseni}, \citenamefont {Mutus}, \citenamefont {Naaman}, \citenamefont
  {Neeley}, \citenamefont {Neill}, \citenamefont {Niu}, \citenamefont {Ostby},
  \citenamefont {Petukhov}, \citenamefont {Platt}, \citenamefont {Quintana},
  \citenamefont {Rieffel}, \citenamefont {Roushan}, \citenamefont {Rubin},
  \citenamefont {Sank}, \citenamefont {Satzinger}, \citenamefont {Smelyanskiy},
  \citenamefont {Sung}, \citenamefont {Trevithick}, \citenamefont
  {Vainsencher}, \citenamefont {Villalonga}, \citenamefont {White},
  \citenamefont {Yao}, \citenamefont {Yeh}, \citenamefont {Zalcman},
  \citenamefont {Neven},\ and\ \citenamefont {Martinis}}]{Arute2019-jy}%
  \BibitemOpen
  \bibfield  {author} {\bibinfo {author} {\bibfnamefont {Frank}\ \bibnamefont
  {Arute}}, \bibinfo {author} {\bibfnamefont {Kunal}\ \bibnamefont {Arya}},
  \bibinfo {author} {\bibfnamefont {Ryan}\ \bibnamefont {Babbush}}, \bibinfo
  {author} {\bibfnamefont {Dave}\ \bibnamefont {Bacon}}, \bibinfo {author}
  {\bibfnamefont {Joseph~C}\ \bibnamefont {Bardin}}, \bibinfo {author}
  {\bibfnamefont {Rami}\ \bibnamefont {Barends}}, \bibinfo {author}
  {\bibfnamefont {Rupak}\ \bibnamefont {Biswas}}, \bibinfo {author}
  {\bibfnamefont {Sergio}\ \bibnamefont {Boixo}}, \bibinfo {author}
  {\bibfnamefont {Fernando G S~L}\ \bibnamefont {Brandao}}, \bibinfo {author}
  {\bibfnamefont {David~A}\ \bibnamefont {Buell}}, \bibinfo {author}
  {\bibfnamefont {Brian}\ \bibnamefont {Burkett}}, \bibinfo {author}
  {\bibfnamefont {Yu}~\bibnamefont {Chen}}, \bibinfo {author} {\bibfnamefont
  {Zijun}\ \bibnamefont {Chen}}, \bibinfo {author} {\bibfnamefont {Ben}\
  \bibnamefont {Chiaro}}, \bibinfo {author} {\bibfnamefont {Roberto}\
  \bibnamefont {Collins}}, \bibinfo {author} {\bibfnamefont {William}\
  \bibnamefont {Courtney}}, \bibinfo {author} {\bibfnamefont {Andrew}\
  \bibnamefont {Dunsworth}}, \bibinfo {author} {\bibfnamefont {Edward}\
  \bibnamefont {Farhi}}, \bibinfo {author} {\bibfnamefont {Brooks}\
  \bibnamefont {Foxen}}, \bibinfo {author} {\bibfnamefont {Austin}\
  \bibnamefont {Fowler}}, \bibinfo {author} {\bibfnamefont {Craig}\
  \bibnamefont {Gidney}}, \bibinfo {author} {\bibfnamefont {Marissa}\
  \bibnamefont {Giustina}}, \bibinfo {author} {\bibfnamefont {Rob}\
  \bibnamefont {Graff}}, \bibinfo {author} {\bibfnamefont {Keith}\ \bibnamefont
  {Guerin}}, \bibinfo {author} {\bibfnamefont {Steve}\ \bibnamefont
  {Habegger}}, \bibinfo {author} {\bibfnamefont {Matthew~P}\ \bibnamefont
  {Harrigan}}, \bibinfo {author} {\bibfnamefont {Michael~J}\ \bibnamefont
  {Hartmann}}, \bibinfo {author} {\bibfnamefont {Alan}\ \bibnamefont {Ho}},
  \bibinfo {author} {\bibfnamefont {Markus}\ \bibnamefont {Hoffmann}}, \bibinfo
  {author} {\bibfnamefont {Trent}\ \bibnamefont {Huang}}, \bibinfo {author}
  {\bibfnamefont {Travis~S}\ \bibnamefont {Humble}}, \bibinfo {author}
  {\bibfnamefont {Sergei~V}\ \bibnamefont {Isakov}}, \bibinfo {author}
  {\bibfnamefont {Evan}\ \bibnamefont {Jeffrey}}, \bibinfo {author}
  {\bibfnamefont {Zhang}\ \bibnamefont {Jiang}}, \bibinfo {author}
  {\bibfnamefont {Dvir}\ \bibnamefont {Kafri}}, \bibinfo {author}
  {\bibfnamefont {Kostyantyn}\ \bibnamefont {Kechedzhi}}, \bibinfo {author}
  {\bibfnamefont {Julian}\ \bibnamefont {Kelly}}, \bibinfo {author}
  {\bibfnamefont {Paul~V}\ \bibnamefont {Klimov}}, \bibinfo {author}
  {\bibfnamefont {Sergey}\ \bibnamefont {Knysh}}, \bibinfo {author}
  {\bibfnamefont {Alexander}\ \bibnamefont {Korotkov}}, \bibinfo {author}
  {\bibfnamefont {Fedor}\ \bibnamefont {Kostritsa}}, \bibinfo {author}
  {\bibfnamefont {David}\ \bibnamefont {Landhuis}}, \bibinfo {author}
  {\bibfnamefont {Mike}\ \bibnamefont {Lindmark}}, \bibinfo {author}
  {\bibfnamefont {Erik}\ \bibnamefont {Lucero}}, \bibinfo {author}
  {\bibfnamefont {Dmitry}\ \bibnamefont {Lyakh}}, \bibinfo {author}
  {\bibfnamefont {Salvatore}\ \bibnamefont {Mandr{\`a}}}, \bibinfo {author}
  {\bibfnamefont {Jarrod~R}\ \bibnamefont {McClean}}, \bibinfo {author}
  {\bibfnamefont {Matthew}\ \bibnamefont {McEwen}}, \bibinfo {author}
  {\bibfnamefont {Anthony}\ \bibnamefont {Megrant}}, \bibinfo {author}
  {\bibfnamefont {Xiao}\ \bibnamefont {Mi}}, \bibinfo {author} {\bibfnamefont
  {Kristel}\ \bibnamefont {Michielsen}}, \bibinfo {author} {\bibfnamefont
  {Masoud}\ \bibnamefont {Mohseni}}, \bibinfo {author} {\bibfnamefont {Josh}\
  \bibnamefont {Mutus}}, \bibinfo {author} {\bibfnamefont {Ofer}\ \bibnamefont
  {Naaman}}, \bibinfo {author} {\bibfnamefont {Matthew}\ \bibnamefont
  {Neeley}}, \bibinfo {author} {\bibfnamefont {Charles}\ \bibnamefont {Neill}},
  \bibinfo {author} {\bibfnamefont {Murphy~Yuezhen}\ \bibnamefont {Niu}},
  \bibinfo {author} {\bibfnamefont {Eric}\ \bibnamefont {Ostby}}, \bibinfo
  {author} {\bibfnamefont {Andre}\ \bibnamefont {Petukhov}}, \bibinfo {author}
  {\bibfnamefont {John~C}\ \bibnamefont {Platt}}, \bibinfo {author}
  {\bibfnamefont {Chris}\ \bibnamefont {Quintana}}, \bibinfo {author}
  {\bibfnamefont {Eleanor~G}\ \bibnamefont {Rieffel}}, \bibinfo {author}
  {\bibfnamefont {Pedram}\ \bibnamefont {Roushan}}, \bibinfo {author}
  {\bibfnamefont {Nicholas~C}\ \bibnamefont {Rubin}}, \bibinfo {author}
  {\bibfnamefont {Daniel}\ \bibnamefont {Sank}}, \bibinfo {author}
  {\bibfnamefont {Kevin~J}\ \bibnamefont {Satzinger}}, \bibinfo {author}
  {\bibfnamefont {Vadim}\ \bibnamefont {Smelyanskiy}}, \bibinfo {author}
  {\bibfnamefont {Kevin~J}\ \bibnamefont {Sung}}, \bibinfo {author}
  {\bibfnamefont {Matthew~D}\ \bibnamefont {Trevithick}}, \bibinfo {author}
  {\bibfnamefont {Amit}\ \bibnamefont {Vainsencher}}, \bibinfo {author}
  {\bibfnamefont {Benjamin}\ \bibnamefont {Villalonga}}, \bibinfo {author}
  {\bibfnamefont {Theodore}\ \bibnamefont {White}}, \bibinfo {author}
  {\bibfnamefont {Z~Jamie}\ \bibnamefont {Yao}}, \bibinfo {author}
  {\bibfnamefont {Ping}\ \bibnamefont {Yeh}}, \bibinfo {author} {\bibfnamefont
  {Adam}\ \bibnamefont {Zalcman}}, \bibinfo {author} {\bibfnamefont {Hartmut}\
  \bibnamefont {Neven}}, \ and\ \bibinfo {author} {\bibfnamefont {John~M}\
  \bibnamefont {Martinis}},\ }\bibfield  {title} {{\selectlanguage {en}\enquote
  {\bibinfo {title} {Quantum supremacy using a programmable superconducting
  processor},}\ }}\href {\doibase 10.1038/s41586-019-1666-5} {\bibfield
  {journal} {\bibinfo  {journal} {Nature}\ }\textbf {\bibinfo {volume} {574}},\
  \bibinfo {pages} {505--510} (\bibinfo {year} {2019})}\BibitemShut {NoStop}%
\bibitem [{\citenamefont {Dmitriev}\ and\ \citenamefont
  {Krivnov}(2004)}]{Dmitriev2004-de}%
  \BibitemOpen
  \bibfield  {author} {\bibinfo {author} {\bibfnamefont {D~V}\ \bibnamefont
  {Dmitriev}}\ and\ \bibinfo {author} {\bibfnamefont {V~Ya}\ \bibnamefont
  {Krivnov}},\ }\bibfield  {title} {\enquote {\bibinfo {title}
  {Quasi-one-dimensional anisotropic heisenberg model in a transverse magnetic
  field},}\ }\href {\doibase 10.1134/1.1825110} {\bibfield  {journal} {\bibinfo
   {journal} {Journal of Experimental and Theoretical Physics Letters}\
  }\textbf {\bibinfo {volume} {80}},\ \bibinfo {pages} {303--307} (\bibinfo
  {year} {2004})}\BibitemShut {NoStop}%
\bibitem [{\citenamefont {Endo}\ \emph {et~al.}(2019)\citenamefont {Endo},
  \citenamefont {Zhao}, \citenamefont {Li}, \citenamefont {Benjamin},\ and\
  \citenamefont {Yuan}}]{Endo2019-rr}%
  \BibitemOpen
  \bibfield  {author} {\bibinfo {author} {\bibfnamefont {Suguru}\ \bibnamefont
  {Endo}}, \bibinfo {author} {\bibfnamefont {Qi}~\bibnamefont {Zhao}}, \bibinfo
  {author} {\bibfnamefont {Ying}\ \bibnamefont {Li}}, \bibinfo {author}
  {\bibfnamefont {Simon}\ \bibnamefont {Benjamin}}, \ and\ \bibinfo {author}
  {\bibfnamefont {Xiao}\ \bibnamefont {Yuan}},\ }\bibfield  {title} {\enquote
  {\bibinfo {title} {Mitigating algorithmic errors in a hamiltonian
  simulation},}\ }\href {\doibase 10.1103/PhysRevA.99.012334} {\bibfield
  {journal} {\bibinfo  {journal} {Phys. Rev. A}\ }\textbf {\bibinfo {volume}
  {99}},\ \bibinfo {pages} {012334} (\bibinfo {year} {2019})}\BibitemShut
  {NoStop}%
\bibitem [{\citenamefont {Tran}\ \emph {et~al.}(2020)\citenamefont {Tran},
  \citenamefont {Su}, \citenamefont {Carney},\ and\ \citenamefont
  {Taylor}}]{Tran2020-ts}%
  \BibitemOpen
  \bibfield  {author} {\bibinfo {author} {\bibfnamefont {Minh~C}\ \bibnamefont
  {Tran}}, \bibinfo {author} {\bibfnamefont {Yuan}\ \bibnamefont {Su}},
  \bibinfo {author} {\bibfnamefont {Daniel}\ \bibnamefont {Carney}}, \ and\
  \bibinfo {author} {\bibfnamefont {Jacob~M}\ \bibnamefont {Taylor}},\
  }\bibfield  {title} {\enquote {\bibinfo {title} {Faster digital quantum
  simulation by symmetry protection},}\ }\href
  {http://arxiv.org/abs/2006.16248} {\bibfield  {journal} {\bibinfo  {journal}
  {arXiv:2006.16248}\ } (\bibinfo {year} {2020})}\BibitemShut {NoStop}%
\bibitem [{\citenamefont {Campbell}(2019)}]{Campbell2019-id}%
  \BibitemOpen
  \bibfield  {author} {\bibinfo {author} {\bibfnamefont {Earl}\ \bibnamefont
  {Campbell}},\ }\bibfield  {title} {{\selectlanguage {en}\enquote {\bibinfo
  {title} {Random compiler for fast hamiltonian simulation},}\ }}\href
  {\doibase 10.1103/PhysRevLett.123.070503} {\bibfield  {journal} {\bibinfo
  {journal} {Phys. Rev. Lett.}\ }\textbf {\bibinfo {volume} {123}},\ \bibinfo
  {pages} {070503} (\bibinfo {year} {2019})}\BibitemShut {NoStop}%
\bibitem [{\citenamefont {Childs}\ \emph {et~al.}(2019)\citenamefont {Childs},
  \citenamefont {Ostrander},\ and\ \citenamefont {Su}}]{Childs2019-dv}%
  \BibitemOpen
  \bibfield  {author} {\bibinfo {author} {\bibfnamefont {Andrew~M}\
  \bibnamefont {Childs}}, \bibinfo {author} {\bibfnamefont {Aaron}\
  \bibnamefont {Ostrander}}, \ and\ \bibinfo {author} {\bibfnamefont {Yuan}\
  \bibnamefont {Su}},\ }\bibfield  {title} {\enquote {\bibinfo {title} {Faster
  quantum simulation by randomization},}\ }\href {\doibase
  10.22331/q-2019-09-02-182} {\bibfield  {journal} {\bibinfo  {journal}
  {Quantum}\ }\textbf {\bibinfo {volume} {3}},\ \bibinfo {pages} {182}
  (\bibinfo {year} {2019})}\BibitemShut {NoStop}%
\bibitem [{\citenamefont {Ouyang}\ \emph {et~al.}(2020)\citenamefont {Ouyang},
  \citenamefont {White},\ and\ \citenamefont {Campbell}}]{Ouyang2020-pp}%
  \BibitemOpen
  \bibfield  {author} {\bibinfo {author} {\bibfnamefont {Yingkai}\ \bibnamefont
  {Ouyang}}, \bibinfo {author} {\bibfnamefont {David~R}\ \bibnamefont {White}},
  \ and\ \bibinfo {author} {\bibfnamefont {Earl~T}\ \bibnamefont {Campbell}},\
  }\bibfield  {title} {\enquote {\bibinfo {title} {Compilation by stochastic
  {H}amiltonian sparsification},}\ }\href {\doibase 10.22331/q-2020-02-27-235}
  {\bibfield  {journal} {\bibinfo  {journal} {Quantum}\ }\textbf {\bibinfo
  {volume} {4}},\ \bibinfo {pages} {235} (\bibinfo {year} {2020})}\BibitemShut
  {NoStop}%
\bibitem [{\citenamefont {Kivlichan}\ \emph {et~al.}(2018)\citenamefont
  {Kivlichan}, \citenamefont {McClean}, \citenamefont {Wiebe}, \citenamefont
  {Gidney}, \citenamefont {Aspuru-Guzik}, \citenamefont {Chan},\ and\
  \citenamefont {Babbush}}]{Kivlichan2018-yz}%
  \BibitemOpen
  \bibfield  {author} {\bibinfo {author} {\bibfnamefont {Ian~D}\ \bibnamefont
  {Kivlichan}}, \bibinfo {author} {\bibfnamefont {Jarrod}\ \bibnamefont
  {McClean}}, \bibinfo {author} {\bibfnamefont {Nathan}\ \bibnamefont {Wiebe}},
  \bibinfo {author} {\bibfnamefont {Craig}\ \bibnamefont {Gidney}}, \bibinfo
  {author} {\bibfnamefont {Al{\'a}n}\ \bibnamefont {Aspuru-Guzik}}, \bibinfo
  {author} {\bibfnamefont {Garnet Kin-Lic}\ \bibnamefont {Chan}}, \ and\
  \bibinfo {author} {\bibfnamefont {Ryan}\ \bibnamefont {Babbush}},\ }\bibfield
   {title} {{\selectlanguage {en}\enquote {\bibinfo {title} {Quantum simulation
  of electronic structure with linear depth and connectivity},}\ }}\href
  {\doibase 10.1103/PhysRevLett.120.110501} {\bibfield  {journal} {\bibinfo
  {journal} {Phys. Rev. Lett.}\ }\textbf {\bibinfo {volume} {120}},\ \bibinfo
  {pages} {110501} (\bibinfo {year} {2018})}\BibitemShut {NoStop}%
\bibitem [{\citenamefont {O'Gorman}\ \emph {et~al.}(2019)\citenamefont
  {O'Gorman}, \citenamefont {Huggins}, \citenamefont {Rieffel},\ and\
  \citenamefont {Birgitta~Whaley}}]{OGorman2019-jj}%
  \BibitemOpen
  \bibfield  {author} {\bibinfo {author} {\bibfnamefont {Bryan}\ \bibnamefont
  {O'Gorman}}, \bibinfo {author} {\bibfnamefont {William~J}\ \bibnamefont
  {Huggins}}, \bibinfo {author} {\bibfnamefont {Eleanor~G}\ \bibnamefont
  {Rieffel}}, \ and\ \bibinfo {author} {\bibfnamefont {K}~\bibnamefont
  {Birgitta~Whaley}},\ }\bibfield  {title} {\enquote {\bibinfo {title}
  {Generalized swap networks for near-term quantum computing},}\ }\href
  {http://arxiv.org/abs/1905.05118} {\  (\bibinfo {year} {2019})},\ \Eprint
  {http://arxiv.org/abs/1905.05118} {arXiv:1905.05118 [quant-ph]} \BibitemShut
  {NoStop}%
\bibitem [{\citenamefont {Koczor}(2020)}]{Koczor2020-ub}%
  \BibitemOpen
  \bibfield  {author} {\bibinfo {author} {\bibfnamefont {B{\'a}lint}\
  \bibnamefont {Koczor}},\ }\bibfield  {title} {\enquote {\bibinfo {title}
  {Exponential error suppression for {Near-Term} quantum devices},}\ }\href
  {http://arxiv.org/abs/2011.05942} {\bibfield  {journal} {\bibinfo  {journal}
  {arXiv:2011.05942}\ } (\bibinfo {year} {2020})}\BibitemShut {NoStop}%
\bibitem [{\citenamefont {Cotler}\ and\ \citenamefont
  {Wilczek}(2020)}]{Cotler2020-xc}%
  \BibitemOpen
  \bibfield  {author} {\bibinfo {author} {\bibfnamefont {Jordan}\ \bibnamefont
  {Cotler}}\ and\ \bibinfo {author} {\bibfnamefont {Frank}\ \bibnamefont
  {Wilczek}},\ }\bibfield  {title} {{\selectlanguage {en}\enquote {\bibinfo
  {title} {Quantum overlapping tomography},}\ }}\href {\doibase
  10.1103/PhysRevLett.124.100401} {\bibfield  {journal} {\bibinfo  {journal}
  {Phys. Rev. Lett.}\ }\textbf {\bibinfo {volume} {124}},\ \bibinfo {pages}
  {100401} (\bibinfo {year} {2020})}\BibitemShut {NoStop}%
\bibitem [{\citenamefont {Bonet-Monroig}\ \emph {et~al.}(2020)\citenamefont
  {Bonet-Monroig}, \citenamefont {Babbush},\ and\ \citenamefont
  {O'Brien}}]{Bonet-Monroig2020-lc}%
  \BibitemOpen
  \bibfield  {author} {\bibinfo {author} {\bibfnamefont {Xavier}\ \bibnamefont
  {Bonet-Monroig}}, \bibinfo {author} {\bibfnamefont {Ryan}\ \bibnamefont
  {Babbush}}, \ and\ \bibinfo {author} {\bibfnamefont {Thomas~E}\ \bibnamefont
  {O'Brien}},\ }\bibfield  {title} {\enquote {\bibinfo {title} {Nearly optimal
  measurement scheduling for partial tomography of quantum states},}\ }\href
  {\doibase 10.1103/PhysRevX.10.031064} {\bibfield  {journal} {\bibinfo
  {journal} {Phys. Rev. X}\ }\textbf {\bibinfo {volume} {10}},\ \bibinfo
  {pages} {031064} (\bibinfo {year} {2020})}\BibitemShut {NoStop}%
\bibitem [{\citenamefont {Yu.~Kitaev}(1995)}]{Yu_Kitaev1995-zv}%
  \BibitemOpen
  \bibfield  {author} {\bibinfo {author} {\bibfnamefont {A}~\bibnamefont
  {Yu.~Kitaev}},\ }\bibfield  {title} {\enquote {\bibinfo {title} {Quantum
  measurements and the abelian stabilizer problem},}\ }\href
  {http://arxiv.org/abs/quant-ph/9511026} {\bibfield  {journal} {\bibinfo
  {journal} {arXiv:quant-ph/9511026}\ } (\bibinfo {year} {1995})}\BibitemShut
  {NoStop}%
\bibitem [{\citenamefont {Kendall}\ and\ \citenamefont
  {{Others}}(1946)}]{Kendall1946-by}%
  \BibitemOpen
  \bibfield  {author} {\bibinfo {author} {\bibfnamefont {Maurice~George}\
  \bibnamefont {Kendall}}\ and\ \bibinfo {author} {\bibnamefont {{Others}}},\
  }\href@noop {} {\emph {\bibinfo {title} {The Advanced Theory of
  Statistics}}}\ (\bibinfo  {publisher} {Charles Griffin and Co., Ltd.,
  London},\ \bibinfo {year} {1946})\BibitemShut {NoStop}%
\bibitem [{\citenamefont {Wang}\ \emph {et~al.}(2010)\citenamefont {Wang},
  \citenamefont {Fowler}, \citenamefont {Stephens},\ and\ \citenamefont
  {Hollenberg}}]{Wang2010-wq}%
  \BibitemOpen
  \bibfield  {author} {\bibinfo {author} {\bibfnamefont {D~S}\ \bibnamefont
  {Wang}}, \bibinfo {author} {\bibfnamefont {A~G}\ \bibnamefont {Fowler}},
  \bibinfo {author} {\bibfnamefont {A~M}\ \bibnamefont {Stephens}}, \ and\
  \bibinfo {author} {\bibfnamefont {L~C~L}\ \bibnamefont {Hollenberg}},\
  }\bibfield  {title} {\enquote {\bibinfo {title} {Threshold error rates for
  the toric and planar codes},}\ }\href
  {https://dl.acm.org/doi/10.5555/2011362.2011368} {\bibfield  {journal}
  {\bibinfo  {journal} {Quantum Inf. Comput.}\ }\textbf {\bibinfo {volume}
  {10}},\ \bibinfo {pages} {456--469} (\bibinfo {year} {2010})}\BibitemShut
  {NoStop}%
\bibitem [{\citenamefont {Horsman}\ \emph {et~al.}(2012)\citenamefont
  {Horsman}, \citenamefont {Fowler}, \citenamefont {Devitt},\ and\
  \citenamefont {Van~Meter}}]{Horsman2012-vd}%
  \BibitemOpen
  \bibfield  {author} {\bibinfo {author} {\bibfnamefont {Clare}\ \bibnamefont
  {Horsman}}, \bibinfo {author} {\bibfnamefont {Austin~G}\ \bibnamefont
  {Fowler}}, \bibinfo {author} {\bibfnamefont {Simon}\ \bibnamefont {Devitt}},
  \ and\ \bibinfo {author} {\bibfnamefont {Rodney}\ \bibnamefont {Van~Meter}},\
  }\bibfield  {title} {{\selectlanguage {en}\enquote {\bibinfo {title} {Surface
  code quantum computing by lattice surgery},}\ }}\href {\doibase
  10.1088/1367-2630/14/12/123011} {\bibfield  {journal} {\bibinfo  {journal}
  {New J. Phys.}\ }\textbf {\bibinfo {volume} {14}},\ \bibinfo {pages} {123011}
  (\bibinfo {year} {2012})}\BibitemShut {NoStop}%
\bibitem [{\citenamefont {Litinski}(2019)}]{Litinski2019-dg}%
  \BibitemOpen
  \bibfield  {author} {\bibinfo {author} {\bibfnamefont {Daniel}\ \bibnamefont
  {Litinski}},\ }\bibfield  {title} {\enquote {\bibinfo {title} {A {G}ame of
  {S}urface {C}odes: {{L}arge-{S}cale} {Q}uantum {C}omputing with {L}attice
  {S}urgery},}\ }\href {\doibase 10.22331/q-2019-03-05-128} {\bibfield
  {journal} {\bibinfo  {journal} {Quantum}\ }\textbf {\bibinfo {volume} {3}},\
  \bibinfo {pages} {128} (\bibinfo {year} {2019})}\BibitemShut {NoStop}%
\bibitem [{\citenamefont {Preskill}(1998)}]{Preskill1998-cb}%
  \BibitemOpen
  \bibfield  {author} {\bibinfo {author} {\bibfnamefont {John}\ \bibnamefont
  {Preskill}},\ }\bibfield  {title} {\enquote {\bibinfo {title} {Lecture notes
  for physics 229: Quantum information and computation},}\ }\href
  {http://www2.fiit.stuba.sk/~kvasnicka/QuantumComputing/PreskilTextbook_all.pdf}
  {\  (\bibinfo {year} {1998})}\BibitemShut {NoStop}%
\bibitem [{\citenamefont {Krantz}\ \emph {et~al.}(2019)\citenamefont {Krantz},
  \citenamefont {Kjaergaard}, \citenamefont {Yan}, \citenamefont {Orlando},
  \citenamefont {Gustavsson},\ and\ \citenamefont {Oliver}}]{Krantz2019-gl}%
  \BibitemOpen
  \bibfield  {author} {\bibinfo {author} {\bibfnamefont {P}~\bibnamefont
  {Krantz}}, \bibinfo {author} {\bibfnamefont {M}~\bibnamefont {Kjaergaard}},
  \bibinfo {author} {\bibfnamefont {F}~\bibnamefont {Yan}}, \bibinfo {author}
  {\bibfnamefont {T~P}\ \bibnamefont {Orlando}}, \bibinfo {author}
  {\bibfnamefont {S}~\bibnamefont {Gustavsson}}, \ and\ \bibinfo {author}
  {\bibfnamefont {W~D}\ \bibnamefont {Oliver}},\ }\bibfield  {title} {\enquote
  {\bibinfo {title} {A quantum engineer's guide to superconducting qubits},}\
  }\href {\doibase 10.1063/1.5089550} {\bibfield  {journal} {\bibinfo
  {journal} {Applied Physics Reviews}\ }\textbf {\bibinfo {volume} {6}},\
  \bibinfo {pages} {021318} (\bibinfo {year} {2019})}\BibitemShut {NoStop}%
\end{thebibliography}
%

\onecolumngrid
\appendix

\section{Measuring Multi-Qubit Observables by Diagonalization}
\label{app:diagonalize_multi_qubit}

In \sec{measurement_by_diagonalization}, we presented a straightforward
strategy for applying the \(M=2\) version of virtual distillation to
single-qubit observables.
Measuring observables with support on more than one qubit is more challenging.
Ref.~\citenum{Cotler2019-rh} solved this issue by using an approach like the one
we describe below in \app{ancilla_assisted_measurement}. Alternatively, one
could apply a circuit to localize an observable of interest to a single qubit
before performing virtual distillation.
Here we present an alternative solution that doesn't require the use of
ancilla-assisted measurement or circuit depth.

The challenge arises due to the use of \eq{symmetrized_O_M_ev}, in particular,
the choice to use the symmetrized version of an observable, a notion defined in
\eq{symmetrized_O}.
Using the symmetrized version of a multi-qubit observable means that it is not
possible to perform the required diagonalization using a tensor product of
separate unitaries across each pair of qubits.
As an example, we consider the operator \(O = Z_iZ_j\). 
Our arguments hold equally well for any other operator composed of a tensor
product of (more than one) single-qubit Pauli operators.
Taking the product of the symmetrized observable and the swap operator yields
\begin{equation}
  O^{(2)}S^{(2)} =
  \frac{1}{2}(Z_i^\textbf{1}Z_j^\textbf{1} + Z_i^\textbf{2}Z_j^\textbf{2})S^{(2)}.
  \label{eq:symmetrized_two_point_example}
\end{equation}
This operator does not factorize into a tensor product of operators with support
on the individual pairs of qubits, nor can it be diagonalized by an operator
that factors this way.

However, instead of using \eq{symmetrized_O_M_ev} to determine the corrected
expectation value of \(O\), we can instead use the non-symmetrized form
introduced in \eq{O_M_ev}.
Returning to our example where \(O = Z_i Z_j\), we see that we need to estimate
the numerator and denominator of
\begin{equation}
  \frac{\tr(Z_i^\textbf{1}Z_j^\textbf{1} S^{(2)} \rho^{\otimes 2})}{\tr(S^{(2)} \rho^{\otimes 2})}.
  \label{eq:nonsymmetrized_two_point_example}
\end{equation}
Unlike the symmetrized observable of \eq{symmetrized_two_point_example}, the
operator \(Z_i^\textbf{1} Z_j^\textbf{1} S^{(2)}\) factorizes into a tensor
product over the \(N\) pairs of qubits (a pair being one qubit from the first system and the corresponding qubit from the second system).
\(Z_i^\textbf{1} Z_j^\textbf{1} S^{(2)}\) is not Hermitian, but because it is
unitary, we can still estimate \(\tr(Z_i^\textbf{1} Z_j^\textbf{1} S^{(2)}
\rho^{\otimes 2})\) by applying a circuit to diagonalize it and measuring in the
computational basis.
As \(Z_i^\textbf{1} Z_j^\textbf{1} S^{(2)}\) factorizes into a product of
two-qubit operators, the circuit that diagonalizes it does as well.

Note that because \(Z_i^\textbf{1}Z_j^\textbf{1}\) does not commute with
\(S^{(2)}\), we will be unable to simultaneously estimate the numerator and
denominator of \eq{nonsymmetrized_two_point_example}. 
We will also be unable to simultaneously measure the corrected expectation value
corresponding to different choices of \(i\) and \(j\). 
More generally, we are able to measure any single tensor product of one-qubit
operators at a time, regardless of the number of qubits it acts on. 
The details of the diagonalization will, of course, depend upon the operator to
be measured.
We do not carefully analyze the number of measurements required by this flavor
of virtual distillation, but we note that the inability to parallelize the
measurement of commuting multi-qubit observables would make it challenging to
profitably combine this approach with sophisticated NISQ measurement strategies,
such as the one presented in Ref.~\citenum{Huggins2019-vu}. 
In particular, individual multi-qubit operator must be measured separately using
this approach, even if they commute. 
This fact increases the overall number of circuit repetitions required for many
applications.

\section{Measurement by Diagonalization with Three or More Copies}
\label{app:diagonalize_three_or_more}

In \sec{measurement_by_diagonalization} and \app{diagonalize_multi_qubit} we
described protocols for measuring the expectation value of \(O\) with respect to
the state \(\frac{\rho^2}{\tr(\rho^2)}\) by diagonalizing \(S^{(2)}\) and either
\(O^\textbf{1}S^{(2)}\) or \(O^{(2)}S^{(2)}\). 
Here we describe how these approaches can be generalized to higher powers of
\(\rho\) in a natural way.
Like the swap operator, the cyclic shift operator between \(M\) \(N\)-qubit
systems, \(S^{(M)}\),
factorizes into a tensor product of \(N\) \(M\)-qubit gates. Specifically, it
factorizes into the tensor product of \(N\) single-qubit cyclic shift operators.
The symmetrized operator \(Z^{(M)}_k = \frac{1}{M} \sum_{i=1}^M Z^\textbf{i}_k\)
commutes with the operator \(S^{(M)}\). 
Therefore, \(Z_k^{(M)}S^{(M)}\) and \(S^{(M)}\) are simultaneously
diagonalizable even though \(S^{(M)}\) is unitary but not Hermitian for \(M >
2\).
Because \(Z_k^{(M)}\) and \(S^{(M)}\) both factorize into tensor products over
\(N\) \(M\)-tuples of qubits, the unitary that diagonalizes these operators then
factorizes into a tensor product of \(M\)-qubit operators in the same way.

The same concerns about correcting the expectation values of multi-qubit
observables that we discussed in \app{diagonalize_multi_qubit} for the two-copy
(\(M=2\)) case apply to this generalized proposal. 
The tools developed so far allow us to simultaneously estimate \(\tr(Z_k
\rho^M)\) for all values of \(m\) and also \(\tr(\rho^M)\). 
If we are interested in reconstructing \(\tr(P \rho^M)\) for some multi-qubit
Pauli operator \(P\), we can do so using a generalization of
\eq{nonsymmetrized_two_point_example}, but we would be limited to measuring the
operators required for one particular \(P\) at a time.

\begin{figure}[t]
  \centering
\includegraphics[width=.5\textwidth]{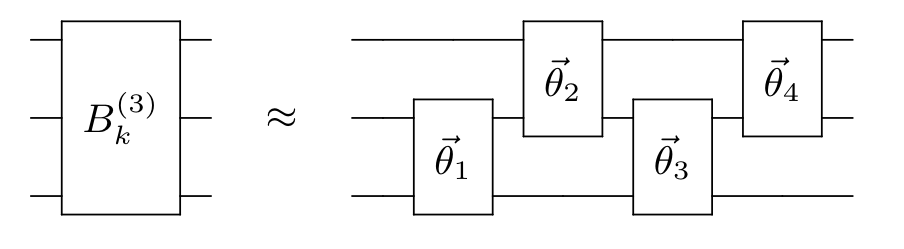}
  \caption{The ansatz that we numerically optimize to approximately diagonalize
    \(S^{(3)}_k\) and \(Z_k^{(3)}\). 
    The two-qubit gates parameterized by the \(\vec{\theta}_i\)s are arbitrary
    two-qubit gates. 
    We performed the numerical optimization using the Julia language.}
  \label{fig:three_qubit_beamsplitter}
\end{figure}

For the specific case of \(M=3\), we have numerically optimized the quantum
circuit of \fig{three_qubit_beamsplitter} to simultaneously diagonalize
\(Z_k^{(3)}S^{(3)}_k\) and \(S^{(3)}_k\). 
We obtained parameters for the four two-qubit gates that allow for an error
(measured in the Frobenius norm of the difference between the exact and
approximate matrices) of approximately \(5E-5\) when the following equations are
used,
\begin{align}
  B^{(3)}_k S^{(3)}_k B^{(3)\dagger}_{k} \rightarrow&  \frac{1}{8} \Big( 2 + \\
                                                    & ~\quad (-3 - \sqrt{3}i) Z_k^\textbf{1} + \nonumber \\ 
                                                    & ~\quad (3 - \sqrt{3}i) Z_k^\textbf{1}Z_k^\textbf{2} + \nonumber \\ 
                                                    & ~\quad (3 + \sqrt{3}i) Z_k^\textbf{3} + \nonumber \\
                                                    & ~\quad 2i \sqrt(3) Z_k^\textbf{1}Z_k^\textbf{3} + \nonumber \\
                                                    & ~\quad (3 - \sqrt{3}i) Z_k^\textbf{2}Z_k^\textbf{3} \Big) , \nonumber \\
  B^{(3)}_k Z^{(3)}_k B^{(3) \dagger}_{k} \rightarrow& \frac{1}{3}(Z_k^\textbf{1} + Z_k^\textbf{2} + Z_k^\textbf{3}).
\end{align}

\section{Ancilla-Assisted Measurement}
\label{app:ancilla_assisted_measurement}

In this section, we present the approach one may take if ancilla-assisted
measurement is feasible in the experimental setup. 
This is a simplified version of the proposal for ancilla-assisted measurement
protocol found in Ref.~\citenum{Cotler2019-rh} for estimating the expectation
value of an observable \(O\) with respect to the state \(\rho^2 / \tr(\rho^2)\).
As with the method we discussed in \sec{measurement_by_diagonalization}, we do
this by approximating the numerator and denominator of \eq{symmetrized_O_M_ev}.
Unlike that method, this approach uses a non-destructive measurement of the swap
operator (\(S^{(2)}\)). 
The main reward for this added complexity is that this variant of virtual
distillation doesn't restrict the form of the operators being measured, nor does
it prevent simultaneous measurement of operators acting on overlapping subsets
of qubits. 
Therefore, it is compatible with some of the recently developed techniques for
efficiently measuring a large collection of commuting
operators~\cite{Cotler2020-xc, Bonet-Monroig2020-lc, Huggins2019-vu}. 
While we focus on the \(M=2\) copy version here, we also briefly discuss the
generalization to \(M\geq 3\) copies. 

To use this method, we begin with two system registers, each in the state
\(\rho\), as well as an ancilla qubit in the \(\ket{0}\) state.  
We then perform a non-destructive measurement of \(S^{(2)}\) in the standard
way, using the so-called swap or Hadamard
test~\cite{Yu_Kitaev1995-zv,Horodecki2002-pf, Ekert2002-gl}.
Specifically, we apply a Hadamard gate to the ancilla qubit, apply \(S^{(2)}\)
conditioned on the ancilla qubit being in the \(\ket{1}\) state, and measure the
ancilla qubit in the \(X\) basis.
The expectation value of \(X\) on the ancilla qubit is then equal to
\(\ev{S^{(2)}}\).
Because \(S^{(2)}\) factorizes into a tensor product of two-qubit swap gates,
its controlled version likewise factorizes into a series of \(N\) Fredkin
(controlled-swap) gates. Compiling this circuit may necessitate some extra steps
(such as expanding the single ancilla qubit into a GHZ state using a series of
CNOT gates) in order to deal with the restricted connectivity of a near-term device.

It isn't technically necessary, but it simplifies the analysis and reduces the
variance of the resulting estimator to focus on the symmetrized form of \(O\),
\(O^{(2)} = \frac{1}{2}(O^\textbf{1} + O^\textbf{2})\). 
As in \sec{measurement_by_diagonalization}, this is beneficial because the
symmetrized observable \(O^{(2)}\) commutes with \(S^{(2)}\). 
We can therefore measure the product \(O^{(2)} S^{(2)}\) by first measuring
\(S^{(2)}\) using the Hadamard test described above and then measuring
\(O^{(2)}\) on the system registers. 
This protocol does not require a separate estimation of \(\tr(\rho O)\) like the
original proposal of Ref.~\citenum{Cotler2019-rh}. 
Furthermore, it allows us to make us to make use of measurements of \(O\) on
both copies of \(\rho\) and also simultaneously estimate the numerator and
denominator of \eq{symmetrized_O_M_ev}, leading to a relatively sample-efficient scheme.


Now let us consider the case with three or more copies of \(\rho\). 
\(S^{(N)}\) is not Hermitian for \(N > 2\) but the natural generalization to the
above strategy still works as expected. 
Specifically, we can use a controlled version of the cyclic shift operator,
\(S^{(N)}\), to sample an observable whose expectation value is equal to
\(Re(\tr(S^{(N)} \rho^{\otimes N}))\)~\cite{Ekert2002-gl, Brun2004-do, Cotler2019-rh}. 

Because the symmetrized observable \(O^{(N)}\) commutes with \(S^{(N)}\), it
also commutes with the observable measured by this generalization of the swap
test. 
Therefore, we can sample from an observable whose expectation value is
\(\tr(S^{(N)} O^{(N)} \rho ^{\otimes N})\) by first performing the higher-order
swap test and then a measurement of \(O^{(N)}\).

\section{Variance of the Corrected Expectation Value Estimator}
\label{app:estimator_variance}

In this section, we calculate the variance of the estimator for the
corrected expectation value obtained by applying \eq{symmetrized_O_M_ev} with
\(M=2\). 
Specifically, we consider the estimation of the expectation value of an
observable \(O\) with respect to the state \(\frac{\rho^2}{\tr(\rho^2)}\)
constructed by repeatedly measuring the operators \(S^{(2)}O^{(2)}\) and
\(S^{(2)}\),
\begin{equation}
  \frac{\tr(O \rho^2)}{\tr(\rho^2)} = \frac{\tr(O^{(2)} S^{(2)} \rho^{\otimes 2})}{\tr(S^{(2)} \rho^{\otimes 2})}.
  \label{eq:symmetrized_O_2_ev}
\end{equation}
We assume that both operators are simultaneously measured by averaging over
\(R\) repetitions of state-preparation and measurement.
This assumption applies to the measurement by diagonalzation method presented in
\sec{measurement_by_diagonalization} of the main text and also to the
ancilla-assisted measurement approach of \app{ancilla_assisted_measurement}.

The outcomes obtained from measurements of these operators are classical random
variables, and we can proceed by determining the variance of these two random
variables and their covariance. 
We begin by calculating the variance of the numerator (with respect to the state
\(\rho^{\otimes 2}\)).
\begin{align}
  \textrm{Var}(S^{(2)}O^{(2)}) &= \ev{(S^{(2)} O^{(2)})^2} - \ev{S^{(2)} O^{(2)}}^2 \\
                      &= \ev{(O^{(2)})^2} - \tr(\rho^2 O)^2 \\
                      &= \frac{1}{4}\tr((\rho \otimes \rho) (O^2 \otimes \mathbb{I} + 2 O \otimes O
                        + \mathbb{I} \otimes O^{2})) - \tr(\rho^2 O)^2 \\
                      &= \frac{1}{2}\tr(\rho O^2) + \frac{1}{2}\tr(\rho O)^2 - \tr(\rho^2 O) ^2
\end{align}
The variance of the random variable in the denominator follows by taking \(O =
\mathbb{I}\),
\begin{align}
  \textrm{Var}(S^{(2)}) &= \ev{(S^{(2)})^2} - \ev{S^{(2)}}^2 \\
               &= 1 - \tr(\rho^2)^2.
\end{align}
We'll also need the covariance between the random variables representing
measurements of the operators which estimate the numerator and the denominator.
\begin{align}
  \textrm{Cov}(S^{(2)}O^{(2)}, S^{(2)}) &= \ev{S^{(2)} O^{(2)} S^{(2)}} - \ev{S^{(2)}O^{(2)}}\ev{S^{(2)}} \\
                               &= \ev{O^{(2)}} - \tr(\rho^2 O)\tr(\rho^2) \\
                               &= \tr(\rho O) - \tr(\rho^2 O)\tr(\rho^2).
\end{align}

There isn't a closed-form expression for the variance of the ratio of two random
variables~\cite{Kendall1946-by}, but we can take the standard approximation
based on a Taylor series expansion,
\begin{equation}
  \textrm{Var}(\frac{A}{B}) \approx \frac{1}{\ev{B}^2} \textrm{Var}(A) - 2 \frac{\ev{A}}{\ev{B}^3} Cov(A, B) + \frac{\ev{A}^2}{\ev{B}^4} \textrm{Var}(B).
  \label{eq:two_random_variable_variance}
\end{equation}
We estimate the expectation values for the numerator and denominator of
\eq{symmetrized_O_2_ev} by averaging over a series of \(R\) experiments. 
This scales the variances calculated above by a factor of \(\frac{1}{R}\). 
If \(R\) is sufficiently large, then the approximation presented in
\eq{two_random_variable_variance} will be a good one.
Applying this expression to determine the variance of the estimator from
\eq{symmetrized_O_2_ev} yields
\begin{align}
  \label{eq:rho_2_estimator_variance}
  \textrm{Var(Estimator)} \approx &
                           \frac{1}{R}\Big(\frac{1}{\tr(\rho^2)^2}\big(\frac{1}{2}\tr(\rho O^2) + \frac{1}{2}\tr(\rho O)^2 - \tr(\rho^2 O)^2\big) \\
                         &- 2 \frac{\tr(\rho^2 O)}{\tr(\rho^2)^3}\big(\tr(\rho O) - \tr(\rho^2 O)\tr(\rho^2)\big) + \frac{\tr(\rho^2 O)^2}{\tr(\rho^2)^4}\big(1 - \tr(\rho^2)^2\big)\Big).
                           \nonumber
\end{align}

\section{Variance of the Proposed Collective Measurement}
\label{app:collective_measurement_variance}

In \sec{sample_efficiency}, we claimed that there exists a joint measurement on
\(2K\) copies of \(\rho\) that allows us to estimate \(\tr(O \rho^2)\) with a
lower variance than performing \(K\) copies of the basic virtual distillation
procedure in parallel. Specifically, we claimed that the operator \(\tilde{O}\)
(whose definition we reproduce below) exhibits a lower variance than the simple
alternative under certain conditions. In this appendix, we prove this claim.
First, we recall the definition,
\begin{equation}
  \tilde{O} = \frac{1}{\binom{2K}{2}}\sum_{i=1}^{2K} \sum_{j > i}\frac{1}{2}( O^\textbf{i} + O^\textbf{j})S^{{(i, j)}},
\end{equation}
where we \(S^{(i, j)}\) denotes the swap operator between subsystems \(i\) and
\(j\) and \(O\) is an arbitrary Pauli operator. 
Linearity ensures that a calculation of the expectation value of this operator
can be reduced to the virtual distillation procedure applied to two copies,
yielding
\begin{equation}
  \tr(\tilde{O} \rho^{\otimes 2K}) = \tr(O \rho^2).
  \label{eq:complicated_expression_same_expectation}
\end{equation}

We now bound the variance of measurements of this operator with
respect to the state \(\rho^{\otimes 2K}\);
\begin{align}
  \textrm{Var}(\tilde{O}) &= \ev{O^2} - \ev{O}^2 \\
                 &= \frac{1}{4\binom{K}{2}^2} \sum_{i=1}^{2K} \sum_{j > i} \sum_{a=1}^{2K} \sum_{b>a} \ev{ (O^\textbf{i} + O^\textbf{j})S^{(i, j)}(O^\textbf{a} + O^\textbf{b}) S^{(a, b)}} - \tr(O \rho^2)^2.
\end{align}
Note that we have made use of \eq{complicated_expression_same_expectation} to
replace \(\ev{O}^2\) by \(\tr(O \rho^2)^2\).
We proceed by breaking the summation up into three cases. 
In the first case, \(i = a\) and \(j = b\). 
In the second case, there are only three distinct values amongst the indices
\(i, j, a, b\). 
In the fourth case, all four of the indices take distinct values.

Consider the first case where \(i=a\) and \(j=b\). 
Then we can simplify and bound the sum,
\begin{align}
  & \frac{1}{4\binom{2K}{2}^2} \sum_{i=1}^{2K} \sum_{j > i} \ev{ (O^\textbf{i} + O^\textbf{j})S^{(i,
    j)}(O^\textbf{i} + O^\textbf{j}) S^{(i, j)}} =
    \frac{1}{4\binom{2K}{2}^2} \sum_{i=1}^{2K} \sum_{j > i} \ev{2 + O^\textbf{i} O^\textbf{j} + O^\textbf{j} O^\textbf{i}} \\
  & \leq \frac{1}{\binom{K}{2}}.
\end{align}
Here we have used the properties that \(S^{(i, j)}\), \(O^{\textbf{i}}\), and
\(O^{\textbf{j}}\) are self-inverse, and that
\(S^{(i, j)}\) commutes with \((O^\textbf{i} + O^{\textbf{j}})\).

Next, let's consider the third case, where all four indices take distinct
values. 
Then the operators \((O^\textbf{i} + O^\textbf{j})S^{(i, j)}\) and
\((O^\textbf{a} + O^\textbf{b})S^{(a, b)}\) act on distinct pairs of systems. 
Therefore, their expectation values with respect to the tensor product
\(\rho^{\otimes 2K}\) can be evaluated separately and multiplied together. 
We can use this fact to simplify and bound this component of the sum,
\begin{align}
  & \frac{1}{4\binom{2K}{2}^2} \sum_{i=1}^{2K} \sum_{j > i} \sum_{a=1, a\neq i, a\neq j}^{2K} \sum_{b>a, b\neq i, b\neq j} \ev{ (O^\textbf{i} + O^\textbf{j})S^{(i, j)}(O^\textbf{a} + O^\textbf{b}) S^{(a, b)}}\\
  =& \frac{1}{4\binom{2K}{2}^2} \sum_{i=1}^{2K} \sum_{j > i} \sum_{a=1, a\neq i,
     a\neq j}^{2K} \sum_{b>a, b\neq i, b\neq j} \ev{ (O^\textbf{i} + O^\textbf{j})S^{(i, j)}}\ev{(O^\textbf{a}
     + O^\textbf{b}) S^{(a, b)}} \\
  =& \frac{1}{\binom{2K}{2}^2}
     \sum_{i=1}^{2K} \sum_{j > i}
     \sum_{a=1, a\neq i, a\neq
     j}^{2K} \sum_{b>a, b\neq i,
     b\neq j}  \tr(O \rho^2)^2\\
  & = \frac{1}{\binom{2K}{2}} \binom{2K - 2}{2} \tr(O \rho^2)^2 < \tr(O\rho^2)^2.
\end{align}

\begin{figure}[t]
  \includegraphics[width=.75\textwidth]{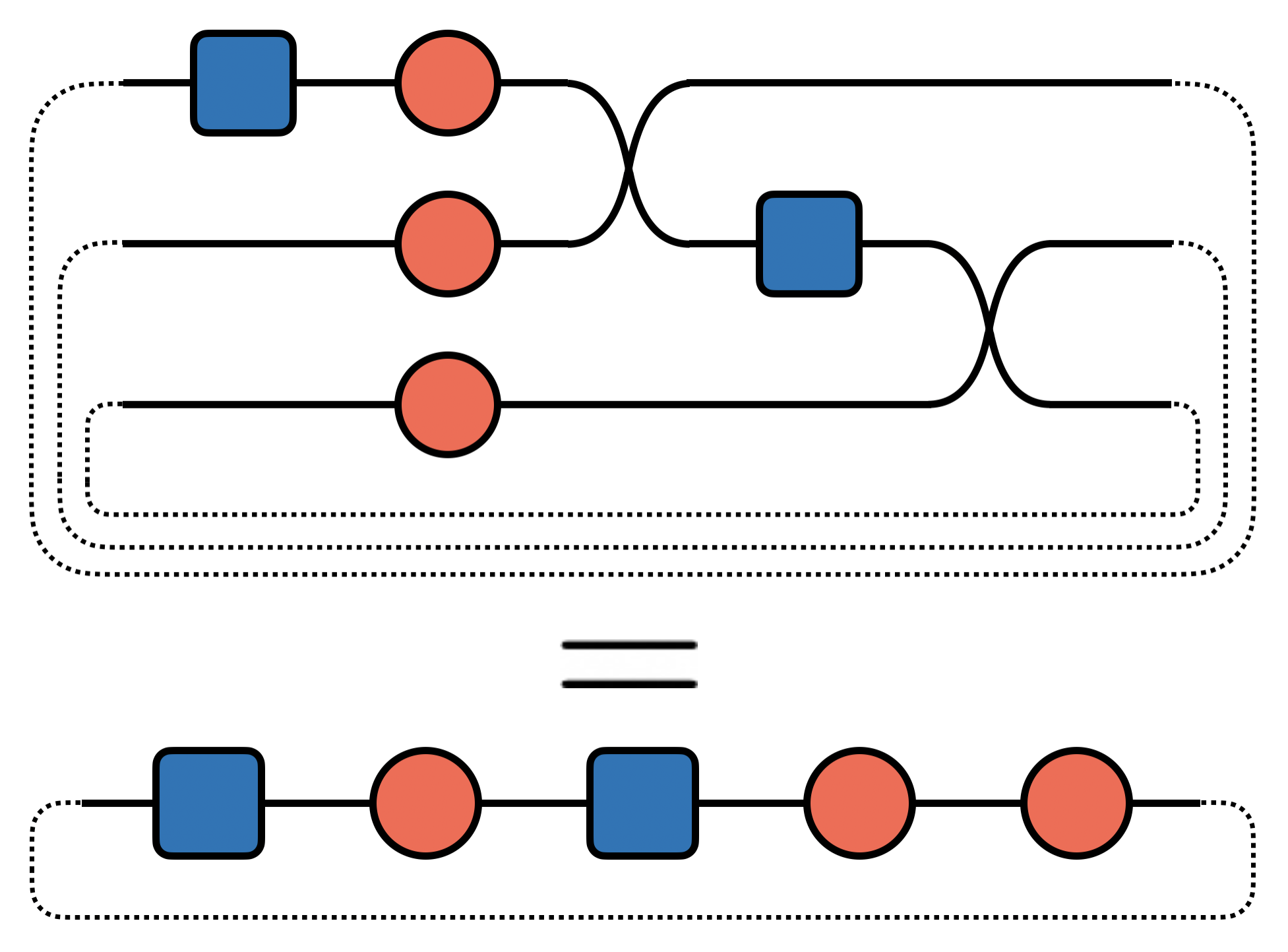}
  \caption{A diagrammatic proof of \eq{second_case_simplification}.}
  \label{fig:tn_diagram_second_case_proof}
\end{figure}

Now we treat the case where the indices take three distinct values. 
Actually, there are four sub-cases here. 
We could have any one of the four possibilities, \(i=a\), \(i=b\), \(j=a\), or
\(j=b\). 
We work out the details for the \(i=a\) case below, noting that the others
behave symmetrically.
\begin{align}
  & \ev{ (O^\textbf{i} + O^\textbf{j})S^{(i,j)}(O^\textbf{i} + O^\textbf{b}) S^{(i, b)}} \\
  = & \ev{(O^\textbf{i} + O^\textbf{j}) (O^\textbf{j} + O^\textbf{b}) S^{(i, j)}S^{(i, b)}} \\
  = & \ev{(O^\textbf{i}O^\textbf{j} + O^\textbf{i}O^\textbf{b} + O^\textbf{j}O^\textbf{b} + 1) S^{(i, j)}S^{(i, b)}}.
\end{align}
Here we have used the property that \(O^\textbf{i}\) is self-inverse. 
Now we note that the product \(S^{(i, j)}S^{(i, b)}\) is a cyclic shift between
the subsystems \(i, j, b\) and that this product commutes with the operator
\((O^\textbf{i}O^\textbf{j} + O^\textbf{i}O^\textbf{b} +
O^\textbf{j}O^\textbf{b} + 1)\). 
Computation using a tensor network diagram (see
\fig{tn_diagram_second_case_proof}) establishes that
\begin{align}
  & \ev{(O^\textbf{i}O^\textbf{j} + O^\textbf{i}O^\textbf{b} + O^\textbf{j}O^\textbf{b} + 1) S^{(i, j)}S^{(i, b)}} \\
  &= 3\tr(O \rho O \rho^2) + \tr(\rho^3).
    \label{eq:second_case_simplification}
\end{align}

In order to bound this quantity, let us denote the projector onto the \(+1\)
eigenspace of \(O\) by \(P_+\) and the projector onto the \(-1\) eigenspace by
\(P_-\). 
Then we can expand \eq{second_case_simplification} in terms of these projectors,
yielding
\begin{align}
  & 3\tr(O \rho O \rho^2) + \tr(\rho^3) \\
  &=3 \tr(P_+ \rho P_+ \rho^2) - 3 \tr(P_- \rho P_+ \rho^2) - 3 \tr(P_+ \rho P_-
    \rho^2) + 3 \tr(P_- \rho P_- \rho^2) + \tr(\rho^3) \\
  &\leq 7 \tr(\rho^3).
\end{align}
For simplicity, let us define the indicator function
\begin{equation}
  W(i, j, a, b) =
\begin{cases}
  1 & \text{if $$i, j, a, b$$ take exactly three distinct values} \\
  0 & \text{otherwise}
\end{cases}
\end{equation}
Now we can bound the component of the sum where the indices take three distinct
values.
For each of the \(\binom{2K}{2}\) values of \(i\) and \(j\), there are exactly
\(4K - 4\) values of \(a\) and \(b\) such that \(I(i, j, a, b) = 1\).
Therefore, we have
\begin{align}
  & \frac{1}{4\binom{2K}{2}^2} \sum_{i=1}^{2K} \sum_{j > i} \sum_{a=1}^{2K} \sum_{b>a} \ev{ (O^\textbf{i} + O^\textbf{j})S^{(i, j)}(O^\textbf{a} + O^\textbf{b}) S^{(a, b)}} I(i, j, a, b) \\
  & \leq \frac{7}{4\binom{2K}{2}^2} \sum_{i=1}^{2K} \sum_{j > i} \sum_{a=1}^{2K} \sum_{b>a} I(i, j, a, b) \tr(\rho^3).
  \\
  & \leq \frac{7(4K - 4)}{4\binom{2K}{2}} \tr(\rho^3) \\
  & \leq \frac{7(2K - 2)}{2K(2K - 1)} \tr(\rho^3).
\end{align}

Now we can combine the bounds from the three different cases and simplify the
expression for the variance to yield
\begin{equation}
  \textrm{Var}(\tilde{O}) \leq \frac{1 + 7 (K-1) \tr(\rho^3)}{K(2K-1)}.
\end{equation}
Note that we have simplified by subtracting the \(\tr(O\rho^2)^2\) term that
arose from evaluating \(\ev{O}^2\).

\section{Details Regarding the Numerical Experiments with Scrambling Circuits}
\label{app:scrambling_circuit_details}

In \sec{scrambling_circuits} we briefly described the random circuits that we
simulated to produce \fig{tensor_product_scrambler_scaling},
\fig{1d_scrambler_depth_scaling}, and \fig{1d_scrambler_error_rate_scaling}.
Here we expand upon that description.

The first class of circuits are essentially the one-dimensional analogues of the
random circuits of Ref.~\citenum{Arute2019-jy}.
They are constructed by alternating between layers of two-qubit gates and
single-qubit gates. 
The two-qubit gate layers consist of `Sycamore gates,' two-qubit gates that
enact the unitary
\begin{equation}
\begin{bmatrix}
  1  & 0  & 0  & 0 \\
  0  & 0  & -i  & 0 \\
  0  & -i  & 0  & 0 \\
  0 & 0 & 0 & e^{\frac{-i\pi}{6}}
\end{bmatrix}.
\end{equation}
The two-qubit gate layers themselves alternate between layers that have Sycamore
gates on every even-odd pair and every odd-even pair.
The ensemble of random circuits is defined by adding a layer of randomly chosen
single-qubit gates between every layer of two-qubit gates in this fixed
structure.
These single-qubit gates are drawn from the set
\begin{equation}
  \Big\{ X,\quad Y,\quad Z,\quad \sqrt{X},\quad \sqrt{Y},\quad \sqrt{Z} \Big\},
\end{equation}
with the square root of a gate being defined by taking the principal square root
in the eigenbasis of the gate.

The second class of random circuits is exactly the same as the first class,
except that we effectively remove the two-qubit gates by replacing the Sycamore
gates with the identity. 
When we perform noisy simulations of these circuits we apply the single-qubit
depolarizing channels in the same locations despite the lack of two-qubit gates. 
Note that because this second class of random circuits contains only
single-qubit gates, the applications of the single-qubit depolarizing noise
channels can be commuted to the end of the circuit and combined together.

We can therefore write an analytical expression for the density matrix at the
end of the noisy computation in terms of the noiseless single qubit states
\(\{\ket{\phi_i}\}\), the single-qubit depolarizing probability, \(p\), and the
depth of the circuit, \(D\).
We define the single-qubit depolarizing channel in the usual way,
\begin{equation}
  \Delta(\rho) = (1-p) \rho + \frac{p}{3} + \frac{p}{3}\big(X\rho X + Y\rho Y + Z\rho Z \big).
  \label{eq:single_qubit_depolarizing}
\end{equation}
An equivalent formulation which will be useful for our purposes is
\begin{equation}
  \Delta(\rho) = (1 - \frac{4}{3}p) \rho + \frac{4}{3}p\frac{\mathbb{I}}{2}.
  \label{eq:single_qubit_depolarizing_totally_mixed}
\end{equation}
For a pure state \(\rho = \ketbra{\phi}\), we have
\begin{equation}
  \Delta(\ketbra{\phi}) = (1 - \frac{2}{3}p) \ketbra{\phi} + \frac{2}{3}p \ketbra{\phi^\perp},
\end{equation}
where \(\ket{\phi^\perp}\) is orthogonal to \(\ket{\phi}\).
\eq{single_qubit_depolarizing_totally_mixed} allows us to easily analyze \(D\)
repeated applications of the channel,
\begin{equation}
  \Delta(\rho)^D = (1 - \frac{4}{3}p)^D \rho + (1- (1 - \frac{4}{3}p)^D)\frac{\mathbb{I}}{2},
\end{equation}
which tells us that \(D\) applications of the channel with an error rate \(p\)
are equivalent to a single application with error rate
\begin{equation}
  \tilde{p} = \frac{3}{4} - \frac{3}{4}(1- \frac{4}{3} p)^D.
  \label{eq:renormalized_p}
\end{equation}

We can now write an expression for the density matrix at the end of the
computation,
\begin{equation}
  \bigotimes_{i=1}^{N} (1 - \frac{2}{3}\tilde{p}_{(i)}) \ketbra{\phi_i} + \frac{2}{3}\tilde{p}_{(i)} \ketbra{\phi_i^\perp}.
  \label{eq:final_state_non_scrambling}
\end{equation}
Here the effective single-qubit depolarizing probability \(\tilde{p}_{(i)}\)
depends on \(i\) because the qubits at the end of the circuit are only subject
to \(\frac{D}{2}\) applications of the single-qubit depolarizing channel instead
of the \(D\) that are applied to qubits in the bulk. 
Therefore, employing \eq{renormalized_p}, we have
\begin{equation}
  \tilde{p}_{(i)}= \left\{ 
    \begin{array}{cll}
      \frac{3}{4} - \frac{3}{4}(1- \frac{4}{3} p)^\frac{D}{2} && \text{for } i = 1 \text{ or } i = N, \\[8pt]
      \frac{3}{4} - \frac{3}{4}(1- \frac{4}{3} p)^D && \text{for } 2 \leq i \leq N - 1.
    \end{array} \right.
  \label{eq:renormalized_prob_non_scrambling}
\end{equation}
We can observe a few things from the combination of
\eq{final_state_non_scrambling} and \eq{renormalized_prob_non_scrambling}. 
First of all, for any value of \(p\) smaller than the maximal \(p =
\frac{3}{4}\), the dominant eigenvector of the density matrix is exactly the
ideal state (\(\ket{\phi_1} \otimes \ket{\phi_2} \otimes \cdots \otimes
\ket{\phi_N}\)). 
Secondly, when \(p\) is small, the next largest eigenvectors of the density
matrix will correspond to states with an error on a single qubit. 
Neglecting the subtlety caused by the two different values of \(\tilde{p}\), we
can see that there will be \(N\) such eigenvectors with eigenvalues \(\approx
\frac{2}{3} \tilde{P}\). 
After these states, there will be \(\binom{N}{2}\) eigenvectors corresponding
two states with two errors. 
This distribution doesn't exactly match the phenomenological noise model we
assumed in \sec{orthogonal_errors}, but it is qualitatively similar.

\section{Interplay with the surface code}\label{app:surface_code}
While the majority of this work has focused on the NISQ regime, one interesting
question to ask is what role this approach can play after some degree of quantum
error correction has been deployed. 
To explore this connection concretely, we imagine that a fault-tolerant surface
code quantum computer is in operation with typical gate error rates on the order
of $10^{-3}$. 
For such systems, it has been determined numerically~\cite{Wang2010-wq} that in conjunction with a
minimum-weight perfect matching decoder, the error rate of a surface code cycle
is roughly
\begin{align}
  \epsilon_c = 10^{-(d+3)/2}
\end{align}
where $d$ is the distance of the code protecting a given logical qubit. 
Including data and measurement qubits, the translation to physical qubits for a
given distance is $n = 2d^2$. 
In order to guarantee protection against measurement (or time-like) errors up to
the same distance without using an excessive number of qubits, one must repeat
measurements a number of cycles proportional to $d$. 
For operations like gates, additional cycles are required to perform the
operation as well. 
For example, many simple Clifford operations may be done in a number of cycles
like $2d$ using lattice surgery techniques~\cite{Horsman2012-vd,Litinski2019-dg}. 
However more complicated arbitrary rotations like the ones used in many NISQ
algorithms, must first be broken down into a combination of discrete gates like
T and Clifford gates through gate synthesis, then those $T$ gates consume on the
order of $20d$ cycles for successful distillation. 
Using a coarse synthesis heuristic of roughly $10$ T gates and $10$ Clifford
gates per arbitrary rotation, this gives approximately $200d$ cycles of the
surface code per arbitrary rotation.
If we average this coarsely, assuming an even distribution of Clifford and
arbitrary rotations, as is common in NISQ approaches, then we can model on
average that we require $100d$ rounds of the surface code per gate we wish to
perform.  While these numbers are subject to refinement and improvements, we believe
these can approximately serve to understand where an advantageous combination
of methods might occur.
As is common with early circuit implementations, we may assume that the gates
are densely packed so that additional idling error is minimal. 
With these assumptions, using $n$ physical qubits to represent a single logical
qubit, we have an overall fidelity of
\begin{align}
  f_1 = \left( 1 - 10^{-(\sqrt{n/2} + 3)/2} \right)^{100 \sqrt{n/2} G}
\end{align}
where $G$ is the number of gates performed. 
The virtual distillation technique uses twice the qubits to effect a large constant
factor improvement over the bare circuit. 
Hence the apt comparison here is to consider the use of twice the qubits within
the virtual distillation technique, or to use twice the qubits to improve the
distance of the surface code logical qubit. 
An asymptotic analysis would argue that the exponential returns of the error
correcting code would be the best option, however the overhead can mean that a
large constant factor could make virtual distillation advantageous in some cases. 
To examine this, consider the error rate of $G$ gates in the surface code using
twice the qubits
\begin{align}
  f_2 = \left( 1 - 10^{-(\sqrt{n} + 3)/2} \right)^{100 \sqrt{n} G}.
\end{align}
A strict analysis would consider that we need to round these to integer distances, 
but for this approximate analysis, this should suffice.  
If we consider the ratio between the implied error rates $c_s = (1-f_1) /
(1-f_2)$, we can find the required constant factor for a given number of qubits
per logical qubit and number of gates to make using virtual distillation
advantageous. 
Past a certain number of qubits, this constant factor is enormous, but we find
that up to distance $10-15$ the empirical improvements measured in the text are
sufficient to justify the use of virtual distillation in place of additional qubit
protection.  
In particular, at distance $10$ with $n=200$ physical qubits per logical qubit, 
performing $G=1000$ gates on the logical qubit, the respective error rates 
are about are $10^{-1}$ and $10^{-5}$, and hence a constant improvement is 
about of about $10^4$ is sufficient to justify the use of virtual distillation, 
which is on par with some improvements seen in the main text.  
To be fair, one might argue that an overall error rate of $10^{-5}$ would already suffice, 
and by a distance of $15$, the required improvement is on the order of $10^7$ 
which is at the upper limit of what we imagine can be achieved with this technique.
At smaller distances and numbers of gates, the required constant factors decrease as well.
If we assume that we will consistently push the limits of the number of logical
qubits we use, reducing the number of physical qubits per logical qubit
available, this may imply a regime in early fault tolerance where this technique
is applicable.
Further studies will be required to identify precisely under what conditions
this may be the case.

\section{Virtual Distillation Applied to Distinct States}
\label{app:different_noise_models}

In the main body of this paper, we applied virtual distillation to a variety of
systems under the assumption that we had access to multiple copies of the same
noisy state.
In reality, even if we attempt to perform the same computation multiple times in
parallel, the noise experienced by each copy will not be identical.
We now consider the case where we apply virtual distillation to two distinct states,
\(\rho_A\) and \(\rho_B\).
It's straightforward to show that we then effectively measure expectation values
with respect to the state
\begin{equation}
    \rho_\textrm{eff} = \frac{\rho_A\rho_B}{\tr(\rho_A\rho_B)}.
\end{equation}
This can be demonstrated in various ways, but the most straightforward is to use a diagrammatic proof of the kind we illustrated in \fig{cyclic_shift_trace_diagram}.
Note that we still rely on the important assumption that the two copies are
unentangled prior to virtual distillation.

Furthermore, some additional care must be taken when virtual distillation is applied to two distinct states. In particular, \(\rho_\textrm{eff}\) is not guaranteed to be a positive semidefinite matrix and does not, in general, correspond to a valid quantum state. This is especially important to note in the context of variational algorithms, where care would have to be taken to ensure that a non-variational answer is not achieved due to the non-phyiscality of \(\rho_\textrm{eff}\). It will be an interesting direction for future work to address this potential challenge.

\begin{figure}[t]
  \centering
  \includegraphics[width=.48\textwidth]{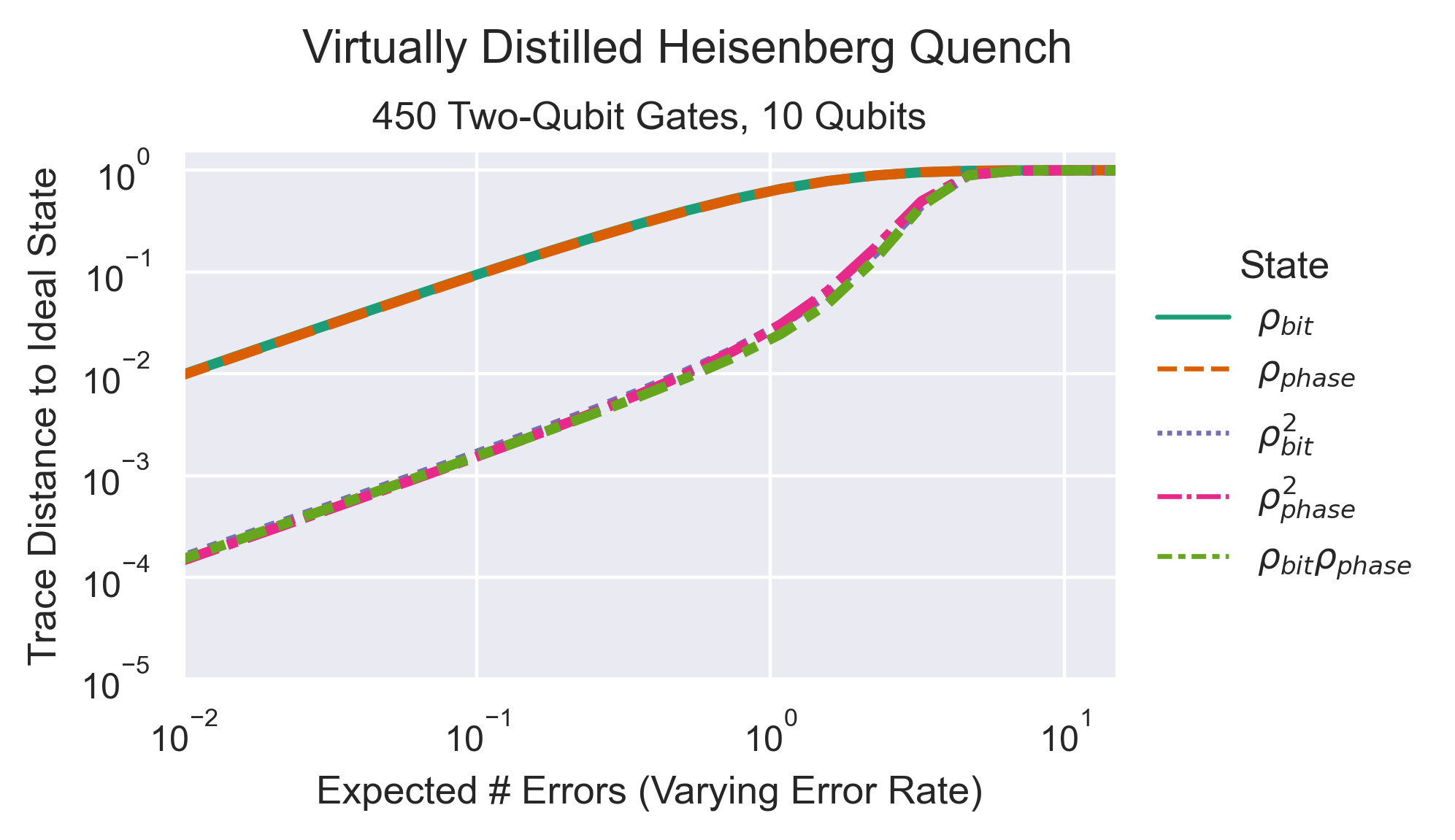}
  \caption{The error in the unmitigated noisy states, and the states accessed by
    virtual distillation, for the same \(10\) qubit Heisenberg evolution under
    two different noise models.
    We consider a bit-flip error model (\(\rho_{bit}\), teal curve) and
    phase-flip error model (\(\rho_{phase}\), orange dashed curve). 
    We show the error for virtual distillation applied in the usual way to two
    identical copies of each noisy state (\(\rho_{bit}^2\) and
    \(\rho_{phase}^2\), purple dotted curve and pink dashed curve). 
    We also consider the error when virtual distillation is applied to one copy
    of each state state (\(\rho_{bit}\rho_{phase}\), green dotted curve). 
    We quantify the error in terms of the trace distance to the state obtained
    from noiseless evolution as a function of the expected number of
    single-qubit gate errors. 
    The expected number of errors is varied by changing the error rate per-gate,
    fixing the number of two-qubit gates to be \(450\). 
    Ultimately, we find that the performance of virtual distillation is barely
    affected when the two input states are generated with different noise
    processes.}
    \label{fig:heisenberg_two_different_error_rate_scaling}
\end{figure}

In order to explore the impact of virtually distilling two different states
together, we present an additional simulation of the Heisenberg evolution that
we considered in \sec{heisenberg_evolution}. 
In \fig{heisenberg_two_different_error_rate_scaling}, rather than employing the
single-qubit depolarizing noise model we used previously, we simulate \(\rho_A
= \rho_{bit}\) using an analogous application of a bit-flip error channel, while
using a phase-flip channel for \(\rho_B = \rho_{phase}\). 
We find that the error in the effective state accessed by performing virtual
distillation to these two different states closely tracks the error we obtain by
using two copies of either state individually.

\section{Performance Under Amplitude Damping and Dephasing Noise}
\label{app:amp_damp_dephasing}

In \sec{numerical_performance}, we used numerical simulations of three different
systems to explore the performance of virtual distillation under a noise model
where we applied a single-qubit depolarizing noise channel after each two-qubit
gate.
It is natural to ask how virtual distllation performs under a more realistic
approximation to the actual noise experienced on NISQ hardware.
In this appendix, we shed light on this question by considering a more
physically motivated model of stochastic errors in two-qubit gates based on
single-qubit amplitude damping and dephasing channels.
As we did in the main text, we follow each two-qubit gate the application of a
single-qubit error channel to each qubit. 
Here that single qubit error channel is the concatenation of the dephasing and
amplitude damping channels described below.

Our amplitude damping channel is parameterized by \(\gamma_1\), which represents
the probability that a qubit in the \(\ket{1}\) state will spontaneously decay
into the \(\ket{0}\) state. 
We can express this channel in a standard (but non-unique) way using the Kraus
operators,
\begin{equation}
  \label{eq:amp_damp}
  M_0 =
  \begin{bmatrix}
    1 & 0 \\
    0 & \sqrt{1 - \gamma_1}
  \end{bmatrix}, \;\;\;
  M_1 = 
  \begin{bmatrix}
    0 & \sqrt{\gamma_1} \\
    0 & 0
  \end{bmatrix}.
\end{equation}
Our dephasing channel is parameterized by \(\gamma_2\), which represents the
probability that an unintended interaction between a qubit and its environment
entangles the two, effectively performing a measurement in the
computational basis. One standard way of expressing this channel in terms of
Kraus operators is given below,
\begin{equation}
  \label{eq:dephasing}
  M_0 =
  \sqrt{1-\gamma_2}
  \begin{bmatrix}
    1 & 0 \\
    0 & 1
  \end{bmatrix}, \;\;\;
  M_1 = 
  \begin{bmatrix}
    \sqrt{\gamma_2} & 0 \\
    0 & 1
  \end{bmatrix}, \;\;\;
  M_2 = 
  \begin{bmatrix}
    0 & 0 \\
    0 & \sqrt{\gamma_2}
  \end{bmatrix}.
\end{equation}
It can also be convenient to use the equivalent representation,
\begin{equation}
  \label{eq:dephasing_cirq}
  M_0 =
  \begin{bmatrix}
    1 & 0 \\
    0 & \sqrt{1 - \tilde{\gamma}}
  \end{bmatrix}, \;\;\;
  M_1 = 
  \begin{bmatrix}
    0 & 0 \\
    0 & \sqrt{\tilde{\gamma}}
  \end{bmatrix}, \;\;\;
  \tilde{\gamma} = 2\gamma_2 - \gamma_2^2.
\end{equation}

The parameters \(\gamma_1\) and \(\gamma_2\) can be related to the related to
the \(T_1\) and \(T_2\) times frequently used to characterize relaxation in
two-level systems~\cite{Preskill1998-cb, Krantz2019-gl}.
In our simulations, we consider \(\lambda_1 = \lambda_2\), a choice with a
physical model where the \(T_1\) and \(T_2\) times are
comparable.
This allows us to plot the error (quantified by the trace distance to the state
that would be obtained in the absence of noise) as a function of the expected
number of error events. 
The expected number of errors (\(E\)) can be expressed in terms of the number of
two-qubit gates in the ciruit (\(G\)) and the two error probabilities,
\begin{equation}
  \label{eq:expected_num_errors_dephasing}
  E = 2G(\lambda_1 + \lambda_2).
\end{equation}

\begin{figure}
  \centering
  \includegraphics[width=.48\textwidth]{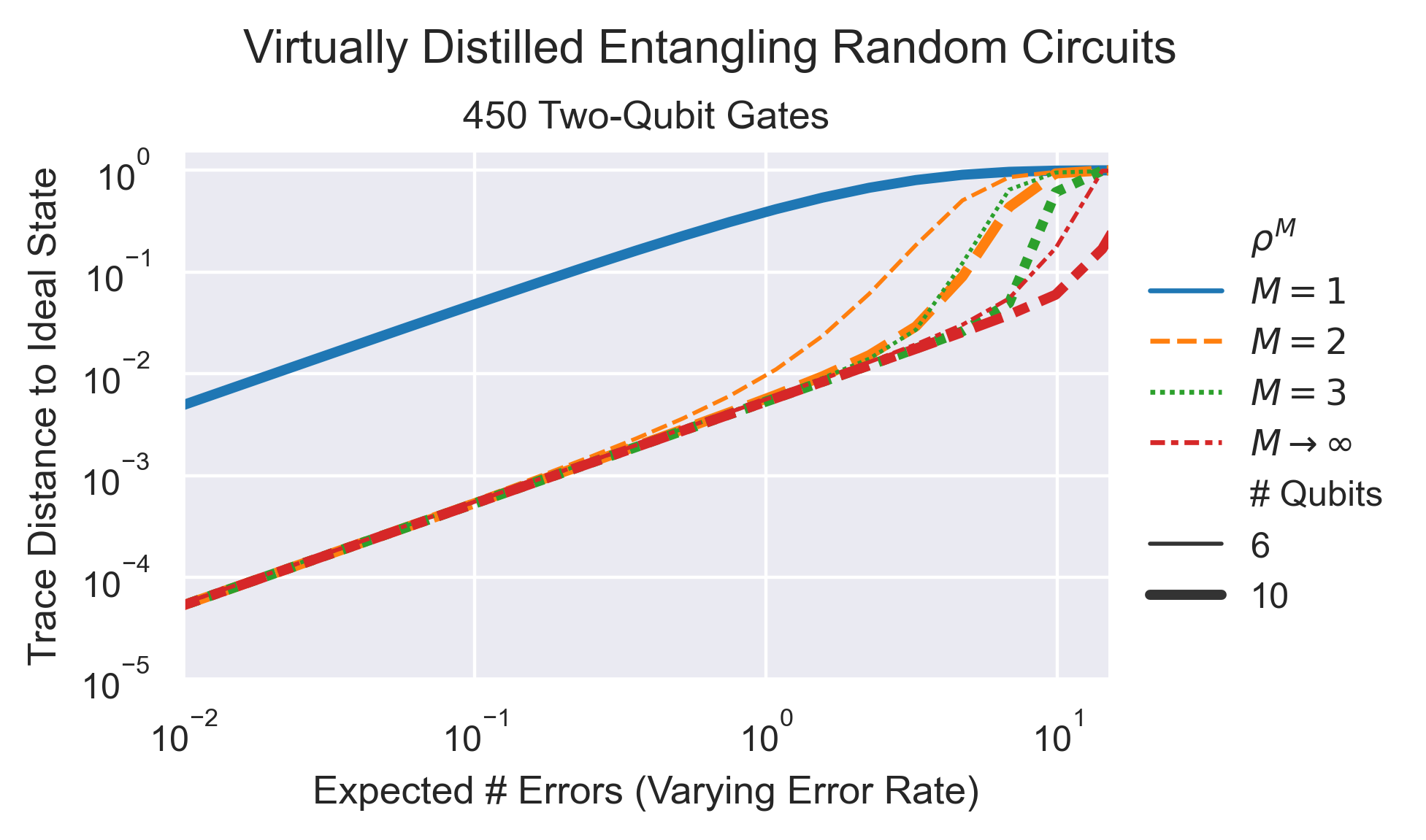}
  \caption{ The error in the unmitigated noisy states (\(M=1\)), the states
    accessed by virtual distillation (\(M=2,3\)), and the dominant eigenvector
    of the density matrices (\(M \rightarrow \infty\)) for a variety of
    entangling random circuits (described in \sec{scrambling_circuits}).
    We plot the trace distance to the state obtained from noiseless evolution,
    as a function of the expected number of single-qubit amplitude damping and
    dephasing errors, for 6 and 10 qubit systems (demarcated by the thickness of
    the symbols).
    We vary the expected number of errors by varying the error rate per-gate,
    fixing the number of two-qubit gates to be \(450\).
    This figure is constructed to parallel
    \fig{1d_scrambler_error_rate_scaling}, except that we consider an error
    model based on single-qubit amplitude damping and dephasing rather than a
    depolarizing channel.
    We see that the system size dependence vanishes at low error rates in this
    case, in contrast with the data from \fig{1d_scrambler_error_rate_scaling}.
  }
    \label{fig:1d_scrambler_error_rate_scaling_realistic}
  \end{figure}
  
\begin{figure}
  \centering
  \includegraphics[width=.48\textwidth]{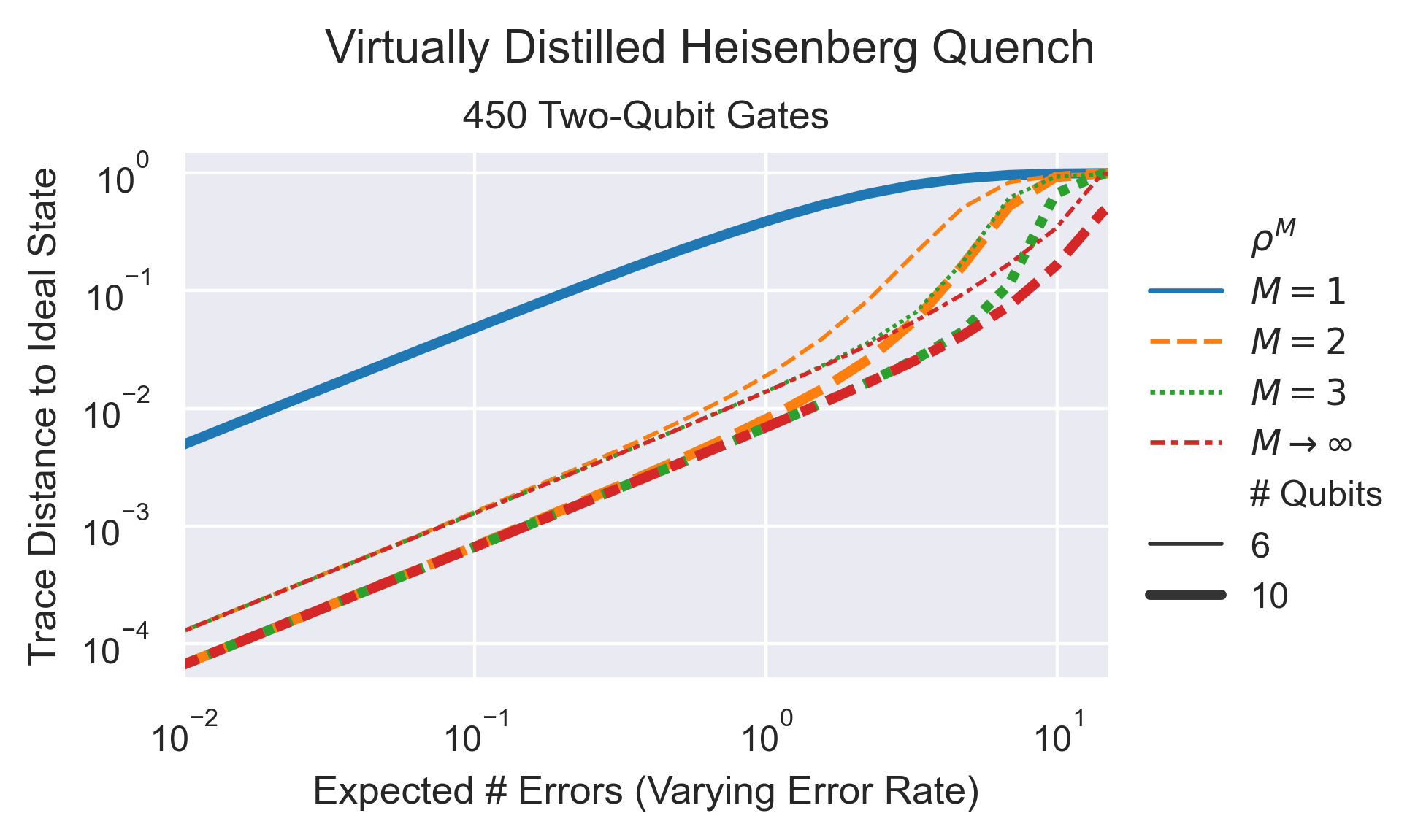}
  \caption{The error in the unmitigated noisy states (\(M=1\)), the states
    accessed by virtual distillation (\(M=2,3\)), and the dominant eigenvectors
    of the density matrices (\(M \rightarrow \infty\)), for states generated by
    the Trotterized time evolution of a Heisenberg model (described in
    \sec{heisenberg_evolution}).
    We plot the trace distance to the state obtained from noiseless evolution as
    a function of the expected number of single-qubit amplitude damping and
    dephasing errors.
    The expected number of errors is varied by changing the error rate per-gate,
    fixing the number of two-qubit gates to be \(450\).
    We show this data for 6 and 10 qubit systems (differentiated by the size of
    the markers).
    This figure mirrors \fig{heisenberg_error_rate_scaling}, except that we
    consider an error model based on single-qubit amplitude damping and
    dephasing rather than depolarizing noise.
    The two figures display similar behaviour, indicating that our conclusions
    about virtual distillation have some robustness to changes in the noise
    model.
  }
  \label{fig:heisenberg_error_rate_scaling_realistic}
\end{figure}

In \fig{1d_scrambler_error_rate_scaling_realistic} and
\fig{heisenberg_error_rate_scaling_realistic} we present the results of two sets
of simulations under this error model.
\fig{1d_scrambler_error_rate_scaling_realistic} closely follows
\fig{1d_scrambler_error_rate_scaling} from the main text, examining the
performance of virtual distillation applied to a random circuit on a
one-dimensional array of qubits (see \sec{scrambling_circuits}). 
Likewise, \fig{heisenberg_error_rate_scaling_realistic} considers the same
Heisenberg evolution treated in \fig{heisenberg_error_rate_scaling} and
described in \sec{heisenberg_evolution}.
In both cases, we show the error in the unmitigated noisy state, the states
accessed by virtual distillation with \(M=2\) and \(M=3\) copies, and the
dominant eigenvector of the noisy density matrix. 
We quantify the error in terms of the trace distance to the ideal state that
would be obtained in the absence of noise, plotting this trace distance as a
function of the expected number of errors.

Comparing \fig{1d_scrambler_error_rate_scaling_realistic} and
\fig{heisenberg_error_rate_scaling_realistic} with
\fig{1d_scrambler_error_rate_scaling} and \fig{heisenberg_error_rate_scaling}
from the main text, we note a few things. 
First of all, the qualitative behaviour of virtual distillation is largely
unaffected by the change in the noise model. 
When the expected number of errors is small enough we still see that virtual
distillation decreases the error by orders of magnitude. 
The performance is still limited by the drift in the dominant eigenvector and we
find that in many regimes \(M=2\) copies is sufficient to maximize the potential
benefit. 
The largest differences are observed when one compares the two figures that
analyze the performance of virtual distillation on random circuits
(\fig{1d_scrambler_error_rate_scaling_realistic} and
\fig{1d_scrambler_error_rate_scaling}).
In this case, the potential benefit of virtual distillation is smaller (roughly
two orders of magnitude instead of three for the ten qubit system) under the
noise model based on amplitude damping and dephasing.
Additionally, and perhaps more interestingly, we see that the dependence on
system size mostly vanishes for the random circuits when we switch to the noise
model considered in this appendix, while it persists when we consider the
Heisenberg evolution.

\end{document}